\documentclass[12pt]{article}
\usepackage{graphicx} % Required for inserting images
\usepackage{subcaption}
\usepackage{amsmath,amssymb,xcolor,physics}
\usepackage{url}
\usepackage{ulem}
\usepackage{hyperref}
\usepackage{soul}  % for strikethrough text \st
\usepackage{tikz}
\usepackage{pgfplots}
\usetikzlibrary{fadings}
\pgfplotsset{compat=1.18}

\hypersetup{
    colorlinks=true,   % Set to false for boxed links
    linkcolor=blue,    % Color for internal links
    citecolor=red,     % Color for citations
    filecolor=magenta, % Color for file links
    urlcolor=cyan      % Color for URLs
}

\allowdisplaybreaks[4]

\begin{document}

\begin{center}
{\large A universal framework for the quantum simulation of Yang--Mills theory }
\end{center}
\vspace{0.1cm}
\vspace{0.1cm}
\begin{center}

Jad C.~Halimeh$^a$, Masanori Hanada$^b$, Shunji Matsuura$^c$,\\
\vspace{2mm}
Franco Nori$^d$, Enrico Rinaldi$^e$, Andreas Sch\"{a}fer$^f$ 

\end{center}
\vspace{0.3cm}
{\small
\begin{center}
$^a$ Max Planck Institute of Quantum Optics, 85748 Garching, Germany
\\
\vspace{1mm}
$^a$Department of Physics and Arnold Sommerfeld Center for Theoretical Physics (ASC)\\
Ludwig Maximilian University of Munich, 80333 Munich, Germany\\
\vspace{1mm}
$^a$Munich Center for Quantum Science and Technology (MCQST), 80799 Munich, Germany
\\
\vspace{1mm}
$^b$School of Mathematical Sciences, Queen Mary University of London\\
Mile End Road, London, E1 4NS, United Kingdom\\
\vspace{1mm}
$^b$qBraid Co., Harper Court 5235, Chicago, IL 60615, United States\\
\vspace{1mm}
$^{c,e}$Interdisciplinary Theoretical \& Mathematical Science Program (iTHEMS)\\
RIKEN, Wako, Saitama 351-0198, Japan\\
\vspace{1mm}
$^c$Department of Electrical and Computer Engineering,
University of British Columbia\\
Vancouver, BC V6T 1Z4, Canada\\
\vspace{1mm}
$^c$Department of Physics, University of Guelph, ON N1G 1Y2, Canada \\
\vspace{1mm}
$^c$Center for Mathematical Science and Advanced Technology, Japan Agency for Marine-Earth Science and Technology,
Yokohama 236-0001, Japan \\
\vspace{1mm}
$^{d,e}$Center for Quantum Computing (RQC), RIKEN,  
Wako, Saitama 351-0198, Japan\\
\vspace{1mm}
$^d$Physics Department, University of Michigan, Ann Arbor, MI 48109, United States\\
\vspace{1mm}
$^e$Quantinuum K.K., Otemachi Financial City Grand Cube 3F\\
1-9-2 Otemachi, Chiyoda-ku, Tokyo, Japan
\\
\vspace{1mm}
$^f$Institute of Theoretical Physics, University of Regensburg\\ 
Universit\"{a}tsstrasse 31, D-93053 Regensburg, Germany
\end{center}
}

\newpage
\vspace{0.5cm}

\begin{center}
  {\bf Abstract}
\end{center}
We provide a universal framework for the quantum simulation of SU($N$) Yang--Mills theories on fault-tolerant digital quantum computers adopting the orbifold lattice formulation. 
As warm-up examples, we also consider simple models, including scalar field theory and the Yang--Mills matrix model, to illustrate the universality of our formulation, which shows up in the fact that the truncated Hamiltonian can be expressed in the same simple form for any $N$, any dimension, and any lattice size, in stark contrast to the popular approach based on the Kogut--Susskind formulation. 
In all these cases, the truncated Hamiltonian can be programmed on a quantum computer using only standard tools well-established in the field of quantum computation. 
As a concrete application of this universal framework, we consider Hamiltonian time evolution by Suzuki--Trotter decomposition. 
This turns out to be a straightforward task due to the simplicity of the truncated Hamiltonian. 
We also provide a simple circuit structure that contains only CNOT and one-qubit gates, independent of the details of the theory investigated.

\newpage
\tableofcontents
%%%%%%%%%%%%%%%%%%%%%%
%%%%%%%%%%%%%%%%%%%%%%
\section{Introduction}
%%%%%%%%%%%%%%%%%%%%%%
%%%%%%%%%%%%%%%%%%%%%%

Imagine waking up from an unexpectedly long sleep to find that fault-tolerant quantum computers have become a reality.\footnote{This sleep might be as short as $5$ years or as long as $O(10)$ years, depending on the quantum computing company of your choice.}
Could you now simulate Quantum Chromodynamics (QCD)? 
Most physicists might be inclined to say no, but here we argue that this is, in fact, entirely possible.

Recent advances in quantum error correction have made the prospect of fault-tolerant quantum computing ever more promising. 
A very exciting prospect of achieving the latter lies in quantum-simulating the holy grail of high-energy physics, QCD~\cite{Georgescu:2013oza,Dalmonte:2016alw,Banuls:2019bmf,Zohar:2015hwa,Aidelsburger:2021mia,Zohar:2021nyc,Klco:2021lap,Bauer:2023qgm,Halimeh:2023ca}. For example, this would provide a complementary venue to dedicated particle colliders for the investigation of QCD, aiding in unraveling many of its outstanding mysteries~\cite{Bauer:2022hpo,diMeglio:2023qc}.

One of the first steps to realize this potential is to write the QCD Hamiltonian explicitly in a form that can be implemented on digital universal quantum computers. 
The standard method is to replace the infinite-volume continuum space with a finite-size lattice, in such a way that the continuum and large-volume result can be obtained systematically by sending the lattice size to infinity~\cite{Byrnes:2003gg}. 
Furthermore, because gluons are bosons, we need to \textit{truncate} the Hilbert space of the lattice theory by introducing a certain \textit{truncation level} $\Lambda$ and taking the limit $\Lambda\to\infty$. 
We must write the \textit{truncated Hamiltonian} explicitly for arbitrary lattice size and arbitrary truncation level, in such a way that the implementation on quantum computers is straightforward. 

QCD is a Yang--Mills theory with an SU($3$) gauge group coupled to fermions in the fundamental representation~\cite{Weinberg_book}. 
The QCD effects are typically the least well-understood for Standard Model processes and thus limit the theoretical precision reached, especially for time-dependent genuinely non-perturbative processes that cannot be treated by either perturbative QCD (pQCD) or Lattice QCD (LQCD).
Currently, phenomenological models using various approximations are used to glean, e.g.,  information about intermediate-time far-from-equilibrium highly non-perturbative quantum processes underlying the formation of the Quark Gluon Plasma.
Quantum simulation offers the unique prospect of probing such processes from a first-principles standpoint, providing snapshots of this dynamics that can yield deep insights into outstanding questions~\cite{Su:2024uuc}.
Furthermore, a major driving force in particle physics is to find and investigate physics beyond the Standard Model, e.g., quantum gravity for which a plethora of suggestions have been made, typically involving different gauge groups, additional symmetries, or novel interaction terms~\cite{Maldacena:2024qlf}.

Thus, to fully profit from the possibilities opened up by future, fault-tolerant quantum computing, it will be crucial to develop universal formulations that can easily be adapted to any member of large groups of theories. 
(In principle, any formulation with the correct continuum limit is eligible.) 
For example, the large-$N$ limit of SU($N$) gauge theories plays a prominent role because it allows us to obtain exact analytic results~\cite{tHooft:1973alw,Lucini:2001ej}. 
Other examples are, e.g., SU(5) and SO(10) candidates for Grand Unified gauge theories. 
Therefore, the study of SU($N$) Yang--Mills theory with $N\ge 3$ is a promising starting point, covering  QCD as well as many models beyond the Standard Model. 

As an almost trivial but important remark, we note that Yang--Mills theory and QCD are merely a small fraction of many important problems.  
It is presently hoped that quantum computing will allow us to solve a long list of computational problems for which classical computers are inefficient, and this list is expected to only get longer with time~\cite{Troyer:2024kra}. 
To meet all these expectations one will need versatile codes that allow treating many of these problems without the need to undertake quantum code development for each of them from scratch. 
The situation will thus be quite different from what it is now, where development focuses on few, highly specific applications, and invests most work on highly specific resource optimization using, e.g., special properties of the chosen problem that do not generalize to other problems of actual interest.
It may be better \emph{not} to rely on special properties such as the simplicity of the representation theory for U(1) or SU(2), or perhaps, any features specific to Yang--Mills theory, so that we can utilize the power of more generic methods developed by the wide research community.

Currently, the most popular choice of lattice Hamiltonian for SU($N$) Yang--Mills theory within the high-energy physics community is the Kogut--Susskind formulation~\cite{Kogut:1974ag}. 
This is the Hamiltonian version of Wilson's Lagrangian formulation~\cite{Wilson:1974sk} that uses unitary link variables. 
Specifically, there are unitary link operators $\hat{U}_{j,\vec{x}}$ living on a link connecting lattice site $\vec{x}$ and $\vec{x}+\hat{j}$, where $j=1,2,3$ are spatial dimensions and $\hat{j}$ is the unit vector along the $j$-th direction. 
In addition, the conjugate momenta $\hat{E}_{j,\vec{x}}$ are introduced. 
To describe the Hilbert space, one can use the coordinate basis (also called magnetic basis) or the momentum basis (also called electric basis). 
The coordinate basis uses the coordinate eigenstates $\ket{U}$ that satisfies $\hat{U}_{j,\vec{x}}\ket{U}=U_{j,\vec{x}}\ket{U}$, where $U_{j,\vec{x}}$ is an $N\times N$ special unitary matrix. 
For a quantum state $\ket{\Phi}$, the wave function $\Phi(U)=\bra{U}\ket{\Phi}$ is defined on $\prod_{j,\vec{x}}[\textrm{SU}(N)]_{j,\vec{x}}$, where $[\textrm{SU}(N)]_{j,\vec{x}}$ is the SU($N$) group manifold corresponding to the link between $\vec{x}$ and $\vec{x}+\hat{j}$. 
It is a nontrivial task to truncate the SU($N$) group manifold systematically so that the truncation effect can be evaluated straightforwardly and at the same time $\hat{E}_{j,\vec{x}}$ takes a simple form.
The momentum basis uses the SU($N$)-analog of the Fourier transform defined by the Peter-Weyl theorem~\cite{Peter-Weyl-original,Zohar:2014qma}. 
This requires complicated group theory, specifically the knowledge of all irreducible representations and their Clebsch--Gordan coefficients. 
Although the momentum basis allows, in principle, a systematic truncation, as shown in a pioneering paper by Byrnes and Yamamoto~\cite{Byrnes:2005qx}, it is technically complicated except for special cases such as SU(2). 
Whether we use the coordinate basis or the momentum basis, it is non-trivial to write down the truncated Hamiltonian, particularly for $N\ge 3$. 
To get some intuition for the level of complications, see the relatively simple cases of the coordinate basis for SU(2)~\cite{Garofalo:2023zkd,Fontana:2024rux}, Fourier transform for a few discrete subgroups of SU(3)~\cite{Murairi:2024xpc}, the use of q-deformation~\cite{Hayata:2023puo,Hayata:2023bgh,Zache:2023dko}, and some simplifications in the large-$N$ limit~\cite{Ciavarella:2024fzw}.

It is fair to say it is very challenging to program SU($N$) Yang--Mills theory with $N\ge 3$ in $2+1$ or $3+1$ on a quantum computer using the Kogut--Susskind Hamiltonian; starting already with the simplest task of writing down the explicit Hamiltonian in terms of Pauli strings.
Perhaps it is possible to write down the Hamiltonian explicitly either on the momentum basis or coordinate basis by using automated computer algebra systems, but still, there is no clear path to resolve other issues associated with the complicated expressions. 
Indeed, currently the only known large-scale experimental realizations of lattice gauge theories on quantum hardware are either in $1+1$ \cite{Yang2020observation,Zhou2022thermalization} or $2+1$ dimensions \cite{Cochran:2024rwe,Gyawali:2024hrz,gonzalezcuadra2024observationstringbreaking2}, with a two-level representation of an Abelian gauge field.
This motivates efforts to find an alternative lattice formulation that is straightforward to generalize to any dimension and gauge group.
We suggest choosing the orbifold lattice formulation which does not suffer from these technical complications because of its use of \textit{non-compact} complex link variables $Z_{j,\vec{x}}$ instead of \textit{compact} unitary link variables $U_{j,\vec{x}}$~\cite{Buser:2020cvn,Bergner:2024qjl}.
\footnote{The original motivation of the orbifold lattice formulation when it was invented by Kaplan, Katz, and \"{U}nsal~\cite{Kaplan:2002wv} was to build lattice gauge theories with exact supersymmetry, having application to quantum gravity via holography in mind~\cite{KU-private-communication}. 
The orbifold lattice construction uses the dimensional deconstruction technique~\cite{Arkani-Hamed:2001kyx} whose original motivation was to generate a fifth dimension from renormalizable, asymptotically-free, four-dimensional gauge theories.} 
Formally, the orbifold lattice is obtained from the Hermitian matrix model via the orbifold projection~\cite{Kaplan:2002wv}. 

Quantum simulations of a matrix model and a gauge theory using the orbifold lattice formulation are very similar. 
Matrix models are interesting in their own right for many reasons, most notably as a non-perturbative definition of quantum gravity via gauge/gravity duality. 
(See Ref.~\cite{Maldacena:2023acv} for a recent review.)
Therefore, by understanding how matrix models and orbifold lattice gauge theories can be studied on quantum computers, we can approach many important problems including QCD and quantum gravity.  

In this paper, we study both SU($N$) Yang--Mills theories on orbifold lattices and SU($N$) Hermitian matrix models, because they are closely related in our universal framework.
We will discuss a concrete realization in terms of basic quantum gates (specifically, CNOT gates and one-qubit gates) and perform a resource estimate for the unitary time evolution task.
The key feature that enables us to do such analyses is that both theories can be expressed by using very standard bosonic variables. 
Namely, by using the coordinate operators $\hat{x}_1,\hat{x}_2,\cdots$ and the conjugate momenta $\hat{p}_1,\hat{p}_2,\cdots$ that satisfy the canonical commutation relations~\cite{Dirac:1925jy} 
\begin{align}
[\hat{x}_a,\hat{p}_b]
=
\mathrm{i}\delta_{ab}\, , 
\qquad
[\hat{x}_a,\hat{x}_b]
=
[\hat{p}_a,\hat{p}_b]
=
0\, , 
\end{align}
the Hamiltonian can be written as
\begin{align}
\hat{H}
=
\frac{1}{2}\sum_a\hat{p}_a^2
+
V(\hat{x})\, , 
\label{eq:simple-form}
\end{align}
where the potential term $V(\hat{x})$ is at most quartic.
An obvious advantage over the Kogut--Susskind formulation is that almost no group theory is required~\cite{Buser:2020cvn}. 
Specifically, nothing more than the structure constant is needed.
As a consequence, we can go back and forth between momentum basis (states $\ket{p}$ that satisfy $\hat{p}_a\ket{p}=p_a\ket{p}$) and coordinate basis (states $\ket{x}$ that satisfy $\hat{x}_a\ket{x}=x_a\ket{x}$) by the standard quantum Fourier transform~\cite{Coppersmith:2002skh}. 
The kinetic term $\sum_a\hat{p}_a^2$ and potential term $V(\hat{x})$ are simple in the momentum basis and coordinate basis, respectively~\cite{Bergner:2024qjl}. 

In Sec.~\ref{sec:qubit-truncation} we will show that, in the coordinate basis, the potential term $V(\hat{x})$ can be written as a sum of Pauli strings consisting of only $\sigma_z$ and having at most length four~\cite{Bergner:2024qjl}. 
For the orbifold lattice formulation of Yang--Mills theory, this is true for any $N$, any spatial dimensions, and any lattice volume. 
For a matrix model, this is true for any $N$ and any number of matrices. 
Regardless of the details of the theory, we can write the truncated Hamiltonian explicitly in the same universal form, and we can discuss the algorithms and estimate the cost of quantum simulations in a unified manner.

We note that the scalar $\phi^4$ theory considered in Ref.~\cite{Jordan:2012xnu} has the same simple form \eqref{eq:simple-form}. 
(We will show this in Sec.~\ref{sec:JLP}.) 
The only difference is in the detail of the polynomial $V(\hat{x})$. 
In this sense, the implementation of the Hamiltonian of SU($N$) Yang--Mills theory on a quantum computer is not more complicated than that of scalar $\phi^4$ theory. 
We can see the essence already in much simpler models, e.g., the anharmonic oscillator. 

In Table~\ref{tab:resource_summary}, we show a summary of qubit and T gate requirements for the Hamiltonian time evolution of the theories we will study in this paper. 
Here, $V_{\rm lattice}$ is the lattice volume (number of lattice sites), $d$ is the spatial dimension (for scalar QFT and orbifold YM) or the number of matrices (for matrix model), $N$ characterizes the gauge group SU($N$), and $Q$ is the number of qubits assigned to each bosonic degree of freedom. 
For fault-tolerant quantum simulations, `the number of qubits' means the number of logical qubits. 
As for the gate counts, we showed only the number for one time step in the Suzuki-Trotter decomposition, and we showed only scaling with the parameters characterizing the system size.  
More details are explained in the following sections. 
Note that the number of gates in this table is proportional to the number of the interaction terms in the Hamiltonian. 
Each interaction term can be implemented efficiently so that the simulation cost does not become unnecessarily large. 

%We consider the number of T gates in units of $T_{\rm typ}=10$ -- 50~\cite{CampbellETFQC,Kliuchnikov:2016ong} for practical purposes.

\begin{table}[ht]
    \centering
    \begin{tabular}{|c||c|c|c|}
    \hline
    & Number of qubits & \# T gates in $V(\hat{x})$ term & \# T gates in $\hat{p}^2$ term \\
    \hline
    \hline
    Scalar QFT & $V_{\rm lattice}Q$ & $ V_{\rm lattice} \genfrac{(}{)}{0pt}{}{Q}{4}$ & $V_{\rm lattice}Q(Q-1)$ \\
    \hline
    Matrix Model & $dN^2Q$ & $ d(d-1)N^4Q^4$ & $ dN^2Q(Q-1)$ \\
    \hline
    Orbifold YM & $2dN^2V_{\rm lattice}Q$ & $d^2V_{\rm lattice}N^4Q^4$ & $ N^2dV_{\rm lattice}Q(Q-1)$ \\
    \hline
    \end{tabular}
    \caption{Summary of qubit and T gate requirements for a single Suzuki--Trotter step of the Hamiltonian time evolution of the theories in this paper, which fall into the universal Hamiltonian form of~\eqref{eq:simple-form}.
    See Section~\ref{sec:Trotter_cost} for details.}
    \label{tab:resource_summary}
\end{table}

The plan of this paper is as follows.
\begin{itemize}
\item
In Sec.~\ref{sec:qubit-truncation}, we discuss how a Hamiltonian of the form \eqref{eq:simple-form} can be simulated on a quantum device. 
We introduce a truncation scheme following Ref.~\cite{Bergner:2024qjl}.

\item
In Sec.~\ref{sec:JLP}, we show how the scalar quantum field theory, which was studied in Ref.~\cite{Jordan:2012xnu}, fits into our universal framework.

\item
In Sec.~\ref{sec:Matrix-Model} and Sec.~\ref{sec:orbifold-lattice}, we explain how the Hamiltonians of matrix model and orbifold lattice reduce to the form \eqref{eq:simple-form}. 
In both cases, the Hamiltonian can be written as sum of products of coordinate operators $\hat{x}_1,\hat{x}_2,\cdots$ and momentum operators $\hat{p}_1,\hat{p}_2,\cdots$ that satisfy the canonical commutation relation $[\hat{x}_a,\hat{p}_b]=i\delta_{ab}$. 
The kinetic terms are simply $\sum_a\hat{p}_a^2$. 
The interaction terms can be written using products of only a few coordinate operators. 
To be precise, the most complicated terms we have to deal with are $\hat{x}_a\hat{x}_b\hat{x}_c\hat{x}_d$. 
We will count how many terms of such form appear in the Hamiltonian. 

\item
  In Sec.~\ref{sec:Trotter_cost} we discuss an explicit construction of a quantum circuit for the Hamiltonian time evolution based on the Suzuki--Trotter decomposition. 

% \item 
% Concluding remarks will be provided in Sec.~\ref{sec:conclusion}. 

\end{itemize}

A significant fraction of the first sections is a review of Refs.~\cite{Buser:2020cvn,Bergner:2024qjl}.
The material is organized in such a way as to help make the main message of this paper clearer for a broader audience of readers.
Specifically, Sec.~\ref{sec:qubit-truncation} is mostly a refinement of part of Ref.~\cite{Bergner:2024qjl}, while Sec.~\ref{sec:Matrix-Model} and Sec.~\ref{sec:orbifold-lattice} contain some new materials such as the counting of quartic terms that will be used to estimate quantum resources for Hamiltonian evolution.

%%%%%%%%%%%%%%%%%%%%%%%%%%%
%%%%%%%%%%%%%%%%%%%%%%%%%%%
\section{Basic idea}\label{sec:qubit-truncation}
%%%%%%%%%%%%%%%%%%%%%%%%%%%
%%%%%%%%%%%%%%%%%%%%%%%%%%%

As we will see in Sec.~\ref{sec:Matrix-Model} and Sec.~\ref{sec:orbifold-lattice}, the Hamiltonians of matrix model and orbifold lattice gauge theory are schematically written as \eqref{eq:simple-form}, where $V(\hat{x})$ is at most a fourth-order polynomial. 
Therefore, we discuss how a Hamiltonian for $N_b$ bosons of the form
\begin{align}
\hat{H}
=
\sum_{a}
\frac{1}{2}
\hat{p}_a^2
+
\sum_{a,b,c,d}
C_{abcd}\, 
\hat{x}_a\hat{x}_b\hat{x}_c\hat{x}_d\, ,
\label{Hamiltonian_quartic_potential}
\end{align}
where $C_{abcd}$ is an arbitrary real number, can be simulated on a quantum device. 
In this section, we focus on getting a simple truncated Hamiltonian, postponing an explicit construction of a quantum circuit for the Hamiltonian time evolution until Sec.~\ref{sec:Trotter_cost}.

The potential part and kinetic part of the Hamiltonian become simple in the coordinate basis and momentum basis, respectively. 
Unlike in the Kogut--Susskind formulation, the Fourier transform between these two bases is straightforward for a generic gauge group SU($N$). 
Truncation can be performed in a way compatible with the quantum Fourier transform. 

By setting $N_b=1$, the summations in \eqref{Hamiltonian_quartic_potential} reduce to single terms, and the interaction part can be simplified until we arrive at the simplest non-trivial example, a single quantum anharmonic oscillator, 
\begin{align}
\hat{H}
=
\frac{\hat{p}^2}{2}
+
\frac{\hat{x}^4}{4}\, . 
\label{anharmonic_oscillator}
\end{align}
We will comment on this example in Sec.~\ref{sec:conclusion}.
We are optimistic that, by then, readers will see how this example captures the essence of our approach and serves as an excellent starting point for quantum simulations of more intriguing systems.

%%%%%%%%%%%%%%%%%%%%%%%%%%%
%%%%%%%%%%%%%%%%%%%%%%%%%%%
\subsection{Truncation in the coordinate basis}\label{sec:review_coordinate_basis}
%%%%%%%%%%%%%%%%%%%%%%%%%%%
%%%%%%%%%%%%%%%%%%%%%%%%%%%

Let us use $\{\ket{\vec{x}}\}$ to denote all $N_b$ bosons simultaneously.
By using this expression we mean that the coordinate eigenstate of the system is 
\begin{align}
\ket{\vec{x}}
=
\otimes_a\ket{x_a}\, ,
\end{align}
where each boson has coordinate eigenstate $\ket{x_a}$ ($a=1,2,\cdots, N_b$).
The coordinate eigenstate $\ket{\vec{x}}$ is defined by 
\begin{align}
\hat{\vec{x}}\ket{\vec{x}}
=
\vec{x}\ket{\vec{x}}\, . 
\end{align}
Moreover, we consider the system Hilbert space as a tensor product of the Hilbert spaces of the individual bosons in the coordinate eigenbasis\footnote{
This $\mathcal{H}$ corresponds to the extended Hilbert space $\mathcal{H}_{\rm ext}$ in later sections. 
}
\begin{align}
\mathcal{H}
=
\otimes_a\mathcal{H}_a\, , 
\qquad
\mathcal{H}_a
=
\mathrm{Span}\{\ket{x_a}|x_a\in\mathbb{R}\}\, . 
\end{align}

So far, each boson has a wavefunction that can be represented in the basis given by $\ket{x_a}$, which lives in an infinite-dimensional Hilbert space.
For example, one can imagine a one-dimensional quantum oscillator being in a superposition of many coordinates/positions.
In order to reduce the problem to a finite-dimensional Hilbert space, for each boson coordinate $x_a$ we introduce a cutoff,
\begin{align}
-R\le x_a\le R\, , 
\end{align}
and discretize $x_a$ by introducing $\Lambda\geq 2$ points,\footnote{It is convenient to change $x_{a,n}$ and $\delta_x$ slightly if we use periodic boundary condition, as we will see shortly.} as depicted in Fig.~\ref{fig:digitization}
\begin{align}
\label{eq:digitization}
x_{a,n_a}
=
-R+n_a\delta_x\, ,
\qquad
\delta_x=\frac{2R}{\Lambda-1}\, , 
\qquad
n_a=0,1,\cdots,\Lambda-1\, .   
\end{align}
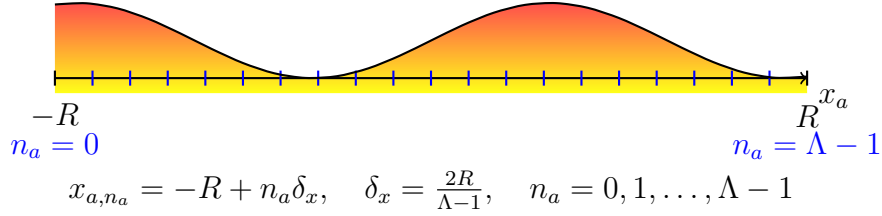
\begin{figure}[ht!]
\centering
    \begin{tikzpicture}
        
        % Gradient fill below the wave
        \begin{scope}
            \clip (-5, -0.5) rectangle (5, 2);
            \shade[top color=red!70, bottom color=yellow!90] 
                plot[domain=-5:5, smooth] (\x, {(\x+5)*(\x+5)*(\x-5)*(\x-5)/400}) -- (5,-0.2) -- (-5,-0.2) -- cycle;
        \end{scope}
        
        % Axis
        \draw[thick,->] (-5,0) -- (5,0) node[below right] {\(x_a\)};
        \draw[thick] (-5,-0.1) -- (-5,0.1);
        \draw[thick] (5,-0.1) -- (5,0.1);
        
        % Vertical ticks
        \foreach \x in {-4.5,-4,...,4.5} {
            \draw[blue, thick] (\x,-0.1) -- (\x,0.1);
        }
        
        % Labels
        \node at (-5, -0.5) {\(-R\)};
        \node at (5, -0.5) {\(R\)};
        \node[blue] at (-5, -0.9) {\(n_a=0\)};
        \node[blue] at (5, -0.9) {\(n_a=\Lambda-1\)};
        
        % Equation
        \node at (0, -1.5) {\(x_{a, n_a} = -R + n_a \delta_x, \quad \delta_x = \frac{2R}{\Lambda-1}, \quad n_a=0,1,\dots,\Lambda-1\)};
        
        % Top label
        \node at (0, 2.2) {\(\hat{x}_{a} | n_a \rangle = x_{a,n_a} | n_a \rangle\)};
        
        % Wave
        \draw[thick] plot[domain=-5:5, smooth] (\x, {(\x+5)*(\x+5)*(\x-5)*(\x-5)/400});
      
    \end{tikzpicture}
    \caption{\label{fig:digitization} Equation~\eqref{eq:digitization} is reproduced in this figure to highlight the schematic construction of the discretized and truncated Hilbert space $\mathcal{H}_a$ of a single boson $a$. 
    The coordinate operator $\hat{x}_a$ can act on a limited number of $\Lambda$ different states $|n_a\rangle$ with discretized eigenvalues $x_{a, n_a}$, labeled by an integer $n_a$.
    A pictorial shaded profile represents a possible wavefunction realization for this single bosonic degree of freedom.}
\end{figure}
We consider $\Lambda$, $\delta_x$ and $R$ as truncation parameters that can and need to be adjusted.
In particular, they each have limiting values that should be reached in order to recover the original infinite dimensional Hilbert space:
$\Lambda$ should be sent to $\infty$, together with $R$, while $\delta_x$ goes to 0.

By using $\ket{n_a}$ to denote $\ket{x_{a,n_a}}$, we can write the operator $\hat{x}_a$ acting diagonally on $\mathcal{H}_a$ as
\begin{align}
\hat{x}_a
=
\sum_{n_a=0}^{\Lambda-1}
x_{a,n_a}
\ket{n_a}\bra{n_a}
=
-R\cdot\textbf{1}
+\delta_x\cdot\hat{n}_a\, , 
\end{align}
where
\begin{align}
\hat{n}_a
\equiv
\sum_{n_a}n_a
\ket{n_a}\bra{n_a}
\end{align}
is the bosonic number operator. 
This operator can be extended to the operator acting on $\mathcal{H}$, assuming that it acts as the identity on $\mathcal{H}_{a'}$ for $a'\neq a$. 

We then write $\hat{n}_a$ as a sum of Pauli operators acting on $Q$ qubits representing a number of states equal to the number of points $\Lambda = 2^{Q}$.
Using the binary form with $b_{a,i}=\{0,1\}$, 
\begin{align}
\ket{n_a} =\ket{b_{a,1}}\ket{b_{a,2}}\cdots\ket{b_{a,Q}}\, , 
\qquad
n_a=b_{a,1}+2b_{a,2}\cdots+2^{Q-1}b_{a,Q}\, ,
\end{align}
the number operator can be written by using Pauli $\sigma_z$ gates, 
\begin{align}
\hat{n}_a
&=
-
\frac{\hat{\sigma}_{z;a,1}-\textbf{1}}{2}
-
2\cdot\frac{\hat{\sigma}_{z;a,2}-\textbf{1}}{2}
-
\cdots
-
2^{Q-1}\cdot\frac{\hat{\sigma}_{z;a,Q}-\textbf{1}}{2}
\nonumber\\
&=
-
\frac{\hat{\sigma}_{z;a,1}}{2}
-
2\cdot\frac{\hat{\sigma}_{z;a,2}}{2}
-
\cdots
-
2^{Q-1}\cdot\frac{\hat{\sigma}_{z;a,Q}}{2}
+
\frac{\Lambda-1}{2}\cdot\textbf{1}\, , 
\end{align}
where $\hat{\sigma}_{z;a,i}$ is the Pauli $\hat{\sigma}_{z}$ operator acting on $\ket{b_{a,i}}$. Note that our convention is 
\begin{align}
    \sigma_z \equiv \ket{0}\bra{0} - \ket{1}\bra{1}\, . 
    \label{eq: pauli Z}
\end{align}Therefore, 
\begin{align}
\hat{x}_a
=
-
\delta_x\cdot
\left(
\frac{\hat{\sigma}_{z;a,1}}{2}
+
2\cdot\frac{\hat{\sigma}_{z;a,2}}{2}
+
\cdots
+
2^{Q-1}\cdot\frac{\hat{\sigma}_{z;a,Q}}{2}
\right)\, . 
\label{eq:x_as_sum_of_z}
\end{align}

There are many four-boson couplings of the form $\hat{x}_a\otimes\hat{x}_b\otimes\hat{x}_c\otimes\hat{x}_d$. 
Each $\hat{x}$ is a sum of $\hat{\sigma}_{z,1}$,..., $\hat{\sigma}_{z,Q}$. 
Therefore, each four-boson coupling consists of $Q^4$ terms, each of them is a tensor product of four $\hat{\sigma}_z$'s.\footnote{
If the same boson appears more than once in a given coupling, e.g., $\hat{x}_a^2\hat{x}_b^2$, terms with less than four $\hat{\sigma}_z$'s appear as well.
} 
The same structure is already present for the harmonic oscillator~\eqref{anharmonic_oscillator} and was explicitly studied in Ref.~\cite{Hanada:2022pps}, and much earlier in Ref.~\cite{Somma:2015bcw}.

%%%%%%%%%%%%%%%%%%%%%%%%%%%
%%%%%%%%%%%%%%%%%%%%%%%%%%%
\subsubsection*{Periodic boundary condition}
%%%%%%%%%%%%%%%%%%%%%%%%%%%
%%%%%%%%%%%%%%%%%%%%%%%%%%%

Technically, it is convenient to use the periodic boundary condition $x+2R\sim x$. 
In this case, a convenient convention is to take
\begin{align}
   \delta_x=\frac{2R}{\Lambda} 
\end{align}
and 
\begin{align}
    x_{a,n_a}
    =
    -\frac{\Lambda-1}{\Lambda}R + n_a\delta_x
    =
    \left(n_a-\frac{\Lambda-1}{2}\right)\delta_x\, . 
    \label{eq:x_pbc}
\end{align} 
Hence, $x_{a,n_a}$ takes values $\pm\frac{\delta_x}{2}$, $\pm\frac{3\delta_x}{2}$, ..., $\pm\frac{(\Lambda-1)\delta_x}{2}$. 
To reduce the truncation effect, we must take $R$ large enough for $x\sim\pm R$ not to be significantly excited, such that the boundary condition does not affect the physics of interest. 
In other words: if the boundary condition matters, $R$ is not large enough. 
The study of truncation effects is usually specific for the system under study, including its parameters, such as the coupling constant.
It is often the case that these systematic effects need to be studied numerically. An example targeting expectation values computed via classical sampling methods was reported in Ref.~\cite{Hanada:2022pps}.

%%%%%%%%%%%%%%%%%%%%%%%%%%%
%%%%%%%%%%%%%%%%%%%%%%%%%%%
\subsection{Quantum Fourier transform and momentum basis}\label{sec:Fourier_transform}
%%%%%%%%%%%%%%%%%%%%%%%%%%%
%%%%%%%%%%%%%%%%%%%%%%%%%%%

With the periodic boundary condition, the shift operator $\hat{S}_a\equiv\sum_{n_a}\ket{n_a+1}\bra{n_a}$ is identified with $e^{\mathrm{i}\delta_X\hat{p}_a}$. 
Therefore, we can approximate $\hat{p}_a$ by $\frac{\hat{S}_a^{1/2}-\hat{S}_a^{-1/2}}{\mathrm{i}\delta_X}$ up to corrections of order $\delta_X$. 
Then, 
\begin{align}
\hat{p}_a^2
=
\frac{2\cdot\textbf{1}-\hat{S}_a-\hat{S}_a^{-1}}{\delta_X^2}
=
\frac{1}{\delta_X^2}
\sum_{n_a=0}^{\Lambda-1}
\left\{
2\ket{n_a}\bra{n_a}
-
\ket{n_a+1}\bra{n_a}
-
\ket{n_a}\bra{n_a+1}
\right\}\, . 
\label{p^2-coordinate-basis}
\end{align}
Because $\sum_{n_a}\ket{n_a}\bra{n_a}$ is the identity, the nontrivial parts of $\hat{p}_a^2$ are $\hat{S}_a=\sum_{n_a}\ket{n_a+1}\bra{n_a}$ and $\hat{S}_a^{-1}$. 

By applying the quantum Fourier transform, we can switch to the momentum eigenstates $\ket{\tilde{n}_a}$ ($\tilde{n}_a=0,1,\cdots,\Lambda-1$):
\begin{align}
    \ket{\tilde{n}_a}
    =
    \frac{1}{\sqrt{\Lambda}}\sum_{n_a} e^{2\pi \mathrm{i}\tilde{n}_a(n_a+1/2)/\Lambda}\ket{n_a}\, . 
\end{align}
The shift operator becomes diagonal in the momentum basis:
\begin{align}
    \hat{S}\ket{\tilde{n}_a}
    =
    e^{2\pi\mathrm{i}\tilde{n}_a/\Lambda}
    \ket{\tilde{n}_a}\, . 
\end{align}
Therefore, $\hat{p}_a$ is diagonal, too:
\begin{align}
    \hat{p}_a\ket{\tilde{n}_a}
    =
    \frac{2}{\delta_X}
    \sin\left(\frac{\pi\tilde{n}_a}{\Lambda}\right)
    \ket{\tilde{n}_a}\, . 
    \label{p-option-1}
\end{align}

Alternatively, we can define $\hat{p}_a$ as
\begin{align}
    \hat{p}_a\ket{\tilde{n}_a}
    =
    \frac{2\pi}{\delta_X\Lambda}
    \left(
\tilde{n}_a +\frac{1}{2}  
    \right)
    \ket{\tilde{n}_a}
    =
    \frac{\pi}{R}
    \left(
\tilde{n}_a +\frac{1}{2}  
    \right)
    \ket{\tilde{n}_a}\,  , 
    \label{p-option-2}
\end{align}
restricting the range of $\tilde{n}_a$ from $-\frac{\Lambda}{2}$ to $+\frac{\Lambda}{2}-1$ instead of \eqref{p-option-1}. 
Here, we used $\tilde{n}_a +1/2$ rather than $\tilde{n}_a$ to respect the symmetry under $\hat{p}_a\to -\hat{p}_a$. Associated with this, the Fourier transform is modified to 
\begin{align}
    \ket{\tilde{n}_a}
    =
    \frac{1}{\sqrt{\Lambda}}\sum_{n_a} e^{2\pi \mathrm{i}(\tilde{n}_a+1/2)(n_a+1/2)/\Lambda}\ket{n_a}\, . 
\end{align}
With this option, the kinetic term is nonlocal in the coordinate basis. 
These two options are the same up to truncation effects. 

Note that the Fourier transform can be performed for each boson in parallel and hence the depth of the circuit depends only on the truncation level $\Lambda$ and \emph{not} on the number of bosons.
Therefore, if we can perform a Fourier transform to a one-boson system, like the anharmonic oscillator \eqref{anharmonic_oscillator}, in principle, we only have to add more qubits that can describe more bosons and add the same circuits for the other bosons.

%%%%%%%%%%%%%%%%%%%%%
%%%%%%%%%%%%%%%%%%%%%%%
%%%%%%%%%%%%%%%%%%%%%%%
\section{Scalar Quantum Field Theory}\label{sec:JLP}
%%%%%%%%%%%%%%%%%%%%%%%
%%%%%%%%%%%%%%%%%%%%%%%

An important example of quantum field theory is a scalar $\phi^4$ theory in $3+1$ spacetime dimensions. 
Despite its simplicity, this theory contains many important features of quantum field theory, and it is often used to demonstrate new concepts or new techniques. 
See e.g., the famous textbooks by Peskin and Schr\"{o}der~\cite{Peskin:1995ev} and by Fradkin~\cite{Fradkin:2021zbi}. 
Naturally, the seminal paper on quantum computing by Jordan, Lee, and Preskill~\cite{Jordan:2012xnu} studies this theory, too.

We regularize this theory on a cubic lattice with equal lattice spacing $a$ in all three directions. 
Following the notations in Ref.~\cite{Hanada:2022pps}, we write the lattice Hamiltonian as
\begin{align}
    \hat{H}
    =
    \sum_{\vec{n}}\left(
    \frac{1}{2}
    \hat{\pi}_{\vec{n}}^2
    +
    \frac{1}{2}
    \sum_{j=1}^3
    \left(
    \hat{\phi}_{\vec{n}+\hat{j}}-\hat{\phi}_{\vec{n}}
    \right)^2
    +
    \frac{m^2}{2}\hat{\phi}_{\vec{n}}^2
    +
    \frac{\lambda}{4}\hat{\phi}_{\vec{n}}^4
    \right) \, . 
    \label{scalar-QFT-Hamiltonian}
\end{align}
The scalar field $\hat{\phi}$ and its conjugate momentum $\hat{\pi}$ are dimensionless. They correspond to fields in the continuum theory according to  $\hat{\phi}=a\hat{\phi}_{\rm cont.}$ and $\hat{\pi}=a^2\hat{\pi}_{\rm cont.}$ (where $a$ is the dimensionfull lattice spacing).
The Hamiltonian and the mass parameter are also made dimensionless, i.e., $\hat{H}=a\times\hat{H}_{\rm cont.}$, $m=a\times m_{\rm cont.}$. 
The lattice sites are labeled by $\vec{n}\in\mathbb{Z}^d$; and $\hat{j}$ is the unit vector along the $j$-th dimension of the spatial lattice ($j=1,2,3$).

The canonical commutation relation is imposed, i.e., 
\begin{align}
    [\hat{\phi}_{\vec{n}},\hat{\pi}_{\vec{n}'}]
    =
    \mathrm{i}\delta_{\vec{n},\vec{n}'} \, .
\end{align}
These operators $\hat{\phi}$ and $\hat{\pi}$ are the same as $\hat{x}$ and $\hat{p}$ in the previous sections, and the Hamiltonian takes the universal form~\eqref{eq:simple-form}. 

As a minor comment on terminology, we note that the quadratic term in $\hat{\phi}$ in the lattice Hamiltonian~\eqref{scalar-QFT-Hamiltonian} corresponds to the spatial derivative term $(\partial_j\hat{\phi})^2$ in the continuum theory, which is usually called a kinetic term. 
However, in the context of the universal form~\eqref{eq:simple-form}, we regard it as a quadratic part of the potential $V(\hat{x})$, because $\hat{\phi}$ is playing the role of a (bosonic) coordinate.

%%%%%%%%%%%%%%%%%%%%%%%
%%%%%%%%%%%%%%%%%%%%%%%
\subsection{Hilbert space}
%%%%%%%%%%%%%%%%%%%%%%%
%%%%%%%%%%%%%%%%%%%%%%%

A convenient way to define the Hilbert space is to use coordinate eigenstates $\ket{\phi}$ that satisfy $\hat{\phi}_{\vec{n}}\ket{\phi}=\phi_{\vec{n}}\ket{\phi}$ as 
\begin{align}
\mathcal{H}
=
\left\{
\ket{\Psi}
\equiv
\int\mathrm{d}^{V_{\rm lattice}}\phi\, \Psi(\phi)\ket{\phi}
\Bigg|
\int\mathrm{d}^{V_{\rm lattice}}\phi\, |\Psi(\phi)|^2<\infty
\right\}\, ,  
\end{align}
where $V_{\rm lattice}$ is the lattice volume (the number of lattice points). 
We can also use the momentum eigenstates $\ket{\pi}$ that satisfy $\hat{\pi}_{\vec{n}}\ket{\pi}=\pi_{\vec{n}}\ket{\pi}$ as 
\begin{align}
\mathcal{H}
=
\left\{
\ket{\Psi}
\equiv
\int\mathrm{d}^{V_{\rm lattice}}\pi\, \tilde{\Psi}(\pi)\ket{\pi}
\Bigg|
\int\mathrm{d}^{V_{\rm lattice}}\pi\, |\tilde{\Psi}(\pi)|^2<\infty
\right\}\, . 
\end{align}
Wave functions $\Psi(\phi)$ and $\tilde{\Psi}(\pi)$ are related by the Fourier transform. Note that
\begin{align}
    \bra{\pi}\ket{\phi}
    =
    \exp\left(-\mathrm{i}\sum_{\vec{n}}\pi_{\vec{n}}\phi_{\vec{n}}\right)\, . 
\end{align}
We often use a notation 
\begin{align}
\mathcal{H}
=
\mathrm{Span}\left\{
\ket{\phi}\Big|\phi\in\mathbb{R}^{V_{\rm lattice}}
\right\}
=
\mathrm{Span}\left\{
\ket{\pi}\Big|\pi\in\mathbb{R}^{V_{\rm lattice}}
\right\}
\end{align}
assuming the square-integrability condition. 

We assign $Q$ qubits to each bosonic degree of freedom. 
Then, the number of qubits needed to describe the Hilbert space is $QV_{\rm lattice}$. 

%%%%%%%%%%%%%%%%%%%%%%%
%%%%%%%%%%%%%%%%%%%%%%%
\section{Matrix Models}\label{sec:Matrix-Model}
%%%%%%%%%%%%%%%%%%%%%%%
%%%%%%%%%%%%%%%%%%%%%%%
Now we turn our attention to the SU($N$) bosonic $d$-matrix model, and again we realize that we are dealing with a Hamiltonian of the form \eqref{eq:simple-form}.
The Lagrangian is
\begin{align}
    L
    =
    \mathrm{Tr}\left(
    \frac{1}{2}(D_tX_I)^2
    -
    \frac{g^2}{4}[X_I,X_J]^2
    \right)\, , 
\end{align}
where $X_{I=1,2.\cdots,d}$ are $N\times N$ Hermitian matrices and $D_tX_I$ is the gauge covariant derivative defined by $D_tX_I=\partial_tX_I-ig[A_t,X_I]$. The integration of gauge field $A_t$ leads to the Gauss-law constraint. 

We can either impose or not impose a traceless condition on $X_I$. Both options are explained below. There is no difference in physics because the trace part is free and decoupled from the rest. 

Below, we confirm that this model belongs to a class of theories whose Hamiltonians take the simple form in \eqref{eq:simple-form}. 
%%%%%%%%%%%%%%%%%%%%%%%
%%%%%%%%%%%%%%%%%%%%%%%
\subsubsection*{With traceless condition}
%%%%%%%%%%%%%%%%%%%%%%%
%%%%%%%%%%%%%%%%%%%%%%%
To have real expansion coefficients, we introduce SU($N$) generators $\tau_\alpha$, where the adjoint index $\alpha$ runs from 1 to $N^2-1$, that are normalized as $\mathrm{Tr}(\tau_\alpha\tau_\beta)=\delta_{\alpha\beta}$. Then the matrix elements are written as 
\begin{align}
X_{I,ij}
=
\sum_{\alpha=1}^{N^2-1} X_I^\alpha\tau_{\alpha,ij}\, 
\qquad
X_I^\alpha\in\mathbb{R}\, . 
\end{align}

By using the structure constant $f_{\alpha\beta}{}^\gamma$, that is related to the generators as $[\tau_\alpha,\tau_\beta]=\mathrm{i}\sum_\gamma f_{\alpha\beta}{}^\gamma\tau_\gamma$, we have
$[X_I,X_J]=\mathrm{i}\sum_{\alpha,\beta,\gamma} f_{\alpha\beta}{}^\gamma X_I^\alpha X_J^\beta\tau_\gamma$. 
See Appendix~\ref{sec:SU(N)-algebra} for an explicit construction of the generators. 

The corresponding Hamiltonian is\footnote{
Note that the commutator $[\hat{X}_I,\hat{X}_J]$ means the commutator of $N\times N$ matrices, i.e., 
\begin{align}
[\hat{X}_I,\hat{X}_J]_{ij}=(\hat{X}_I\hat{X}_J)_{ij}-(\hat{X}_J\hat{X}_I)_{ij}
=
\sum_{k=1}^N
\left(
\hat{X}_{I,ik}\hat{X}_{J,kj}
-
\hat{X}_{J,ik}\hat{X}_{I,kj}
\right)\, . 
\nonumber
\end{align}
``$\mathrm{Tr}$" means the trace as an $N\times N$ matrix. 
When indices are written explicitly in the commutator, it means a commutator as operator, 
\begin{align}
[\hat{X}_{I,ij},\hat{X}_{J,kl}]
=
\hat{X}_{I,ij}\hat{X}_{J,kl}
-
\hat{X}_{J,kl}\hat{X}_{I,ij}\, . 
\nonumber
\end{align}
} 
\begin{align}
    \hat{H}
    =
    \mathrm{Tr}\left(
    \frac{1}{2}\hat{P}_I^2
    -
    \frac{g^2}{4}[\hat{X}_I,\hat{X}_J]^2
    \right)\, . 
    \label{MM-Hamiltonian}
\end{align}
Note that $(\hat{X}_{I,ij})^\dagger=\hat{X}_{I,ji}$, $(\hat{P}_{I,ij})^\dagger=\hat{P}_{I,ji}$. 
We can introduce the operators with the adjoint index $\alpha$ as   
\begin{align}
\hat{X}_{I,ij}
=
\sum_\alpha\hat{X}_I^\alpha\tau_{\alpha,ij}\, , 
\qquad
\hat{P}_{I,ij}
=
\sum_\alpha\hat{P}_I^\alpha\tau_{\alpha,ij}\, . 
\end{align}
$\hat{X}_I^\alpha$ and $\hat{P}_I^\alpha$ are self-adjoint, i.e., $(\hat{X}_I^\alpha)^\dagger=\hat{X}_I^\alpha$, $(\hat{P}_I^\alpha)^\dagger=\hat{P}_I^\alpha$, 
and they satisfy the canonical commutation relation\footnote{These commutators are operators, e.g., $[\hat{X}_I^\alpha,\hat{P}_J^\beta]=\hat{X}_I^\alpha\hat{P}_J^\beta-\hat{P}_J^\beta\hat{X}_I^\alpha$.} 
\begin{align}
[\hat{X}_I^\alpha,\hat{P}_J^\beta]
=
\mathrm{i}\delta_{IJ}\delta_{\alpha\beta}\, , 
\qquad
[\hat{X}_I^\alpha,\hat{X}_J^\beta]
=
[\hat{P}_I^\alpha,\hat{P}_J^\beta]
=
0\, . 
\end{align}
By using $\hat{X}_I^\alpha$ and $\hat{P}_I^\alpha$, the Hamiltonian \eqref{MM-Hamiltonian} can be written in the simple form \eqref{eq:simple-form} with $d(N^2-1)$ real bosonic degrees of freedom. See Sec.~\ref{sec:MM-Hamiltonian-closer-look} for more details. 
%%%%%%%%%%%%%%%%%%%%%%%
%%%%%%%%%%%%%%%%%%%%%%%
\subsubsection*{Without traceless condition}
%%%%%%%%%%%%%%%%%%%%%%%
%%%%%%%%%%%%%%%%%%%%%%%
If we do not impose the traceless condition, we do not need to use generators. We could simply use the diagonal entries $\hat{X}_{I,ii}$, which are real, and the real and imaginary parts of the off-diagonal entries, 
\begin{align}
\hat{X}_{I,ij}^{\rm (R)}
\equiv
\frac{1}{\sqrt{2}}(\hat{X}_{I,ij}+\hat{X}_{I,ji})\, 
\qquad
\hat{X}_{I,ij}^{\rm (I)}
\equiv
\frac{-\mathrm{i}}{\sqrt{2}}(\hat{X}_{I,ij}-\hat{X}_{I,ji})
\end{align}
as real bosonic operators with canonical normalization. 

The trace part is free and decouples from the SU($N$) sector under time evolution, if the initial momentum of the trace part is zero, it just stays zero. To stabilize the trace part regardless of the initial condition, we can add a mass term proportional to $(\mathrm{Tr}\hat{X}_I)^2$. 

%%%%%%%%%%%%%%%%%%%%%%%%%%
%%%%%%%%%%%%%%%%%%%%%%%%%%
\subsection{Hilbert space}
%%%%%%%%%%%%%%%%%%%%%%%%%%
%%%%%%%%%%%%%%%%%%%%%%%%%%

%%%%%%%%%%%%%%%%%%%%%%%
%%%%%%%%%%%%%%%%%%%%%%%
\subsubsection*{With traceless condition}
%%%%%%%%%%%%%%%%%%%%%%%
%%%%%%%%%%%%%%%%%%%%%%%
A convenient way to define the Hilbert space is to use coordinate eigenstates $\ket{X}$ that satisfy $\hat{X}_I^\alpha\ket{X}=X_I^\alpha\ket{X}$ as 
\begin{align}
\mathcal{H}_{\rm ext}
=
\left\{
\ket{\Psi}
\equiv
\int\mathrm{d}^{d(N^2-1)}X\, \Psi(X)\ket{X}
\Bigg|
\int\mathrm{d}^{d(N^2-1)}X\, |\Psi(X)|^2<\infty
\right\}\, . 
\end{align}
Here the subscripts \textrm{ext} indicate that $\mathcal{H}_{\rm ext}$ is the \textit{extended} Hilbert space that contains SU($N$) non-singlets. 
We can also use the momentum eigenstates $\ket{P}$ that satisfy $\hat{P}_I^\alpha\ket{P}=P_I^\alpha\ket{P}$ as  
\begin{align}
\mathcal{H}_{\rm ext}
=
\left\{
\ket{\Psi}
\equiv
\int\mathrm{d}^{d(N^2-1)}P\, \tilde{\Psi}(P)\ket{P}
\Bigg|
\int\mathrm{d}^{d(N^2-1)}P\, |\tilde{\Psi}(P)|^2<\infty
\right\}\, . 
\end{align}
Wave functions $\Psi(X)$ and $\tilde{\Psi}(P)$ are related by the Fourier transform. Note that
\begin{align}
    \bra{P}\ket{X}
    =
    \exp\left(-\mathrm{i}\sum_{I,\alpha}P_I^\alpha X_I^\alpha\right)
    =
    \exp\left(-\mathrm{i}\sum_I\mathrm{Tr}(P_IX_I)\right)\, . 
\end{align}
We often use the notation 
\begin{align}
\mathcal{H}_{\rm ext}
=
\mathrm{Span}\left\{
\ket{X}\Big|X\in\mathbb{R}^{d(N^2-1)}
\right\}
=
\mathrm{Span}\left\{
\ket{P}\Big|P\in\mathbb{R}^{d(N^2-1)}
\right\}
\end{align}
assuming the square-integrability condition. 

Under the SU($N$) gauge transformation, these states transform according to
\begin{align}
    \ket{X}
    \to
    \ket{\Omega^{-1}X\Omega}\, , 
    \qquad
    \ket{P}
    \to
    \ket{\Omega^{-1}P\Omega}\, ,     
\end{align}
while the operators transform as
\begin{align}
    \hat{X}_{I,ij}
    &\to
    (\Omega\hat{X}_I\Omega^{-1})_{ij}
    =
    \sum_{k,l}\Omega_{ik}\hat{X}_{I,kl}\Omega^{-1}_{lj}\, ,
    \nonumber\\
    \hat{P}_{I,ij}
    &\to
    (\Omega\hat{P}_I\Omega^{-1})_{ij}
    =
    \sum_{k,l}\Omega_{ik}\hat{P}_{I,kl}\Omega^{-1}_{lj}\, . 
\end{align}

Gauge-invariant states span a subspace of $\mathcal{H}_{\rm ext}$ which we denote by $\mathcal{H}_{\rm inv}$.
To take into account the gauge-singlet constraint, one can either restrict the Hilbert space to $\mathcal{H}_{\rm inv}$, or one can identify the states in $\mathcal{H}_{\rm ext}$ that transform to each other under SU($N$) transformations. 
$\mathcal{H}_{\rm ext}$ admits the truncation scheme discussed in Sec.~\ref{sec:qubit-truncation}. 

We assign $Q$ qubits to each bosonic degree of freedom. 
Then, the number of qubits needed to describe the Hilbert space is $d(N^2-1)Q$. 

%%%%%%%%%%%%%%%%%%%%%%%
%%%%%%%%%%%%%%%%%%%%%%%
\subsubsection*{Without traceless condition}
%%%%%%%%%%%%%%%%%%%%%%%
%%%%%%%%%%%%%%%%%%%%%%%
We can repeat the same construction, just by replacing $\mathbb{R}^{d(N^2-1)}$ with $\mathbb{R}^{dN^2}$. 

%%%%%%%%%%%%%%%%%%%%%%%%%%
%%%%%%%%%%%%%%%%%%%%%%%%%%
\subsection{A closer look at the Hamiltonian}\label{sec:MM-Hamiltonian-closer-look}
%%%%%%%%%%%%%%%%%%%%%%%%%%
%%%%%%%%%%%%%%%%%%%%%%%%%%
Let us study the Hamiltonian \eqref{MM-Hamiltonian} more closely. 
This section serves as a preparation for the cost estimate in later sections. 
%%%%%%%%%%%%%%%%%%%%%%%%%%
%%%%%%%%%%%%%%%%%%%%%%%%%%
\subsubsection*{Kinetic term}\label{sec:MM_kinetic}
%%%%%%%%%%%%%%%%%%%%%%%%%%
%%%%%%%%%%%%%%%%%%%%%%%%%%
When the traceless condition is imposed, the kinetic term becomes
\begin{align}
\frac{1}{2}\sum_{I=1}^d\mathrm{Tr}\hat{P}_I^2
=
\frac{1}{2}\sum_{I=1}^d\sum_{\alpha=1}^{N^2-1}(\hat{P}_I^\alpha)^2\, . 
\end{align}
Therefore, there are $d(N^2-1)$ terms in the kinetic part. 
When the traceless condition is not imposed, there are $dN^2$ terms:
\begin{align}
\frac{1}{2}\sum_{I=1}^d\mathrm{Tr}\hat{P}_I^2
=
\frac{1}{2}\sum_{I=1}^d
\left[
\sum_{i<j}\left(
(\hat{P}_{I,ij}^{\rm (R)})^2
+
(\hat{P}_{I,ij}^{\rm (I)})^2
\right)
+
\sum_{i}
(\hat{P}_{I,ii})^2
\right]\, . 
\end{align}
Either way, the kinetic term takes the standard form $\sum_a\hat{p}_a^2/2$. 

A simple way to treat the kinetic term is to use the quantum Fourier transform and change the basis to the momentum basis.
The operation of the gates can be completely parallelized because $\hat{p}_a^2$ is diagonal in the momentum basis.
%%%%%%%%%%%%%%%%%%%%%%%%%%
%%%%%%%%%%%%%%%%%%%%%%%%%%
\subsubsection*{Potential term}\label{sec:MM_interaction}
%%%%%%%%%%%%%%%%%%%%%%%%%%
%%%%%%%%%%%%%%%%%%%%%%%%%%
When a traceless condition is imposed, we can write the potential term as 
\begin{align}
    \mathrm{Tr}[\hat{X}_I,\hat{X}_J]^2
    &=
    \sum_\gamma
    \left(
    \mathrm{i}\sum_{\alpha\beta}\hat{X}_I^\alpha\hat{X}_J^\beta f_{\alpha\beta}{}^\gamma
    \right)
    \left(
    \mathrm{i}\sum_{\alpha',\beta'}\hat{X}_I^{\alpha'}\hat{X}_J^{\beta'}f_{\alpha'\beta'}{}^\gamma
    \right)
    \nonumber\\    
    &\equiv
    \sum_{\alpha,\beta,\alpha',\beta'}
    C^{\alpha\beta\alpha'\beta'}
    \hat{X}_I^{\alpha}\hat{X}_J^{\beta}
    \hat{X}_I^{\alpha'}\hat{X}_J^{\beta'}\, ,
    \label{MM-interaction}
\end{align}
where
\begin{align}
C^{\alpha\beta\alpha'\beta'}
    \equiv
    -
    \sum_\gamma
    f_{\alpha\beta}{}^\gamma
    f_{\alpha'\beta'\gamma}\, . 
\end{align}
There are order $N^4$ nonzero components of $C$. 
The number of combinations for $I,J$ is $d(d-1)/2$. Therefore,  the number of quartic interaction terms in \eqref{MM-interaction} scales as $d(d-1)N^4$. It is straightforward to write $C^{\alpha\beta\alpha'\beta'}$ for a given choice of generators using any computer algebra system.

When we do not impose the traceless condition, we go back to \eqref{MM-Hamiltonian} and look at the commutator term. We can examine the terms $\mathrm{Tr}(X_IX_JX_IX_J)$ and $\mathrm{Tr}(X_IX_IX_JX_J)$ separately. 
As example, let us see the former, which can be written as 
\begin{align}
\sum_{i,j,k,l}
X_{I,ij}X_{J,jk}X_{I,kl}X_{J,li}\, . 
\label{eq:XIXJXIXJ}
\end{align}
For the counting to the leading order in $N$, we can assume that $i,j,k,l$ are all different. 
Hence there are $N^4$ terms in the sum. 
In this way, we can see that the number of terms in the potential term scales as $d(d-1)N^4$. 

%%%%%%%%%%%%%%%%%%%%%%%%%%
%%%%%%%%%%%%%%%%%%%%%%%%%%
\subsubsection*{Penalty term to impose singlet constraint}\label{sec:MM_gauge_penalty}
%%%%%%%%%%%%%%%%%%%%%%%%%%
%%%%%%%%%%%%%%%%%%%%%%%%%%
By using the structure constant $f_{\alpha\beta\gamma}$ that is totally antisymmetric and related to the generators by $[\tau_\alpha,\tau_\beta]=\mathrm{i}f_{\alpha\beta\gamma}\tau_\gamma$, generators of gauge transformations can be written as
\begin{align}
\hat{G}_\alpha=\mathrm{i}\sum_{I,\beta,\gamma}f_{\alpha\beta\gamma}\hat{X}_{I,\beta}\hat{P}_{I,\gamma}\, . 
\end{align}
Note that there is no ambiguity in the operator ordering on the right-hand side because $f_{\alpha\beta\gamma}=0$ if $\beta=\gamma$. 

One way to forbid SU($N$) non-singlet states explicitly is to add to the Hamiltonian a penalty term $c\sum_\alpha\hat{G}_\alpha^2$ with a large positive coefficient $c$~\cite{Halimeh2020reliability,Halimeh2022gauge,VanDamme2023reliability}. 
In this paper, we do not consider this option. 
If the Hamiltonian time evolution is precise, then it respects SU($N$) invariance, and hence the gauge invariant sector of the Hilbert space will not be left.

%%%%%%%%%%%%%%%%%%%%%%%%%%
%%%%%%%%%%%%%%%%%%%%%%%%%%
\section{Orbifold Lattice}\label{sec:orbifold-lattice}
%%%%%%%%%%%%%%%%%%%%%%%%%%
%%%%%%%%%%%%%%%%%%%%%%%%%%
Next, we study $(d+1)$-dimensional SU($N$) Yang--Mills theory defined on the orbifold lattice~\cite{Buser:2020cvn}.\footnote{
By default, orbifold lattice theory has U($N$) gauge fields rather than SU($N$). 
For pure Yang--Mills theory, the U(1) part decouples and the SU($N$) part is not affected.
See Refs.~\cite{Buser:2020cvn,Bergner:2024qjl} for the removal of the U(1) part in QCD. 
} 
The present discussion focuses on $d=2$ and $d=3$, but we keep $d$ arbitrary as we can equally easily treat larger dimensions.
The number of lattice sites is $L^d=V_{\rm lattice}$, and periodic boundary conditions are assumed. 

If you already know the orbifold lattice construction, an easy way to understand why its use simplifies simulations is to notice that the orbifold lattice is obtained from the SU($NL^d$) $2d$-matrix model via the orbifold projection~\cite{Kaplan:2002wv}. 
Therefore, if we can simulate the matrix model with the Hamiltonian discussed in the previous section, we can also simulate the orbifold lattice gauge theory. 

The orbifold lattice Hamiltonian can be written in terms of the complex link variables $Z_{j,\vec{n}}$, and their canonical conjugates $P_{j,\vec{n}}$. For $d=3$, the Hamiltonian is\footnote{For $d=2$, we need to replace $g_{\rm 4d}^2$ with $ag_{\rm 3d}^2$, where $a$ is the lattice spacing.}
\begin{align}
\hat{H}
&=
\sum_{\vec{n}}
{\rm Tr}\Biggl(
\sum_{j=1}^3 \hat{P}_{j,\vec{n}}\hat{\bar{P}}_{j,\vec{n}}
+
\frac{g_{\rm 4d}^2}{2a^3}\left|\sum_{j=1}^3
\left(
\hat{Z}_{j,\vec{n}} \hat{\bar{Z}}_{j,\vec{n}} -\hat{\bar{Z}}_{j,\vec{n}-\hat{j}}\hat{Z}_{j,\vec{n}-\hat{j}}
\right)
\right|^2 
\nonumber\\
&\qquad\qquad\qquad
+
\frac{2g_{\rm 4d}^2}{a^3}\sum_{j<k}
\left|
\hat{Z}_{j,\vec{n}}\hat{Z}_{k,\vec{n}+\hat{j}}
-
\hat{Z}_{k,\vec{n}}\hat{Z}_{j,\vec{n}+\hat{k}}
\right|^2
 \Biggl)
 + 
 \Delta\hat{H}\, . 
\label{eq:orbifold-Hamiltonian}
\end{align}
where
\begin{align}
\Delta\hat{H}
&\equiv
\frac{m^2g_{\rm 4d}^2}{2a}\sum_{\vec{n}}\sum_{j=1}^3{\rm Tr}
\left|
\hat{Z}_{j,\vec{n}}
\hat{\bar{Z}}_{j,\vec{n}}
-
\frac{a}{2g_{\rm 4d}^2}
\right|^2
\nonumber\\
&\quad
+
\frac{N\mu^2g_{\rm 4d}^2}{2a}\sum_{\vec{n}}\sum_{j=1}^3
\left| \frac{1}{N}{\rm Tr}(\hat{Z}_{j,\vec{n}}\hat{\bar{Z}}_{j,\vec{n}}) -\frac{a}{2g_{\rm 4d}^2}\right|^2\, . 
\label{orbifold-Hamiltonian-mass}
\end{align}
Note that $\bar{Z}_{j,\vec{n}}$ and $\bar{P}_{j,\vec{n}}$ stand for Hermitian conjugates of $N\times N$ matrices, i.e., 
\begin{align}
\bar{Z}_{j,\vec{n};ab}=(Z_{j,\vec{n};ba})^\ast\, , 
\qquad
\bar{P}_{j,\vec{n};ab}=(P_{j,\vec{n};ba})^\ast\, . 
\end{align}
We do not use dagger $\dagger$ because we save it for conjugate operators acting on the Hilbert space. For the operators this implies  
\begin{align}
\hat{\bar{Z}}_{j,\vec{n};ab}=\left(\hat{Z}_{j,\vec{n};ba}\right)^\dagger\, , 
\qquad
\hat{\bar{P}}_{j,\vec{n};ab}=\left(\hat{P}_{j,\vec{n};ba}\right)^\dagger\, ,  
\end{align}
The canonical commutation relation is
\begin{align}
    [\hat{Z}_{j,\vec{n};ab},\hat{\bar{P}}_{k,\vec{n}';cd}]
    =
    [\hat{Z}_{j,\vec{n};ab},\left(\hat{P}_{k,\vec{n}';dc}\right)^\dagger]
    =\mathrm{i}\delta_{jk}\delta_{\vec{n}\vec{n}'}\delta_{ad}\delta_{bc}\, . 
\end{align}
We denote the real and imaginary parts by the superscripts $\textrm{(R)}$ and $\textrm{(I)}$, respectively, to rewrite the variables and commutation relations as
\begin{align}
    \hat{Z}_{j,\vec{n};ab}
    =
    \frac{\hat{Z}^{\rm (R)}_{j,\vec{n};ab}
    +
    \mathrm{i}
    \hat{Z}^{\rm (I)}_{j,\vec{n};ab}}{\sqrt{2}}\, , 
    \qquad
    \hat{P}_{j,\vec{n};ab}
    =
    \frac{\hat{P}^{\rm (R)}_{j,\vec{n};ab}
    +
    \mathrm{i}
    \hat{P}^{\rm (I)}_{j,\vec{n};ab}}{\sqrt{2}}\, . 
\end{align}
The real and imaginary parts are taken to be self-adjoint, i.e.,
\begin{align}
    \left(\hat{Z}^{\rm (R)}_{j,\vec{n};ab}\right)^\dagger
    =
    \hat{Z}^{\rm (R)}_{j,\vec{n};ab}\, , 
    \qquad
    \left(\hat{Z}^{\rm (I)}_{j,\vec{n};ab}\right)^\dagger
    =
    \hat{Z}^{\rm (I)}_{j,\vec{n};ab}\, , 
\end{align}
and the same for $\hat{P}$. Therefore, in terms of complex operators, 
\begin{align}
    \hat{Z}^{\rm (R)}_{j,\vec{n};ab}
    =
    \frac{
    \hat{Z}_{j,\vec{n};ab}
    +
    \left(\hat{Z}_{j,\vec{n};ab}\right)^\dagger
    }{\sqrt{2}}
    =
    \frac{
    \hat{Z}_{j,\vec{n};ab}
    +
    \hat{\bar{Z}}_{j,\vec{n};ba}
    }{\sqrt{2}}\, , 
\end{align}
and so on. 
The commutation relation is 
\begin{align}
    [\hat{Z}^{\rm (R)}_{j,\vec{n};ab},\hat{P}^{\rm (R)}_{k,\vec{n}';cd}]
    =
    [\hat{Z}^{\rm (I)}_{j,\vec{n};ab},\hat{P}^{\rm (I)}_{k,\vec{n}';cd}]
    =
    \mathrm{i}\delta_{jk}\delta_{\vec{n}\vec{n}'}\delta_{ac}\delta_{bd}\, .
\end{align}
In terms of $\hat{Z}^{\rm (R)}$ and $\hat{Z}^{\rm (I)}$, the Hamiltonian reduces to the form \eqref{eq:simple-form}. 

So far, the Hamiltonian is merely a quiver matrix model. (See e.g., Ref.~\cite{Douglas:1996sw} for a review on quivers in the context of quantum field theory and string theory.) A crucial step is to generate a lattice by using dimensional deconstruction~\cite{Arkani-Hamed:2001kyx}. To see how Yang--Mills theory is obtained from this Hamiltonian, we write $Z_{j,\vec{n}}$ as
\begin{align}
Z_{j,\vec{n}}
=
\sqrt{\frac{a}{2g_{\rm 4d}^2}}
W_{j,\vec{n}}
U_{j,\vec{n}}\, 
\end{align}
where $U_{j,\vec{n}}$ is unitary and $W_{j,\vec{n}}\equiv\sqrt{\frac{2d_{\rm 4d}^2}{a}}\sqrt{Z_{j,\vec{n}}Z_{j,\vec{n}}^\dagger}$ is a positive-definite Hermitian matrix. By writing $U_{j,\vec{n}}$ and $W_{j,\vec{n}}$ as $U_{j,\vec{n}}=\exp\left(\mathrm{i}ag_{\rm 4d}A_{j,\vec{n}}\right)$ and $W_{j,\vec{n}}=\exp\left(ag_{\rm 4d}\phi_{j,\vec{n}}\right)$, respectively, we can interpret $A_j$ and $\phi_{j}$ as the gauge fields and adjoint scalars~\cite{Kaplan:2002wv}. To justify this interpretation, we can stabilize scalars by introducing a large mass in the additional term\footnote{Nonzero vacuum expectation values of scalars effectively shift the lattice spacing. Optionally, one could add a quadratic term $\mathrm{Tr}(\hat{Z}_{j,\vec{n}}\hat{\bar{Z}}_{j,\vec{n}})$ to control the vacuum expectation value of scalars without adding extremely large bare mass. } $\Delta\hat{H}$. Then, we obtain Yang--Mills theory coupled to scalar fields. Indeed, if we turn off the scalars (i.e., choosing $W$ to be the identity), the term   $\mathrm{Tr}\left|\sum_{j=1}^3\left(\hat{Z}_{j,\vec{n}} \hat{\bar{Z}}_{j,\vec{n}} -\hat{\bar{Z}}_{j,\vec{n}-\hat{j}}\hat{Z}_{j,\vec{n}-\hat{j}}\right)\right|^2$  in (\ref{eq:orbifold-Hamiltonian}) becomes zero while the term,
$\sum_{j<k}\mathrm{Tr}\left|\hat{Z}_{j,\vec{n}}\hat{Z}_{k,\vec{n}+\hat{j}}-\hat{Z}_{k,\vec{n}}\hat{Z}_{j,\vec{n}+\hat{k}}\right|^2$ gives the plaquette term, which leads to the magnetic term of Yang--Mills theory. 
Focusing on the leading order in $\phi$, 
the first term gives $\mathrm{Tr}(\sum_jD_j\phi_j)^2$ while the second term gives 
$\sum_{j<k}\mathrm{Tr}(D_j\phi_k-D_k\phi_j)^2$, which sum up to $\sum_{j,k}\mathrm{Tr}(D_j\phi_k)^2$ up to total derivatives. The quartic interaction $\mathrm{Tr}[\phi_j,\phi_k]^2$ appears as well. See Refs.~\cite{Buser:2020cvn,Bergner:2024qjl} for details. 
At low energy, large mass scalars decouple and pure Yang--Mills theory is obtained. The mass of scalars can be as large as the lattice cutoff scale. 

%%%%%%%%%%%%%%%%%%%%%%%%%%
%%%%%%%%%%%%%%%%%%%%%%%%%%
\subsection{Hilbert space}
%%%%%%%%%%%%%%%%%%%%%%%%%%
%%%%%%%%%%%%%%%%%%%%%%%%%%
SU($N$) gauge transformations are defined by $Z_{j,\vec{n}}\to\Omega^{-1}_{\vec{n}}Z_{j,\vec{n}}\Omega_{\vec{n}+\hat{j}}$, which is equivalent to $U_{j,\vec{n}}\to\Omega^{-1}_{\vec{n}}U_{j,\vec{n}}\Omega_{\vec{n}+\hat{j}}$ and $W_{j,\vec{n}}\to\Omega^{-1}_{\vec{n}}W_{j,\vec{n}}\Omega_{\vec{n}}$. 
We use the extended Hilbert space $\mathcal{H}_{\rm ext}$, which can be defined by using the coordinate eigenstates $\ket{Z}$ that satisfy $\hat{Z}_{j,\vec{n}}\ket{Z}=Z_{j,\vec{n}}\ket{Z}$ as
\begin{align}
\mathcal{H}_{\rm ext}
=
\left\{
\ket{\Psi}
\equiv
\int\mathrm{d}^{2dN^2V_{\rm lattice}}Z\, \Psi(Z)\ket{Z}
\Bigg|
\int\mathrm{d}^{2dN^2V_{\rm lattice}}Z\, |\Psi(Z)|^2<\infty
\right\}\, . 
\end{align}
We can also use the momentum eigenstates $\ket{P}$ that satisfy $\hat{P}_{j,\vec{n}}\ket{P}=P_{j,\vec{n}}\ket{P}$ as
\begin{align}
\mathcal{H}_{\rm ext}
=
\left\{
\ket{\Psi}
\equiv
\int\mathrm{d}^{2dN^2V_{\rm lattice}}P\, \tilde{\Psi}(P)\ket{P}
\Bigg|
\int\mathrm{d}^{2dN^2V_{\rm lattice}}P\, |\tilde{\Psi}(P)|^2<\infty
\right\}\, . 
\end{align}
Wave functions $\Psi(Z)$ and $\tilde{\Psi}(P)$ are related by Fourier transform. 
Under the SU($N$) gauge transformation, these states transform as
\begin{align}
    \ket{Z}
    \to
    \ket{\Omega^{-1}Z\Omega}\, , 
    \qquad
    \ket{P}
    \to
    \ket{\Omega^{-1}P\Omega}\, ,     
\end{align}
while the operators transform as
\begin{align}
    \hat{Z}_{j,\vec{n};ab}
    &\to
    (\Omega_{\vec{n}}\hat{Z}_{j,\vec{n}}\Omega^{-1}_{\vec{n}+\hat{j}})_{ab}
    =
    \sum_{c,d}\Omega_{\vec{n};ac}\hat{Z}_{j,\vec{n};cd}\Omega^{-1}_{\vec{n}+\hat{j};db}\, ,
    \nonumber\\
    \hat{P}_{j,\vec{n};ab}
    &\to
    (\Omega_{\vec{n}}\hat{P}_{j,\vec{n}}\Omega^{-1}_{\vec{n}+\hat{j}})_{ab}
    =
    \sum_{c,d}\Omega_{\vec{n};ac}\hat{P}_{j,\vec{n};cd}\Omega^{-1}_{\vec{n}+\hat{j};db}\, . 
\end{align}
Gauge-invariant states span a subspace of $\mathcal{H}_{\rm ext}$ which we denote by $\mathcal{H}_{\rm inv}$.

In this paper, we do not impose an SU($N$)-singlet constraint on the Hilbert space. 
There are $dV_{\rm lattice}$ links and each link carries $2N^2$ real bosonic degrees of freedom.
Therefore, there are $2N^2dV_{\rm lattice}$ real bosonic degrees of freedom in total.
We assign $Q$ qubits to each of them, such that the number of qubits needed to describe the Hilbert space is $2N^2dV_{\rm lattice}Q$.
%%%%%%%%%%%%%%%%%%%%%%%%%%
%%%%%%%%%%%%%%%%%%%%%%%%%%
\subsection{A closer look at the Hamiltonian}
%%%%%%%%%%%%%%%%%%%%%%%%%%
%%%%%%%%%%%%%%%%%%%%%%%%%%
Below, we investigate the Hamiltonian more closely. 
This section serves as a preparation for the cost estimate in later sections. The readers who are not interested in the details of the cost estimate can skip this section.  
%%%%%%%%%%%%%%%%%%%%%%%%%%
%%%%%%%%%%%%%%%%%%%%%%%%%%
\subsubsection*{Kinetic term}\label{sec:orbifold_kinetic}
%%%%%%%%%%%%%%%%%%%%%%%%%%
%%%%%%%%%%%%%%%%%%%%%%%%%%
By construction, the kinetic term is 
\begin{align}
\hat{H}
&=
\frac{1}{2}
\sum_{\vec{n}}
\sum_{j=1}^3
\sum_{a,b=1}^N
\left(
(\hat{P}^{\rm (R)}_{j,\vec{n};ab})^2
+
(\hat{P}^{\rm (I)}_{j,\vec{n};ab})^2
\right)\, , 
\end{align}
which is the same as the standard form in \eqref{eq:simple-form}.

There are $2N^2dV_{\rm lattice}$ terms, which can be easily treated by using a quantum Fourier transform and going to the momentum basis. 
%%%%%%%%%%%%%%%%%%%%%%%%%%
%%%%%%%%%%%%%%%%%%%%%%%%%%
\subsubsection*{Potential term}\label{sec:orbifold_interaction}
\begin{figure}[htbp]
  \begin{center}
   \includegraphics[width=80mm]{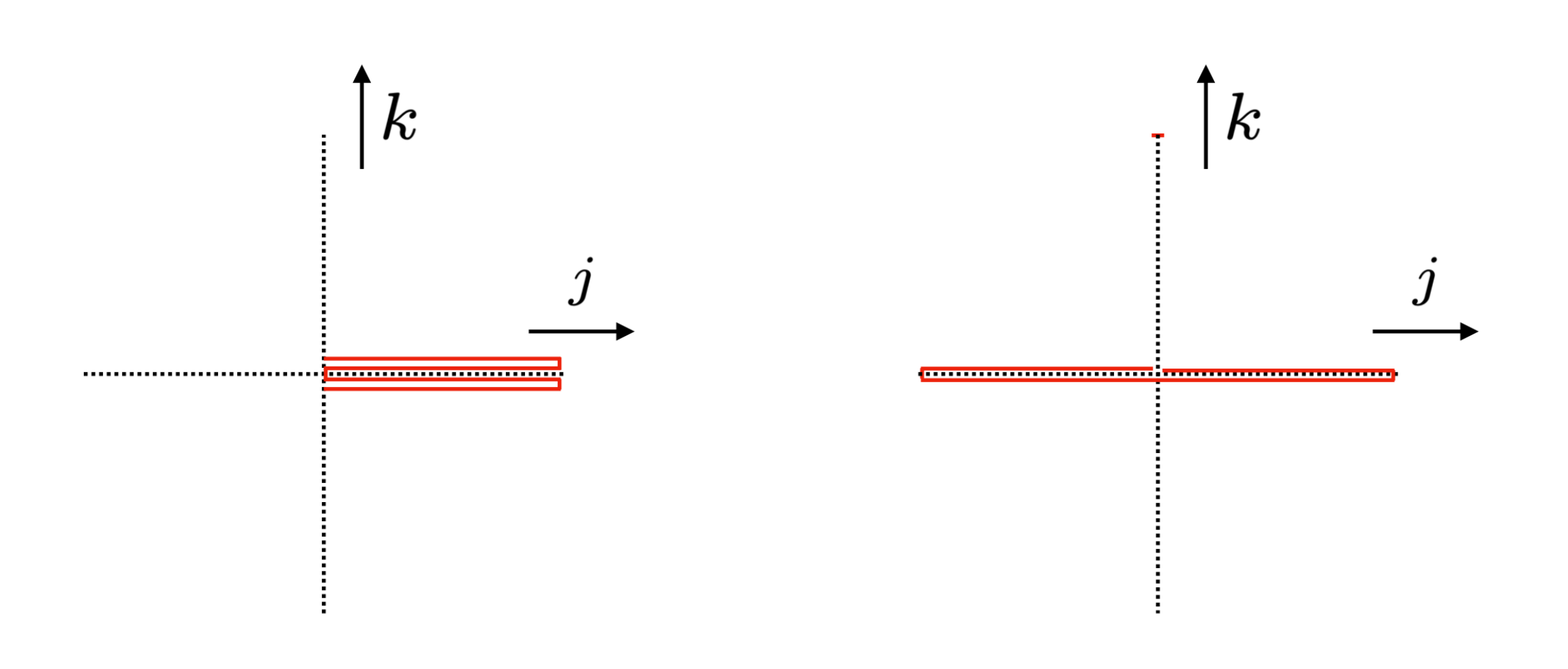}
  \end{center}
  \caption{Visual representation of 
$\hat{Z}_{j,\vec{n}} \hat{\bar{Z}}_{j,\vec{n}} \hat{Z}_{j,\vec{n}} \hat{\bar{Z}}_{j,\vec{n}}$ (left) and 
$\hat{Z}_{j,\vec{n}} \hat{\bar{Z}}_{j,\vec{n}} 
\hat{\bar{Z}}_{j,\vec{n}-\hat{j}}\hat{Z}_{j,\vec{n}-\hat{j}}$ (right). 
Red lines represent links. 
  }\label{fig:H-additional-1}
\end{figure}

\begin{figure}[htbp]
  \begin{center}
   \includegraphics[width=80mm]{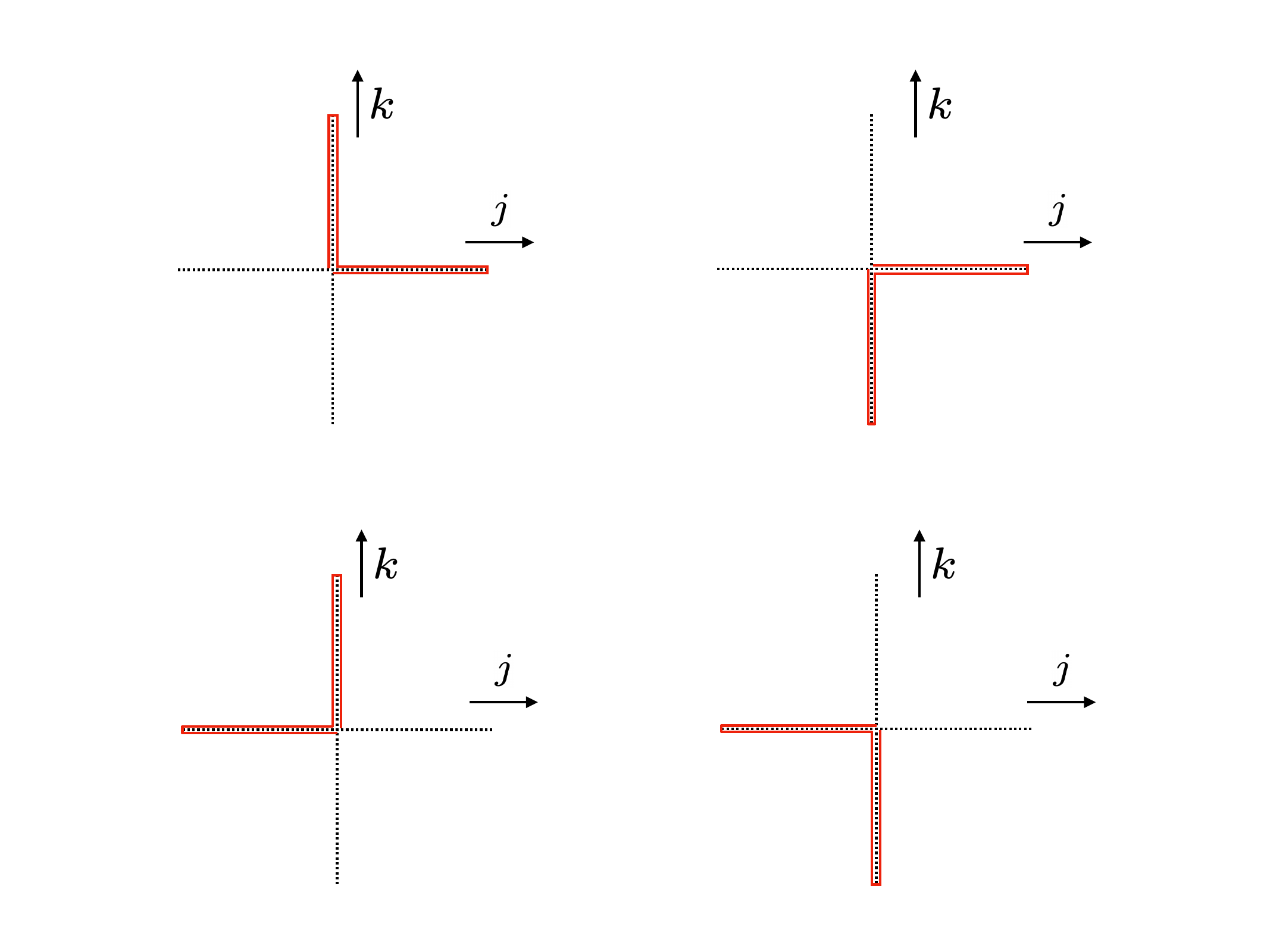}
  \end{center}
  \caption{Visual representation of  $\hat{Z}_{j,\vec{n}} \hat{\bar{Z}}_{j,\vec{n}} \hat{Z}_{k,\vec{n}} \hat{\bar{Z}}_{k,\vec{n}}$ (Top, Left),
$\hat{Z}_{j,\vec{n}} \hat{\bar{Z}}_{j,\vec{n}} 
\hat{\bar{Z}}_{k,\vec{n}-\hat{k}}\hat{Z}_{k,\vec{n}-\hat{k}}$ (Top, Right),
$\hat{\bar{Z}}_{j,\vec{n}-\hat{j}}\hat{Z}_{j,\vec{n}-\hat{j}}\hat{Z}_{k,\vec{n}} \hat{\bar{Z}}_{k,\vec{n}}$ (Bottom, Left),
$\hat{\bar{Z}}_{j,\vec{n}-\hat{j}}\hat{Z}_{j,\vec{n}-\hat{j}}
\hat{\bar{Z}}_{k,\vec{n}-\hat{k}}\hat{Z}_{k,\vec{n}-\hat{k}}$ (Bottom, Right). Red lines represent links. 
  }\label{fig:H-additional-2}
\end{figure}

%%%%%%%%%%%%%%%%%%%%%%%%%%
%%%%%%%%%%%%%%%%%%%%%%%%%%
The potential term in $\hat{H}$ can be written as
\begin{align}
&
     \frac{g_{\rm 4d}^2}{a^3}
  \sum_{\vec{n}}
{\rm Tr}
\sum_{j}
\Biggl(
\hat{Z}_{j,\vec{n}} \hat{\bar{Z}}_{j,\vec{n}} \hat{Z}_{j,\vec{n}} \hat{\bar{Z}}_{j,\vec{n}}
-
\hat{Z}_{j,\vec{n}} \hat{\bar{Z}}_{j,\vec{n}} 
\hat{\bar{Z}}_{j,\vec{n}-\hat{j}}\hat{Z}_{j,\vec{n}-\hat{j}}
 \Biggl)
 \nonumber\\
&
\quad
+
  \frac{g_{\rm 4d}^2}{a^3}
  \sum_{\vec{n}}
{\rm Tr}
\sum_{j<k}
\Biggl(
\hat{Z}_{j,\vec{n}} \hat{\bar{Z}}_{j,\vec{n}} \hat{Z}_{k,\vec{n}} \hat{\bar{Z}}_{k,\vec{n}}
+
\hat{Z}_{j,\vec{n}} \hat{\bar{Z}}_{j,\vec{n}} 
\hat{\bar{Z}}_{k,\vec{n}-\hat{k}}\hat{Z}_{k,\vec{n}-\hat{k}}
\nonumber\\
&\qquad
+
\hat{\bar{Z}}_{j,\vec{n}-\hat{j}}\hat{Z}_{j,\vec{n}-\hat{j}}
\hat{Z}_{k,\vec{n}} \hat{\bar{Z}}_{k,\vec{n}}
+
\hat{\bar{Z}}_{j,\vec{n}-\hat{j}}\hat{Z}_{j,\vec{n}-\hat{j}}
\hat{\bar{Z}}_{k,\vec{n}-\hat{k}}\hat{Z}_{k,\vec{n}-\hat{k}}
\nonumber\\
&\qquad
-
2\hat{Z}_{j,\vec{n}}\hat{Z}_{k,\vec{n}+\hat{j}}\hat{\bar{Z}}_{j,\vec{n}+\hat{k}}\hat{\bar{Z}}_{k,\vec{n}}
-
2\hat{Z}_{k,\vec{n}}\hat{Z}_{j,\vec{n}+\hat{k}}\hat{\bar{Z}}_{k,\vec{n}+\hat{j}}\hat{\bar{Z}}_{j,\vec{n}}
 \Biggl)\, .  \label{orbifold-H-potential}
\end{align}
In \eqref{orbifold-H-potential}, in addition to plaquettes (the final line), there are terms of the forms Fig.~\ref{fig:H-additional-1} and Fig.~\ref{fig:H-additional-2}. By using the real and imaginary parts of $\hat{Z}$, it is straightforward to rewrite them in the standard form of \eqref{eq:simple-form}. The number of terms scales as $d^2N^4V_{\rm lattice}$. 

The additional term $\Delta\hat{H}$ does not change this conclusion. 
The first term on the right-hand side of \eqref{orbifold-Hamiltonian-mass} is 
\begin{align}
\sum_{j=1}^3{\rm Tr}
\left|
\hat{Z}_{j,\vec{n}}
\hat{\bar{Z}}_{j,\vec{n}}
-
\frac{a}{2g_{\rm 4d}^2}
\right|^2
=
\sum_{j=1}^3{\rm Tr}\left(
\hat{Z}_{j,\vec{n}}
\hat{\bar{Z}}_{j,\vec{n}}
\hat{Z}_{j,\vec{n}}
\hat{\bar{Z}}_{j,\vec{n}}
-
\frac{a}{g_{\rm 4d}^2}\hat{Z}_{j,\vec{n}}
\hat{\bar{Z}}_{j,\vec{n}}
\right)
+
\textrm{const}\, . 
\end{align}
Note that the first term $\hat{Z}_{j,\vec{n}}
\hat{\bar{Z}}_{j,\vec{n}}
\hat{Z}_{j,\vec{n}}
\hat{\bar{Z}}_{j,\vec{n}}$ is in $\hat{H}$ as well.
$\mathrm{Tr}(\hat{Z}_{j,\vec{n}}
\hat{\bar{Z}}_{j,\vec{n}})$ has $2N^2\times d$ terms of the form $\hat{x}^2$. 
From the second term on the right hand side of \eqref{orbifold-Hamiltonian-mass}, we obtain $\mathrm{Tr}(\hat{Z}_{j,\vec{n}}
\hat{\bar{Z}}_{j,\vec{n}})$ and $\left[\mathrm{Tr}(\hat{Z}_{j,\vec{n}}
\hat{\bar{Z}}_{j,\vec{n}})\right]^2$. 
The latter can be written as a sum of $4N^4\times dV_{\rm lattice}$ terms of the form $\hat{x}_1^2\hat{x}_2^2$. 

%%%%%%%%%%%%%%%%%%%%%%%%%%
%%%%%%%%%%%%%%%%%%%%%%%%%%
\subsubsection*{Penalty term to impose singlet constraint}\label{sec:orbifold_gauge_penalty}
%%%%%%%%%%%%%%%%%%%%%%%%%%
%%%%%%%%%%%%%%%%%%%%%%%%%%
Generators of gauge transformations at a spatial lattice site $\vec{n}$ can be written as 
\begin{align}
\hat{G}_{\vec{n},pq}
\equiv
\mathrm{i}\sum_{j=1}^3
\left(
-
\hat{Z}_{j,\vec{n}}\hat{\bar{P}}_{j,\vec{n}}
+
\hat{P}_{j,\vec{n}}\hat{\bar{Z}}_{j,\vec{n}}
-
\hat{\bar{Z}}_{j,\vec{n}-\hat{j}}\hat{P}_{j,\vec{n}-\hat{j}}
+
\hat{\bar{P}}_{j,\vec{n}-\hat{j}}\hat{Z}_{j,\vec{n}-\hat{j}}
\right)_{pq}. 
\label{eq:gauge-generators-orbifold}
\end{align}

As we already mentioned in Sec.~\ref{sec:Matrix-Model}, it is possible in principle to remove SU($N$) non-singlet states explicitly from the spectrum by adding a penalty term proportional to $\sum_\alpha\hat{G}_\alpha^2$~\cite{Halimeh2020reliability,Halimeh2022gauge,VanDamme2023reliability}. In this paper, we do not consider this option. If the Hamiltonian time evolution is precise, then it respects SU($N$) invariance, and hence such a penalty term is not necessary.

%%%%%%%%%%%%%%%%%%%%%%%%%%%
%%%%%%%%%%%%%%%%%%%%%%%%%%%
\section{Resource estimate for Hamiltonian time evolution: Suzuki--Trotter decomposition algorithm}\label{sec:Trotter_cost}
%%%%%%%%%%%%%%%%%%%%%%%%%%%
%%%%%%%%%%%%%%%%%%%%%%%%%%%
In this section, we estimate the resources (gate count) needed for Hamiltonian time evolution on a digital quantum computer using the Suzuki--Trotter decomposition.
For gauge theories (i.e., matrix model and orbifold lattice) we adopt the extended Hilbert space.

Given the rapid pace of innovation in quantum computing hardware and software, accurately predicting the performance of an orbifold lattice-based code on a quantum computer by, say, 2030, is challenging. 
Nonetheless, we aim to provide a preliminary upper-bound estimate of its resource requirements to demonstrate how straightforward it is to make such estimates for an orbifold lattice across various lattice geometries, symmetry groups, and matrix models.

As stated above, we do not include a penalty term to enforce the singlet constraint such as $\hat{G}_\alpha^2$ and focus on just one step in the Suzuki--Trotter decomposition.
One should keep in mind that often we would need to implement multiple Suzuki--Trotter steps, with their number scaling with the system size and details of interaction terms in order to keep discretization errors under a given threshold~\cite{Shaw:2020udc}.
For example, the first order product formula will need a number of steps scaling quadratically with the total simulation time and inversely with the discretization error we want to achieve~\cite{Childs:2019hts}.
We neglect this complication in the following analysis, and provide resources for a single step.

%%%%%%%%%%%%%%%%%%%%%%%%%%%
%%%%%%%%%%%%%%%%%%%%%%%%%%%
\subsection{Basic pieces}\label{sec:Trotter_cost_piece}
%%%%%%%%%%%%%%%%%%%%%%%%%%%
%%%%%%%%%%%%%%%%%%%%%%%%%%%
We can consider the momentum part and interaction part separately. 
Because $[\hat{p}_j,\hat{p}_k]=0$, the momentum part factorizes as 
\begin{align}
\exp\left(-\textrm{i}\theta\sum_j\hat{p}^2_j\right)
=
\prod_{j}\exp\left(-\textrm{i}\theta\hat{p}^2_j\right)\, . 
\end{align}
Therefore, the total cost is that for one boson times the number of bosons.
The interaction part factorizes as well.
Schematically, it takes the form 
\begin{align}
\prod_{j,k,l,m}\exp\left(-\textrm{i}\theta C_{jklm}\, 
\hat{x}_j\hat{x}_k\hat{x}_l\hat{x}_m\right)\, . 
\end{align}
Here we ignored the cost of the quadratic and cubic couplings which are computationally cheaper than the quartic couplings.

%%%%%%%%%%%%%%%%%%%%%%%%%%%
%%%%%%%%%%%%%%%%%%%%%%%%%%%
\subsubsection{Estimate for \texorpdfstring{$\exp\left(-\mathrm{i}\theta\hat{x}_j\hat{x}_k\hat{x}_l\hat{x}_m\right)$}{exp(-itx1x2x3x4)}}\label{sec:cost_interaction}
%%%%%%%%%%%%%%%%%%%%%%%%%%%
%%%%%%%%%%%%%%%%%%%%%%%%%%%
As mentioned before, $\hat{x}_j\hat{x}_k\hat{x}_l\hat{x}_m$ is a sum of tensor products of four Pauli $\sigma_z$. 
Schematically, the interaction term is a product of the Pauli rotations, 
\begin{align}
\prod_{pqrs}\exp\left(-\textrm{i}\theta C'_{pqrs}\hat{\sigma}_{z,p}\hat{\sigma}_{z,q}\hat{\sigma}_{z,r}\hat{\sigma}_{z,s}\right)\, . 
\label{eq:interaction-generic-Z}
\end{align}
Below, we will first demonstrate how each of these Pauli rotations can be implemented using CNOT gates and a single-qubit rotation. 
For simulations on the noisy intermediate-scale quantum (NISQ) devices, it is important to reduce the number of CNOT gates. 
Next, we will discuss how many T gates are needed to simulate single-cubit rotation. This is because, for fault-tolerant quantum computing (FTQC), it is important to reduce the number of T gates, which have the largest cost. 
%%%%%%%%%%%%%%%%%%%%%%%%%%%%%%%%%
%%%%%%%%%%%%%%%%%%%%%%%%%%%%%%%%%
\subsubsection*{Counting of CNOT gates}
%%%%%%%%%%%%%%%%%%%%%%%%%%%%%%%%%
%%%%%%%%%%%%%%%%%%%%%%%%%%%%%%%%%
In the following, we demonstrate how these Pauli rotations can be implemented using CNOT gates and a single qubit rotation.
Note that these exponentials are also known as Phase gadgets, or Pauli gadgets~\cite{Cowtan:2019loc}, and are well-recognized structures in quantum circuits.
They can be manipulated and synthesized efficiently by quantum compilers (e.g. thanks to ZX-calculus~\cite{Coecke:2011gh}).
To make this paper self-contained, we will show some useful relations that will help diagonalize these exponentials and decompose them into two-qubit and one-qubit operations.
First, we show that
\begin{align}
\hat{\sigma}_{z,p}\hat{\sigma}_{z,q}\hat{\sigma}_{z,r}\hat{\sigma}_{z,s}
=
\textrm{CNOT}_{p,q}
\textrm{CNOT}_{q,r}
\textrm{CNOT}_{r,s}
\hat{\sigma}_{z,s}
\textrm{CNOT}_{r,s}
\textrm{CNOT}_{q,r}
\textrm{CNOT}_{p,q}\, , 
\label{eq:Z-and-CNOT}
\end{align}
where $\textrm{CNOT}_{p,q}$ is the CNOT gate which uses the qubit $p$ as the controlled qubit and the qubit $q$ as the target qubit:
\begin{align}
    &
    \textrm{CNOT}_{p,q}(\ket{0}_p\ket{0}_q)
    =
    \ket{0}_p\ket{0}_q\, , 
    \qquad
    \textrm{CNOT}_{p,q}(\ket{0}_p\ket{1}_q)
    =
    \ket{0}_p\ket{1}_q\, ,\nonumber\\ 
    &   
    \textrm{CNOT}_{p,q}(\ket{1}_p\ket{0}_q)
    =
    \ket{1}_p\ket{1}_q\, , 
    \qquad
    \textrm{CNOT}_{p,q}(\ket{1}_p\ket{1}_q)
    =
    \ket{1}_p\ket{0}_q\, . 
\end{align}
Equivalently, 
\begin{align}
\textrm{CNOT}_{p,q}\ket{b_p}_p\ket{b_q}_q
=
\ket{b_p}_p\ket{b_p\oplus b_q}_q\, ,  
\label{eq: def of CNOT}
\end{align}
where $\oplus$ represents exclusive OR, i.e., $b_p\oplus b_q = b_p+b_q$ mod 2. 

Let us see how \eqref{eq:Z-and-CNOT} can be obtained. 
From \eqref{eq: def of CNOT}, it is straightforward to show
\begin{align}
&
\textrm{CNOT}_{r,s}
\textrm{CNOT}_{q,r}
\textrm{CNOT}_{p,q}
\ket{b_p}_p
\ket{b_q}_q
\ket{b_r}_r
\ket{b_s}_s
\nonumber\\
&\qquad=
\ket{b_p}_p
\ket{b_p\oplus b_q}_q
\ket{b_p\oplus b_q\oplus b_r}_r
\ket{b_p\oplus b_q\oplus b_r\oplus b_s}_s\, .
\end{align}
Since \eqref{eq: pauli Z} is equivalent to
$
\hat{\sigma}_{z}\ket{b}
=
(-1)^b\ket{b}\, ,$ 
\begin{align}
&
\hat{\sigma}_{z,s}
\textrm{CNOT}_{r,s}
\textrm{CNOT}_{q,r}
\textrm{CNOT}_{p,q}
\ket{b_p}_p
\ket{b_q}_q
\ket{b_r}_r
\ket{b_s}_s
\nonumber\\
&\qquad=
(-1)^{b_p\oplus b_q\oplus b_r\oplus b_s}
\ket{b_p}_p
\ket{b_p\oplus b_q}_q
\ket{b_p\oplus b_q\oplus b_r}_r
\ket{b_p\oplus b_q\oplus b_r\oplus b_s}_s
\end{align}
by combining the former and the latter, and by further multiplying with CNOT gates, we obtain 
\begin{align}
&
\textrm{CNOT}_{p,q}\textrm{CNOT}_{q,r}\textrm{CNOT}_{r,s}
\hat{\sigma}_{z,s}
\textrm{CNOT}_{r,s}
\textrm{CNOT}_{q,r}
\textrm{CNOT}_{p,q}
\ket{b_p}_p
\ket{b_q}_q
\ket{b_r}_r
\ket{b_s}_s
\nonumber\\
&\qquad=
(-1)^{b_p\oplus b_q\oplus b_r\oplus b_s}
\ket{b_p}_p
\ket{b_q}_q
\ket{b_r}_r
\ket{b_s}_s\, . 
\label{z-CNOT-proof-2}
\end{align}
On the other hand,
\begin{align}
\hat{\sigma}_{z,p}
\hat{\sigma}_{z,q}
\hat{\sigma}_{z,r}
\hat{\sigma}_{z,s}
\ket{b_p}_p
\ket{b_q}_q
\ket{b_r}_r
\ket{b_s}_s
=
(-1)^{b_p\oplus b_q\oplus b_r\oplus b_s}
\ket{b_p}_p
\ket{b_q}_q
\ket{b_r}_r
\ket{b_s}_s\, . 
\label{z-CNOT-proof-1}
\end{align}
Comparing \eqref{z-CNOT-proof-2} and \eqref{z-CNOT-proof-1}, we conclude \eqref{eq:Z-and-CNOT}. 

\begin{figure}[ht]
    \centering
        \centering
        \includegraphics[width=0.5\textwidth]{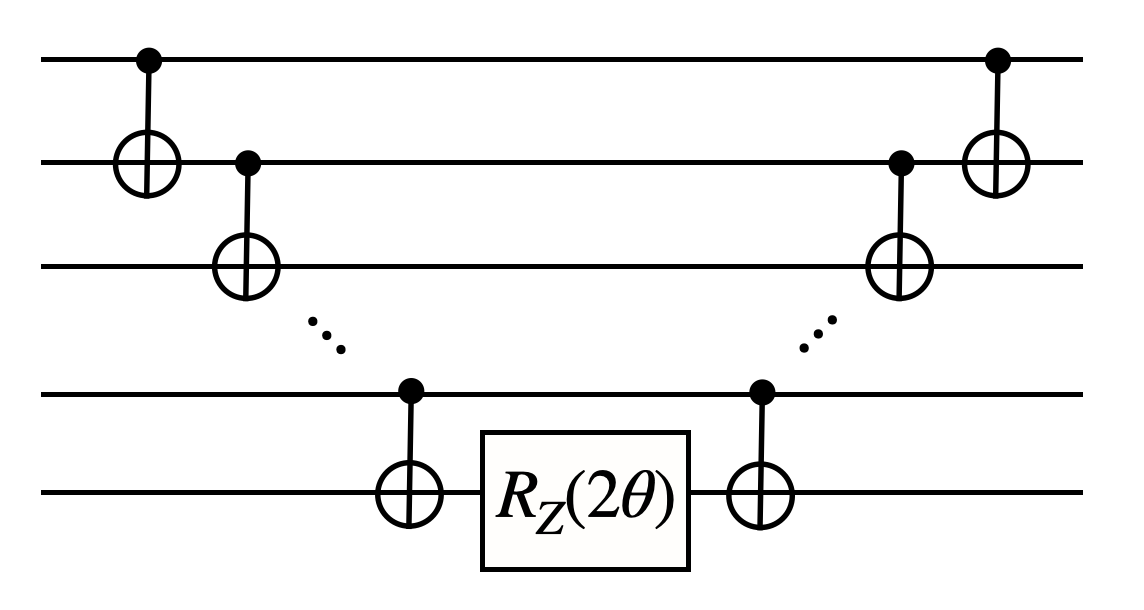}
    \caption{Circuit demonstrating $\exp\left(-\textrm{i}\theta\hat{\sigma}_z\cdots \hat{\sigma}_z\right)$. Here, $R_Z(\theta)=\exp(-\frac{\mathrm{i}\theta}{2}\hat{\sigma}_z)$.}
        \label{fig:cnot_ladder}
\end{figure}

From \eqref{eq:Z-and-CNOT}, we obtain 
\begin{align}
&
\exp\left(-\textrm{i}\theta C'_{pqrs}\hat{\sigma}_{z,p}\hat{\sigma}_{z,q}\hat{\sigma}_{z,r}\hat{\sigma}_{z,s}\right)
\nonumber\\
&\quad=
\textrm{CNOT}_{p,q}
\textrm{CNOT}_{q,r}
\textrm{CNOT}_{r,s}
\exp\left(-\textrm{i}\theta C'_{pqrs}\hat{\sigma}_{z,s}\right)
\textrm{CNOT}_{r,s}
\textrm{CNOT}_{q,r}
\textrm{CNOT}_{p,q}\, ,
\nonumber\\
\label{eq: Pauli rotation and cnot ladder}
\end{align}
which shows how the Suzuki--Trotter step can be implemented by using
CNOT gates and one-qubit rotation gates. See Fig.~\ref{fig:cnot_ladder} for the pictorial representation of \eqref{eq: Pauli rotation and cnot ladder}. 
Note that one can construct rotations with respect to any Pauli operator from \eqref{eq: Pauli rotation and cnot ladder}.
For instance, if a Pauli operator contains $\hat{\sigma}_x$ or
$\hat{\sigma}_y$, one can simply use the change of basis, i.e.,
$\hat{h} \hat{\sigma}_z\hat{h}=\hat{\sigma}_x$ and 
$\hat{s}^\dagger\hat{h}\hat{\sigma}_z\hat{h}\hat{s}=-\hat{\sigma}_y$, where $\hat{h}$ is the Hadamard gate and $\hat{s}$ is the phase gate, i.e., 
\begin{align}
    \hat{h}
    =
    \frac{1}{\sqrt{2}}
    \left(
        \begin{array}{cc}
            1 & 1\\
            1 & -1
        \end{array}
    \right)\, , 
    \qquad
    \hat{s}
    =
    \left(
        \begin{array}{cc}
            1 & 0\\
            0 & \mathrm{i}
        \end{array}
    \right)\, .      
\end{align}

The number of CNOT gates needed to realize \eqref{eq:interaction-generic-Z} is at most 6 times the number of combinations $p,q,r,s$, which is $6Q^4$ times the number of interaction vertices ($\sim d(d-1)N^4$ for a matrix model and $\sim d^2N^4V_{\rm lattice}$ for an orbifold lattice). Here, $Q$ is the number of qubits assigned to each boson. Note that we can reduce the number by taking the products in an appropriate order. For example, 
\begin{align}
&
\exp\left(-\textrm{i}\theta C'_{pqrs}\hat{\sigma}_{z,p}\hat{\sigma}_{z,q}\hat{\sigma}_{z,r}\hat{\sigma}_{z,s}\right)
\cdot
\exp\left(-\textrm{i}\theta C'_{pqrs'}\hat{\sigma}_{z,p}\hat{\sigma}_{z,q}\hat{\sigma}_{z,r}\hat{\sigma}_{z,s'}\right)
\nonumber\\
&=
\textrm{CNOT}_{p,q}
\textrm{CNOT}_{q,r}
\textrm{CNOT}_{r,s}
\exp\left(-\textrm{i}\theta C'_{pqrs}\hat{\sigma}_{z,s}\right)
\textrm{CNOT}_{r,s}
\textrm{CNOT}_{q,r}
\textrm{CNOT}_{p,q}
\nonumber\\
&\quad\times
\textrm{CNOT}_{p,q}
\textrm{CNOT}_{q,r}
\textrm{CNOT}_{r,s'}
\exp\left(-\textrm{i}\theta C'_{pqrs'}\hat{\sigma}_{z,s'}\right)
\textrm{CNOT}_{r,s'}
\textrm{CNOT}_{q,r}
\textrm{CNOT}_{p,q} 
\nonumber\\
&=
\textrm{CNOT}_{p,q}
\textrm{CNOT}_{q,r}
\textrm{CNOT}_{r,s}
\exp\left(-\textrm{i}\theta C'_{pqrs}\hat{\sigma}_{z,s}\right)
\textrm{CNOT}_{r,s}
\nonumber\\
&\quad\times
\textrm{CNOT}_{r,s'}
\exp\left(-\textrm{i}\theta C'_{pqrs'}\hat{\sigma}_{z,s'}\right)
\textrm{CNOT}_{r,s'}
\textrm{CNOT}_{q,r}
\textrm{CNOT}_{p,q}\, . 
\end{align}
In this way, we can eliminate four CNOT gates. 
This simple relation is very useful. Each $\exp\left(-i\theta C_{jklm}\hat{x}_j\hat{x}_k\hat{x}_l\hat{x}_m\right)$ can be written schematically as
\begin{align}
\exp\left(-\textrm{i}\theta C_{jklm}\hat{x}_j\hat{x}_k\hat{x}_l\hat{x}_m\right)
=
\prod_{pqr}
\left(
\prod_s
\exp\left(-\textrm{i}\theta C_{pqrs}\hat{\sigma}_{z,p}\hat{\sigma}_{z,q}\hat{\sigma}_{z,r}\hat{\sigma}_{z,s}\right)
\right)
\label{eq:CNOT-cancellation}
\end{align}
and $\textrm{CNOT}_{p,q}$s and $\textrm{CNOT}_{q,r}$s in 
$\prod_s
\exp\left(-\textrm{i}\theta C_{pqrs}\hat{\sigma}_{z,p}\hat{\sigma}_{z,q}\hat{\sigma}_{z,r}\hat{\sigma}_{z,s}\right)$ cancel out leaving only two $\textrm{CNOT}_{p,q}$s and two $\textrm{CNOT}_{q,r}$s; 
for each set of $(p,q,r)$, the number of CNOT gates left in $\left(\prod_s\cdots\right)$ is $2Q+4$ rather than $6Q$. 

The combination of CNOT gates and single-qubit rotation gates forms a universal gate set, enabling the construction of any quantum algorithm using only these gates. 
This gate set is especially important in
the NISQ era, where quantum computations are performed on physical qubits without the benefit of error correction. In NISQ devices, all gates are implemented directly on the physical qubits. Note that there exist different physical implementations of qubit gates and some platforms, like the H-series hardware by Quantinuum~\cite{quantinuum2024}, can implement directly arbitrary angle two-qubit gates $e^{-\textrm{i}\theta_{pq}\hat{\sigma}_{z,p}\hat{\sigma}_{z,q}}$ using a single laser pulse.

Without error corrections, the fidelity of quantum gates becomes a critical factor in determining how well a computation can be performed. 
Notably, the fidelity of two-qubit gates, like the CNOT gate, is typically much lower than that of single-qubit gates. 
As a result, errors accumulate more rapidly when a quantum circuit relies heavily on two-qubit gates, which can significantly degrade the overall performance of the algorithm.

Given the lower fidelity associated with two-qubit gates, it is reasonable to consider the number of CNOT gates as a key computational resource when assessing the efficiency and accuracy of quantum simulations in the NISQ regime.
The CNOT gate count serves as a useful metric for estimating how error-prone a quantum algorithm might be on current hardware. 
By minimizing the number of CNOT gates in a circuit, we can reduce the potential for error accumulation, thereby improving the fidelity of the computation. 

%%%%%%%%%%%%%%%%%%%%%%%%%%%%%%%%%
%%%%%%%%%%%%%%%%%%%%%%%%%%%%%%%%%
\subsubsection*{Counting of T gates}
%%%%%%%%%%%%%%%%%%%%%%%%%%%%%%%%%
%%%%%%%%%%%%%%%%%%%%%%%%%%%%%%%%%
In FTQC, the computational paradigm shifts significantly compared to NISQ systems. 
In FTQC, error correction is employed through the use of logical qubits, which are constructed from multiple physical qubits. 
This allows for the protection of quantum information from noise and errors, enabling more robust and scalable quantum computations. 
As a result, the focus on computational cost changes: the number of CNOT gates, which plays a critical role in NISQ systems, is no longer the primary factor determining the computational expense.

The most commonly used universal gate set in FTQC is the combination of Clifford gates and the T gates.\footnote{
The T gate is defined by 
\begin{align}
    \hat{T}
    =
    \left(
        \begin{array}{cc}
            1 & 0\\
            0 & \exp(\frac{\mathrm{i}\pi}{4})
        \end{array}
    \right)\, .      
    \nonumber
\end{align}
}
Clifford gates, including the Hadamard, phase, and CNOT gates, are generally not resource-intensive in various error-correcting codes, such as the surface code~\cite{FowlerSurfaceReview}. 
They can often be implemented efficiently through techniques like code deformation and lattice surgery~\cite{Horsman_2012,Herr_2017}. 
Additionally, Clifford gates can be pushed to the end of a quantum circuit, where they can be seamlessly absorbed into Pauli measurements~\cite{Litinski2019gameofsurfacecodes}, further optimizing resource usage.

In contrast, non-Clifford gates, such as the T gate, are typically much more resource-intensive. 
A fault-tolerant implementation of the T gate often involves gate teleportation, which relies on specialized resource states known as magic states~\cite{BravyiKitaevMagic, KnillMagic}.
However, producing high-fidelity magic states is considerably costly. 
As a result, the T gate count has become a standard metric for estimating resource requirements in FTQC.\footnote{There has been significant progress in improving the efficiency of magic state distillation. 
For instance, see the recent advancements in Ref.~\cite{Gidney_Cultivation}.}

In order to switch from the universal gate set $\{$CNOT, single qubit rotations$\}$ to $\{$Clifford, T$\}$, each rotation gate -- specifically the $R_Z$ gate, $R_Z(\theta)=\exp\left(-\frac{\mathrm{i}\theta}{2}\hat{\sigma}_z\right)$ -- needs to be approximated using a combination of T-gates and single-qubit Clifford gates.
For instance, Ref.~\cite{RossTgate} demonstrated that each $R_Z$ rotation can typically be approximated with a T gate count of $3\log\left(1/\epsilon\right)+O\left(\log \log\left(1/\epsilon\right)\right)$, where $\epsilon$ represents the desired accuracy of the approximation
\footnote{We also point out that an intermediate framework between NISQ and FTQC can be implemented and it is partially fault tolerant~\cite{Toshio:2024hut}. 
This framework can be used to compile the type of Trotter circuits we are considering in this section~\cite{Akahoshi:2024yme}.}. 
The $R_Z$ gate can be written in terms of more elementary one-qubit gates, among which the T gate is usually the most costly one. 
The typical T gate count per $R_Z$ gate will fall within the range of 10-50~\cite{CampbellETFQC},  depending on factors such as the rotation angle, desired accuracy, and the specific algorithm used to decompose the  $R_Z$  gate into T-gates and Clifford gates. 
In the following, we denote the T-gate count per $R_Z$ as $T_{\rm typ}$ and use 
%. For the $R_Z$ gate with a generic rotation angle, a typical number is 
$T_{\rm typ}=10$ -- 50~\cite{CampbellETFQC,Kliuchnikov:2016ong}. 

%%%%%%%%%%%%%%%%%%%%%%%%%%%
%%%%%%%%%%%%%%%%%%%%%%%%%%%
\subsubsection{Estimate for \texorpdfstring{$\exp\left(-\mathrm{i}\theta\hat{p}^2\right)$}{exp(-iep2}}\label{sec:cost_kinetic}
%%%%%%%%%%%%%%%%%%%%%%%%%%%
%%%%%%%%%%%%%%%%%%%%%%%%%%%

One of the advantages of the orbifold lattice over the Kogut--Susskind formulation is that the Fourier transform is straightforward. 
By switching from the coordinate basis to the momentum basis via Fourier transform, we can diagonalize the kinetic terms.
Earlier in this paper, we showed two options, \eqref{p-option-1} and \eqref{p-option-2}.
The second option \eqref{p-option-2} is obtained by omitting the higher order terms of the Taylor expansion of \eqref{p-option-1}.
Here, we choose the second option, because these two options give the same results when the truncation is removed, but higher powers of Pauli $\sigma_z$ appear and more gates are needed for the first option.
Then, $\hat{p}$ in the momentum basis takes essentially the same form as $\hat{x}$ in the coordinate basis. 
Specifically, $\hat{p}$ is a linear sum of $Q$ Pauli  $\sigma_z$ (we call them $\hat{\sigma}_{z,1},\cdots,\hat{\sigma}_{z,Q}$), and therefore $\hat{p}^2$ is a combination of $\hat{\sigma}_{z,j}\hat{\sigma}_{z,k}$ ($1\le j < k\le Q$). 
We can write $e^{-\mathrm{i}\theta\hat{p}^2}$ as
\begin{align}
    \exp\left(-\mathrm{i}\theta\hat{p}^2\right)
    =
    \prod_{j<k}
    \exp\left(-\mathrm{i}\theta C_{jk}\hat{\sigma}_{z,j}\hat{\sigma}_{z,k}\right)\, . 
    \label{eq: diagonal kinetic terms}
\end{align}
Each term in the product can be written in terms of CNOT gates and one-qubit gates as before: 
\begin{align}
\exp\left(-\mathrm{i}\theta C_{jk}\hat{\sigma}_{z,j}\hat{\sigma}_{z,k}\right)
=
\textrm{CNOT}_{j,k}
\exp\left(-\textrm{i}\theta C_{jk}\hat{\sigma}_{z,k}\right)
\textrm{CNOT}_{j,k}\, . 
\end{align}
In this case, there is no cancellation between CNOT gates. $2\times(\genfrac{(}{)}{0pt}{}{Q}{2})=Q(Q-1)$ CNOT gates are needed for each boson. 
The number of $R_Z$ rotations is $Q(Q-1)/2$ and the number of T-gates in FTQC is $O\left({Q(Q-1)}\right)$.

\begin{figure}[ht]
\centering
\includegraphics[width=1.0\textwidth]{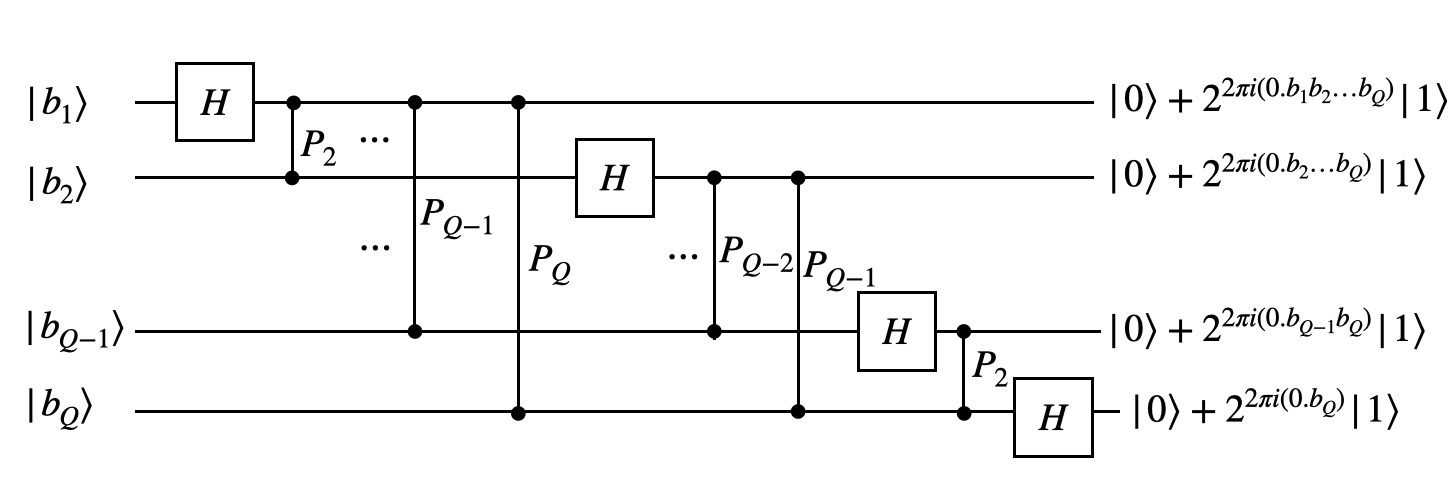}
\caption{Microscopic Quantum Fourier Transform circuit}
\label{fig:qft_circuit}
\end{figure}

\begin{figure}[ht]
\centering
\includegraphics[width=0.7\textwidth]{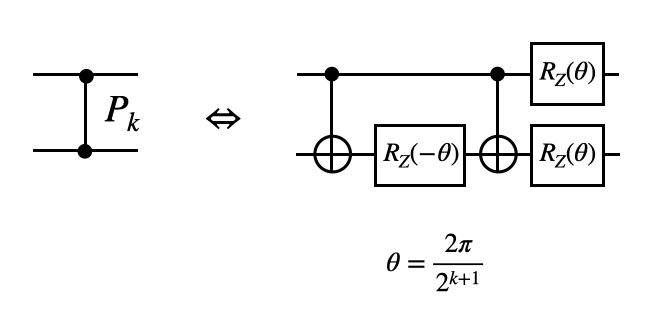}
\caption{Controlled phase gate in terms of CNOT gates and one qubit rotation gates up to a global phase.}
\label{fig:cphase}
\end{figure}

The cost of implementing the diagonal kinetic terms \eqref{eq: diagonal kinetic terms} is dominated by the need to perform quantum Fourier transforms between the potential terms and the kinetic terms.
In Figure~\ref{fig:qft_circuit}, we show the exact quantum Fourier transform circuit.
The quantum Fourier transform circuit consists of Hadamard gates and Controlled-Phase gates.
A Controlled-Phase gate is defined as
\begin{align}
P_k=
\begin{pmatrix}
1 & 0 & 0 & 0 \\
0 & 1 & 0 & 0 \\
0 & 0 & 1 & 0 \\
0 & 0 & 0 & \exp\left(\frac{2\pi\textrm{i}}{2^k}\right) \\
\end{pmatrix}\, .
\end{align}
Up to a global phase, a Controlled-Phase gate is decomposed into CNOT gates and 1 qubit rotation gates $R_Z(\theta)$ as in Fig.~\ref{fig:cphase}.

The quantum Fourier transform circuit (Fig.~\ref{fig:qft_circuit}) has $Q(Q-1)/2$ Controlled-Phase gates. 
In terms of CNOT gates, the two-qubit gate count is ${Q(Q-1)}$ and the number of $R_Z$
rotation gates is $3Q(Q-1)/2$.
An approximate quantum Fourier transform can also be used by truncating rotations with angles below a specified threshold. With this approach, both the T-gate and CNOT gate counts can be reduced to $O(Q\log Q)$ \cite{NamAQFT}.

%%%%%%%%%%%%%%%%%%%%%%%%%%%%%%
%%%%%%%%%%%%%%%%%%%%%%%%%%%%%%
\subsubsection*{Implementation in the coordinate basis}
%%%%%%%%%%%%%%%%%%%%%%%%%%%%%%
%%%%%%%%%%%%%%%%%%%%%%%%%%%%%%
In the NISQ era, it may be advantageous to avoid using the Fourier transform altogether and instead operate solely on the coordinate basis.
For the coordinate basis, 
$\hat{\mathcal{S}}\equiv\sum_{n}\left(\ket{n+1}\bra{n}+\ket{n}\bra{n+1}\right)$ can be written as 
\begin{align}
\hat{\mathcal{S}}=\hat{\sigma}_x\, 
\end{align}
for $Q=1$, 
\begin{align}
\hat{\mathcal{S}}=\hat{\sigma}_x\otimes\textbf{1}_2+\frac{\hat{\sigma}_x\otimes\hat{\sigma}_x+\hat{\sigma}_y\otimes\hat{\sigma}_y}{2}\, . 
\end{align}
for $Q=2$, 
\begin{align}
\hat{\mathcal{S}}
&=\hat{\sigma}_x\otimes\textbf{1}_2\otimes\textbf{1}_2+\frac{\hat{\sigma}_x\otimes\hat{\sigma}_x\otimes\textbf{1}_2+\hat{\sigma}_y\otimes \hat{\sigma}_y\otimes\textbf{1}_2}{2}
\nonumber\\
&\quad
+
\frac{\hat{\sigma}_x\otimes\hat{\sigma}_x\otimes\hat{\sigma}_x-\hat{\sigma}_x\otimes\hat{\sigma}_y\otimes\hat{\sigma}_y+\hat{\sigma}_y\otimes\hat{\sigma}_x\otimes\hat{\sigma}_y+\hat{\sigma}_y\otimes\hat{\sigma}_y\otimes\hat{\sigma}_x}{4}
\end{align}
for $Q=3$, 
\begin{align}
\hat{\mathcal{S}}
&=\hat{\sigma}_x\otimes\textbf{1}_2\otimes\textbf{1}_2\otimes\textbf{1}_2+\frac{\hat{\sigma}_x\otimes\hat{\sigma}_x\otimes\textbf{1}_2\otimes\textbf{1}_2+\hat{\sigma}_y\otimes\hat{\sigma}_y\otimes\textbf{1}_2\otimes\textbf{1}_2}{2}
\nonumber\\
&
+
\frac{\hat{\sigma}_x\otimes\hat{\sigma}_x\otimes\hat{\sigma}_x\otimes\textbf{1}_2-\hat{\sigma}_x\otimes\hat{\sigma}_y\otimes\hat{\sigma}_y\otimes\textbf{1}_2+\hat{\sigma}_y\otimes\hat{\sigma}_x\otimes\hat{\sigma}_y\otimes\textbf{1}_2+\hat{\sigma}_y\otimes\hat{\sigma}_y\otimes\hat{\sigma}_x\otimes\textbf{1}_2}{4}
\nonumber\\
&
+
\frac{1}{8}\left\{
\hat{\sigma}_x\otimes\hat{\sigma}_x\otimes\hat{\sigma}_x\otimes\hat{\sigma}_x
+
\hat{\sigma}_x\otimes\hat{\sigma}_x\otimes\hat{\sigma}_y\otimes\hat{\sigma}_y
+
\hat{\sigma}_x\otimes\hat{\sigma}_y\otimes\hat{\sigma}_x\otimes\hat{\sigma}_y
-
\hat{\sigma}_x\otimes\hat{\sigma}_y\otimes\hat{\sigma}_y\otimes\hat{\sigma}_x
\right.
\nonumber\\
&
\left.
+
\hat{\sigma}_y\otimes\hat{\sigma}_x\otimes\hat{\sigma}_x\otimes\hat{\sigma}_y
-
\hat{\sigma}_y\otimes\hat{\sigma}_x\otimes\hat{\sigma}_y\otimes\hat{\sigma}_x
-
\hat{\sigma}_y\otimes\hat{\sigma}_y\otimes\hat{\sigma}_x\otimes\hat{\sigma}_x
-
\hat{\sigma}_y\otimes\hat{\sigma}_y\otimes\hat{\sigma}_y\otimes\hat{\sigma}_y
\right\}
\nonumber\\
\end{align}
for $Q=4$, and so on. 
In general, we can write $\hat{\mathcal{S}}$ as a sum of $2^{q-1}$ Pauli strings of length $q$ consisting of $\hat{\sigma}_x$ and $\hat{\sigma}_y$, where $q$ runs from 1 to $Q$.
This decomposition can be practical when  $Q$ is relatively small.

A potential advantage of the implementation in the coordinate basis is that the truncation effect can be quantitatively estimated by using a classical sampling method~\cite{Hanada:2022pps}. 
%%%%%%%%%%%%%%%%%%%%%%%%%%%
%%%%%%%%%%%%%%%%%%%%%%%%%%%
\subsection{Resource estimate for Hamiltonian time evolution}\label{sec:Trotter_cost_details}
%%%%%%%%%%%%%%%%%%%%%%%%%%%
%%%%%%%%%%%%%%%%%%%%%%%%%%%
Now, we combine all the pieces and estimate the cost of one Suzuki--Trotter step for the scalar QFT, matrix model, and Yang--Mills theory.
We focus on the asymptotic behavior with respect to the system size parameters, i.e., matrix size $N$, lattice volume $V_{\rm lattice}=L^d$, and truncation parameter $Q$.
Again, we neglect the cost related to reaching a constant accuracy.

%%%%%%%%%%%%%%%%%%%%%%%%%%%%%%%%%%%%%%%%%
%%%%%%%%%%%%%%%%%%%%%%%%%%%
\subsubsection{Scalar quantum field theory}\label{sec:Trotter_cost_scalar}
%%%%%%%%%%%%%%%%%%%%%%%%%%%
%%%%%%%%%%%%%%%%%%%%%%%%%%%
The number of bosons is $V_{\rm lattice}$  with one boson on each lattice site. 
By assigning $Q$ qubits to each boson, a total of $V_{\rm lattice}Q$  qubits are utilized.
Recently, Ref.~\cite{Zemlevskiy:2024vxt} has studied this case with $Q=2$ and $V_{\rm lattice}= L = 60$ in (1+1) dimensions, for a total of 120 qubits on a near-term quantum device.

%%%%%%%%%%%%%%%%%%%%%%%%%%%
%%%%%%%%%%%%%%%%%%%%%%%%%%%
\subsubsection*{Interaction terms}
%%%%%%%%%%%%%%%%%%%%%%%%%%%
%%%%%%%%%%%%%%%%%%%%%%%%%%%
The potential term of the Hamiltonian \eqref{scalar-QFT-Hamiltonian} consists of quadratic and quartic terms. 
The former consists of two-$\sigma_z$ couplings, while the latter consists of two-$\sigma_z$ couplings and four-$\sigma_z$ couplings.

Let us focus on the four-$\sigma_z$ couplings.
The number of such couplings is $\genfrac{(}{)}{0pt}{}{Q}{4}\times V_{\rm lattice}\sim Q^4V_{\rm lattice}$. 
As explained in Sec.~\ref{sec:cost_interaction}, Suzuki--Trotter time evolution with respect to each of these couplings can be written by using six CNOT gates and one $R_Z$ rotation acting on one of the four qubits. 
By taking appropriate ordering, about $2/3$ of CNOT gates cancel (see a comment right after \eqref{eq:CNOT-cancellation}). 
The $R_Z$ gate can be built using more elementary one-qubit gates, including $T_{\rm typ}=10$ -- 50 T gates. 
Therefore, we are left with $O(Q^4 V_{\rm lattice})$ CNOT gates, $O(T_{\rm typ}Q^4 V_{\rm lattice})$ T gates, and $O(Q^4 V_{\rm lattice})$ one-qubit gates simpler than the T gate.  
%%%%%%%%%%%%%%%%%%%%%%%%%%%
%%%%%%%%%%%%%%%%%%%%%%%%%%%
\subsubsection*{Kinetic terms}
%%%%%%%%%%%%%%%%%%%%%%%%%%%
%%%%%%%%%%%%%%%%%%%%%%%%%%%
Next, we consider the kinetic terms. 
If we use an approximate quantum Fourier transform to change to the momentum basis, the cost in terms of both CNOT gates and T gates is $O(V_{\rm lattice}Q\log Q)$.
As we saw in Sec.~\ref{sec:cost_kinetic}, $Q(Q-1)$ CNOT gates and about $T_{\rm typ}Q(Q-1)/2$ T gates are needed in the momentum basis for each boson. 
Multiplying by the number of bosons, we obtain the total cost as $O(Q^2V_{\rm lattice})$ CNOT gates, $O(T_{\rm typ} Q^2V_{\rm lattice})$ T gates, and $O(Q^2V_{\rm lattice})$ one-qubit gates simpler than the T gate.

%%%%%%%%%%%%%
%%%%%%%%%%%%%%%%%%%%%%%%%%%
\subsubsection{Matrix model}\label{sec:Trotter_cost_matrix}
%%%%%%%%%%%%%%%%%%%%%%%%%%%
%%%%%%%%%%%%%%%%%%%%%%%%%%%
The number of bosons scales as $dN^2$.~\footnote{Typically, we are interested in the large-$N$ limit, where the difference between $N^2$ and $N^2-1$ is not important.}
We assign $Q$ qubits to each boson, and hence $dN^2Q$ qubits are used in total.  
%%%%%%%%%%%%%%%%%%%%%%%%%%%
%%%%%%%%%%%%%%%%%%%%%%%%%%%
\subsubsection*{Interaction terms}
%%%%%%%%%%%%%%%%%%%%%%%%%%%
%%%%%%%%%%%%%%%%%%%%%%%%%%%
Firstly, we consider the interaction terms. 
As we saw in Sec.~\ref{sec:MM_interaction}, the number of quartic interactions of the form $\hat{x}_j\hat{x}_k\hat{x}_l\hat{x}_m$ increases as $d(d-1)N^4$. Each $\hat{x}$ can be expressed by using $Q$ Pauli $\hat{\sigma}_z$'s, and hence, there are $\sim d(d-1)N^4Q^4$ quartic couplings of $\hat{\sigma}_z$'s. 

Furthermore, as explained in Sec.~\ref{sec:cost_interaction}, Suzuki--Trotter time evolution with respect to each of these four-$\sigma_z$ couplings can be written by using six CNOT gates and about 1 one-qubit gate (or $T_{\rm typ}$ T-gates). We need  $\sim d(d-1)N^4Q^4$ CNOT gates and $\sim T_{\rm typ} \times d(d-1)N^4Q^4$ T gates. 

Because the number of both one- and two-qubit gates scales as $d(d-1)N^4Q^4$, we can estimate the depth of the circuit by simply dividing this number by the number of qubits $dN^2Q$, which leads to the conclusion that the depth scales like $(d-1)N^2Q^3$.

%%%%%%%%%%%%%%%%%%%%%%%%%%%
%%%%%%%%%%%%%%%%%%%%%%%%%%%
\subsubsection*{Kinetic terms}
%%%%%%%%%%%%%%%%%%%%%%%%%%%
%%%%%%%%%%%%%%%%%%%%%%%%%%%
We consider the implementation in the momentum basis with quantum Fourier transform. 
As we saw in Sec.~\ref{sec:cost_kinetic}, $Q(Q-1)$ CNOT gates and $T_{\rm typ}Q(Q-1)$ T gates are needed for each boson, once we are in the momentum basis. 
Multiplying with the number of bosons, we obtain the cost as $dN^2Q(Q-1)$ CNOT gates and $T_{\rm typ} dN^2Q(Q-1)$ T gates.
In addition, we add the cost of the quantum Fourier transform as we did before: if we use an approximate quantum Fourier transform, this cost is $O(dN^2Q\log Q)$ CNOT (and T) gates.  

In total, the cost for the kinetic term is negligible compared to the cost for the interaction terms.

%%%%%%%%%%%%%%%%%%%%%%%%%%%
%%%%%%%%%%%%%%%%%%%%%%%%%%%
\subsubsection{Orbifold lattice}\label{sec:Trotter_cost_orbifold}
%%%%%%%%%%%%%%%%%%%%%%%%%%%
%%%%%%%%%%%%%%%%%%%%%%%%%%%
Next, we consider the orbifold lattice. 
The number of bosons and logical qubits used for encoding them are $2N^2dV_{\rm lattice}$ and $2N^2dV_{\rm lattice}Q$, respectively.  
%%%%%%%%%%%%%%%%%%%%%%%%%%%
%%%%%%%%%%%%%%%%%%%%%%%%%%%
\subsubsection*{Interaction terms}
%%%%%%%%%%%%%%%%%%%%%%%%%%%
%%%%%%%%%%%%%%%%%%%%%%%%%%%
To see the cost for one Suzuki-Trotter step of the interaction part, we combine the results from Sec.~\ref{sec:cost_interaction} and Sec.~\ref{sec:orbifold_interaction}. 
There are $\sim d^2V_{\rm lattice}N^4$ terms quartic in $\hat{x}$. We can write them using $\hat{\sigma}_z$, leading to $\sim Q^4$ CNOT gates and $\sim T_{\rm typ}\times Q^4$ T gates for each quartic interaction. In total, we need $\sim d^2V_{\rm lattice}N^4Q^4$ CNOT gates and $\sim T_{\rm typ}\times d^2V_{\rm lattice}N^4Q^4$ T gates. 
%%%%%%%%%%%%%%%%%%%%%%%%%%%
%%%%%%%%%%%%%%%%%%%%%%%%%%%
\subsubsection*{Kinetic terms}
%%%%%%%%%%%%%%%%%%%%%%%%%%%
%%%%%%%%%%%%%%%%%%%%%%%%%%%
Again, we consider the implementation in the momentum basis with quantum Fourier transform. 
Then, we need $N^2dV_{\rm lattice}Q(Q-1)$ CNOT gates and $\sim T_{\rm typ}\times N^2dV_{\rm lattice}Q(Q-1)$ T gates once we are in the momentum basis.
The cost of the quantum Fourier transform itself is $O(N^2dV_{\rm lattice}Q\log Q)$ CNOT (and T) gates. 

In total, the cost for the kinetic term is negligible compared to the cost for the interaction terms. 

%%%%%%%%%%%%%%%%%%%%%%%%%%%
%%%%%%%%%%%%%%%%%%%%%%%%%%%
\section{Conclusion and discussion}\label{sec:conclusion}
%%%%%%%%%%%%%%%%%%%%%%%%%%%
%%%%%%%%%%%%%%%%%%%%%%%%%%%

In this paper, we provided a universal framework for the quantum simulation of SU($N$) Yang--Mills theories with arbitrary $N$, arbitrary spatial dimensions, and arbitrary lattice sizes adopting the orbifold lattice formulation, taking the Hamiltonian time evolution as an example. 
Technical difficulties associated with the Kogut--Susskind Hamiltonian, which researchers struggled for years to solve on a case-by-case basis -- such as the definition of the coordinate basis (magnetic basis), the quantum Fourier transform, and the realization of complicated interactions in the momentum basis (electric basis) expressed by the Clebsh--Gordan coefficient in terms of quantum gates -- simply do not exist in the orbifold lattice formulation.
This is a consequence of a simple and universal form \eqref{eq:simple-form} that is common among many theories.
\footnote{We did not consider theories with fermions in this paper. See Ref.~\cite{Bergner:2024qjl} for the orbifold lattice construction of QCD, i.e., Yang--Mills theory with fermions in the fundamental representation. 
We would like to discuss quantum simulation with fermions soon.} Note that we focused on bosonic theories without fermions. The inclusion of fermion is straightforward and we plan to have a separate paper in the near future.

We know explicitly how the orbifold Hamiltonian can be programmed on a quantum computer for any $N$, any dimensions, and any lattice size. 
Furthermore, we need only standard, well-established tools in the field of quantum computing.
As a warm-up example, we also considered the Yang--Mills matrix model and the more standard scalar quantum field theory on a lattice. 
It was straightforward to write a circuit for the unitary  Hamiltonian evolution operator explicitly in terms of CNOT gates and one-qubit gates and count the number of gates. 

We used the extended Hilbert space $\mathcal{H}_{\rm ext}$ that contains SU($N$) non-singlets. 
This is a standard approach, and we believe one should not stick to the projection to singlet Hilbert space $\mathcal{H}_{\rm inv}$ because both $\mathcal{H}_{\rm ext}$ and $\mathcal{H}_{\rm inv}$ lead to mathematically equivalent formulations and, on $\mathcal{H}_{\rm inv}$, one cannot even define the coordinate and momentum operators. 
It is often claimed that the use of $\mathcal{H}_{\rm ext}$ is too costly because the dimension is exponentially larger. 
Such a claim might be missing an important point: although it is true that brute-force computations on a \emph{classical} computer are hard if the dimension of the Hilbert space is exponentially larger, on a \emph{quantum} computer it only requires a moderate number of additional qubits. 
Adding longitudinal modes just increases the number of qubits by 50\% in three spatial dimensions. 
However, with this overhead, the structure of the Hilbert space and quantum circuits simplify drastically, outweighing the increase in \emph{space} resources. 
Because our target is a quantum simulation and not a classical simulation, it is therefore advantageous to use the extended Hilbert space. 
Note also that, if one wants to remove non-singlet modes explicitly from the spectrum, one can add a penalty term such as $\sum_{\vec{n}}\mathrm{Tr}\hat{G}_{\vec{n}}^2$ to the Hamiltonian \cite{Halimeh2020reliability,Halimeh2022gauge,VanDamme2023reliability}. 
We also note that it is rather straightforward to construct many singlet states in the extended Hilbert space because the confined vacuum is a singlet and any state that is obtained by acting with singlet operators on it is also a singlet. 
Many singlet operators can be constructed by using Wilson loops, which are traces of products of link variables $\hat{Z}_{j,\vec{n}}$, $\hat{\bar{Z}}_{j,\vec{n}}$ along a closed contour. 
Regarding the Hamiltonian time evolution, for example, the only thing we need to keep the states in the singlet sector is the precision of the unitary time evolution, which is made significantly more tractable by the simplicity of the simulation scheme in the orbifold lattice formulation.

In this context, it is important to note that the violation of gauge symmetry is suppressed exponentially with regard to the truncation level $\Lambda$; see Appendix~\ref{sec:gauge_symmetry}. Furthermore, note that the violation of gauge symmetry is only one of many possible truncation effects. Our universal approach leads to efficient suppression of all kinds of truncation effects, because we can more easily remove the truncation. On the other hand, imposing exact gauge symmetry at truncated level makes the Hamiltonian complicated and makes it harder to remove truncation. Also, as a trade off, other kinds of truncation effects can become severer~\cite{Hanada2025}. Finding a good gauge-invariant basis that circumvents such an unwanted trade off would be as challenging as finding the energy eigenstates.

We also note that the key role of gauge symmetry is to introduce massless vector fields compatible with Lorentz symmetry without having negative-norm states. In lattice Hamiltonian formulations, even when the truncation in the Hilbert space is removed, Lorentz symmetry is broken at finite lattice spacing. Adherence to the singlet Hilbert space can make it difficult to go to sufficiently small lattice spacing and hence can lead to larger breaking of Lorentz symmetry, spoiling the original motivation of gauge symmetry.

Given the striking simplicity of the Hamiltonians, it is important to investigate efficient simulation techniques for orbifold lattice theory and matrix model systematically. 
Note that even the Hamiltonian of the scalar $\phi^4$ theory belongs to the same class (specifically, it takes the same simple form~\eqref{eq:simple-form}) and hence a very large class of theories can be studied in a unified manner. 
It is also interesting to try simulations on real quantum devices in the short term.
A good starting point is randomly-coupled spin systems~\cite{Erdos:2014zgc,Berkooz:2018qkz,Baldwin:2019dki,Baldwin:2019dki,Hanada:2023rkf,Anschuetz:2023igd,Swingle:2023nvv}. 
Among them, the spin-XY4 model~\cite{Hanada:2023rkf} consists of four-body random coupling of $\hat{\sigma}_{x}$ and $\hat{\sigma}_{y}$. 
We can replace $\hat{\sigma}_{y}$ with $\hat{\sigma}_{z}$ without any change to the physical properties, and $\hat{\sigma}_{x}$ can be written by using $\hat{\sigma}_{z}$ and the Hadamard gate $\hat{h}$ according to $\hat{h}\hat{\sigma}_z\hat{h}=\hat{\sigma}_x$. 
Therefore, quantum simulation of the spin-XY4 model can be a good exercise toward the simulation of matrix models and gauge theories on an orbifold lattice. 
A simplified version of the matrix model studied in Ref.~\cite{Buividovich:2022jgv} or the anharmonic oscillator \eqref{anharmonic_oscillator} (see Ref.~\cite{Somma:2015bcw} for the analysis of this model in the context of quantum simulation) would also be a good target for the first quantum simulation on real devices. 

If we use four logical qubits for a single anharmonic oscillator (one boson), for example, then the coordinate operator is 
\begin{align}
\hat{x}
=
-
\delta_x\cdot
\left(
\frac{\hat{\sigma}_{z;1}}{2}
+
2\cdot\frac{\hat{\sigma}_{z;2}}{2}
+
4\cdot\frac{\hat{\sigma}_{z;3}}{2}
+
8\cdot\frac{\hat{\sigma}_{z;4}}{2}
\right)\, ,  
\end{align}
and $\hat{x}^4$ contains $\hat{\sigma}_{z;1}\otimes\hat{\sigma}_{z;2}\otimes\hat{\sigma}_{z;3}\otimes\hat{\sigma}_{z;4}$. 
This is exactly the interaction we need to describe a wide class of theories including Yang--Mills theory, as we have argued in this paper. 
To study a larger system with more bosons, we need to add more logical qubits.
However, we only have to add the same 4-qubit gates acting on different sets of qubits. 
Furthermore, to switch to the momentum basis, we only have to perform the same quantum Fourier transform to different sets of qubits that describe different bosons. 
In the literal sense, we know how to scale up such a simulation systematically when more logical qubits are available. 
Therefore, it would be already meaningful if we could use a few logical qubits and demonstrate the precise Hamiltonian time evolution, even if the system size is small and there is no quantum advantage. In the context of NISQ, we have recently witnessed how the simple protocol we propose can be enhanced by variational circuits and used to study particle scattering in a $(1+1)$ dimensional Scalar Field Theory with up to 120 qubits~\cite{Zemlevskiy:2024vxt}.
This is a testament to the fact that there are no theoretical obstacles in going to more complicated theories and to higher dimensions using our universal framework.

It is also important to identify the types of hardware suitable for quantum simulations. 
The large-$N$ limit of the matrix model involves nonlocal interactions between matrix entries, and hence, the truncated theory has nonlocal interaction between qubits. 
Trapped-ion quantum computers would be suitable for such a Hamiltonian because any pair of qubits ($=$ ions) can be brought close to each other and nonlocal interaction can naturally be realized. 
On the other hand, the orbifold lattice Hamiltonian has a local structure associated with the spatial lattice, although there is some non-locality that increases with the number of colors $N$ and truncation level $\Lambda = 2^Q$, and hence, quantum hardware with qubits at fixed locations, such as the superconducting qubits machines, may also perform well. 
We also note that the standard lattice simulation on classical devices is a powerful tool to study the non-perturbative features on the orbifold lattice Hamiltonian, because we can get the Euclidean counterpart by using the spacetime lattice and taking the continuum limit along the time direction. 
Such a method was applied for the Kogut--Susskind Hamiltonian, by using the Wilson's action on an anisotropic lattice. 
Just as an incomplete list:  
Refs.~\cite{Carena:2021ltu,Carena:2022hpz} related Hamiltonian time evolution with finite Suzuki--Trotter step to simulations on Euclidean anisotropic lattices. 
Their main interest was in the discretization effects. 
If the Suzuki--Trotter step is sent to zero, then their setup is the same as ours. 
Ref.~\cite{Funcke:2022opx} studied the renormalization of the anisotropy ratio in $(2+1)$-dimensional QED on an anisotropic lattice. 
Such a renormalization is directly related to the tuning of the coupling as a function of spatial lattice spacing, which is needed for the restoration of Lorentz symmetry. 
See also Refs.~\cite{Loan:2005ff}. 

In this paper, we focused on digital simulations with qubits. 
However, the simplifications coming from the use of non-compact variables are not limited to this particular setup. 
For example, it is straightforward to write the truncated Hamiltonian in terms of qudits.
Furthermore, quantum simulation with continuous variables (see Ref.~\cite{Braunstein:2005zz} for a review and Ref.~\cite{Abel:2024kuv,Ale:2024uxf} for a recent application to quantum field theory) is potentially a good framework to simulate orbifold lattices and matrix models. 
Another potentially promising route is to engineer an analog simulator to be described by the same Hamiltonian. 
It would be an interesting research avenue to identify the right setup that allows quantum simulations before the arrival of fault-tolerant quantum computers. 

As an important side comment, we note that the simplicity of the orbifold lattice is connected to the emergent geometry. 
The orbifold lattice emerges from a matrix model with a certain background, via dimensional deconstruction~\cite{Arkani-Hamed:2001kyx,Kaplan:2002wv}. 
A related example is the emergence of D2-branes from D0-branes via the Myers effect~\cite{Myers:1999ps} that enables us to describe a $(2+1)$-dimensional theory by using a $(0+1)$-dimensional theory, whose prototype dates back to the non-commutative torus in the Twisted Eguchi--Kawai model~\cite{Gonzalez-Arroyo:1982hyq}. 
In these examples, theories on emergent spaces inherit simple structures from the original theories~\cite{Gharibyan:2020bab}. 
We could say that \emph{nature} is smarter than humans and hence dynamically-generated spatial dimensions can have better properties than a lattice crafted by humans. 
Holographic dualities provide us with even more profound examples: gravitational geometries emerge from non-gravitational theories, providing us with \emph{simple} Hamiltonians of the non-gravitational theories. 
We contend that the quantum simulation of quantum field theory should be considered within the broader context of the web of dualities and emergent geometry.

To conclude this paper, we would like to refer to a famous essay on AI research \textit{The Bitter Lesson}  written by Rich Sutton in 2019. 
The opening phrase of this essay is \textit{``The biggest lesson that can be read from 70 years of AI research is that general methods that leverage computation are ultimately the most effective, and by a large margin.''} 
Then, it continues as \textit{``[...] Seeking an improvement that makes a difference in the shorter term, researchers seek to leverage their human knowledge of the domain, but the only thing that matters, in the long run, is the leveraging of computation. [...] And the human-knowledge approach tends to complicate methods in ways that make them less suited to taking advantage of general methods leveraging computation. [...] ''}  
The same lesson applied to classical computing for quantum field theory. 
Indeed, revolutionary algorithms such as Metropolis~\cite{Metropolis:1953am} and Hybrid Monte Carlo~\cite{Duane:1987de} do not assume too many details of the theories, and the same methods can be used for a wide class of theories. 
Similar lessons may apply to quantum computing, and hence, it is important to develop general methods that do not rely on the details of the systems and can straightforwardly be scaled up on universal quantum computers.

%%%%%%%%%%%%%%%%%%%%%%
%%%%%%%%%%%%%%%%%%%%%%
\section*{Acknowledgment}
%%%%%%%%%%%%%%%%%%%%%%
%%%%%%%%%%%%%%%%%%%%%%

The authors would like to thank Georg Bergner and Hrant Gharibyan for discussions about the minimal model, and Nathan Fitzpatrick, Yuta Kikuchi, and Emanuele Mendicelli  for useful comments on the manuscript.
M.~H., F.~N.~and E.~R.~thank the Royal Society International Exchanges award IEC/R3/213026.
M.~H.~thanks the STFC for the support through the consolidated grant ST/Z001072/1.
F.~N.~is supported in part by Nippon Telegraph and Telephone Corporation (NTT) Research,
the Japan Science and Technology Agency (JST)
(via the CREST Quantum Frontiers program Grant No. JPMJCR24I2,
the Quantum Leap Flagship Program (Q-LEAP), and the Moonshot R\&D Grant Number JPMJMS2061),
and also by the Office of Naval Research (ONR) Global (via Grant No.~{N62909-23-1-2074}). 
J.~C.~H.~acknowledges funding by the Max Planck Society, the Deutsche Forschungsgemeinschaft (DFG, German Research Foundation) under Germany’s Excellence Strategy --- EXC-2111 ---  390814868, and the European Research Council (ERC) under the European Union’s Horizon Europe research and innovation program (Grant Agreement No.~101165667) --- ERC Starting Grant QuSiGauge. Views and opinions expressed are, however, those of the author(s) only and do not necessarily reflect those of the European Union or the European Research Council Executive Agency. Neither the European Union nor the granting authority can be held responsible for them. This work is part of the Quantum Computing for High-Energy Physics (QC4HEP) working group.

\appendix
%%%%%%%%%%%%%%%%%%%%%%%%%%%
%%%%%%%%%%%%%%%%%%%%%%%%%%%
\section{\texorpdfstring{SU($N$)}{SU(N)} generators and structure constant}\label{sec:SU(N)-algebra}
%%%%%%%%%%%%%%%%%%%%%%%%%%%
%%%%%%%%%%%%%%%%%%%%%%%%%%%
In this appendix, we show explicit examples of SU($N$) generators $\tau_\alpha$ introduced in Sec.~\ref{sec:Matrix-Model}.
Our normalization is 
\begin{align}
\mathrm{Tr}(\tau_\alpha\tau_\beta)=\delta_{\alpha\beta}\, .   
\end{align}
For the SU(2) theory, an explicit example of such generators is obtained by rescaling Pauli matrices as $\frac{\sigma_1}{\sqrt{2}}$, $\frac{\sigma_2}{\sqrt{2}}$, and $\frac{\sigma_3}{\sqrt{2}}$. 
For the SU($3$) theory, we can use the Gell--Mann matrices $\lambda_{1,2,\cdots,8}$ such that
\begin{align}
\tau_{\alpha}
=
\frac{\lambda_\alpha}{\sqrt{2}}
\qquad
(\alpha=1,2,\cdots,8)\, . 
\end{align}
The Gell--Mann matrices are defined by 
\begin{align}
&
\lambda_1
=
\left(
\begin{array}{ccc}
0 & 1 & 0\\
1 & 0 & 0\\
0 & 0 & 0
\end{array}
\right)\, , 
\quad
\lambda_2
=
\left(
\begin{array}{ccc}
0 & -\mathrm{i} & 0\\
\mathrm{i} & 0 & 0\\
0 & 0 & 0
\end{array}
\right)\, , 
\quad
\lambda_3
=
\left(
\begin{array}{ccc}
1 & 0 & 0\\
0 & -1 & 0\\
0 & 0 & 0
\end{array}
\right)\, , 
\nonumber\\
&
\lambda_4
=
\left(
\begin{array}{ccc}
0 & 0 & 1\\
0 & 0 & 0\\
1 & 0 & 0
\end{array}
\right)\, , 
\quad
\lambda_5
=
\left(
\begin{array}{ccc}
0 & 0 & -\mathrm{i}\\
0 & 0 & 0\\
\mathrm{i} & 0 & 0
\end{array}
\right)\, , 
\nonumber\\
&
\lambda_6
=
\left(
\begin{array}{ccc}
0 & 0 & 0\\
0 & 0 & 1\\
0 & 1 & 0
\end{array}
\right)\, , 
\quad
\lambda_7
=
\left(
\begin{array}{ccc}
0 & 0 & 0\\
0 & 0 & -\mathrm{i}\\
0 & \mathrm{i} & 0
\end{array}
\right)\, , 
\quad\lambda_8
=
\frac{1}{\sqrt{3}}
\left(
\begin{array}{ccc}
1 & 0 & 0\\
0 & 1 & 0\\
0 & 0 & -2
\end{array}
\right)\, . 
\end{align}
For SU($N$) theory, we can use $\frac{S_{ab}}{\sqrt{2}}$, $\frac{A_{ab}}{\sqrt{2}}$ ($a<b$), and $\frac{D_{n}}{\sqrt{n(n+1)}}$ ($n=1,\cdots,N-1$), 
where
\begin{align}
(S_{ab})_{ij}\equiv\delta_{ai}\delta_{bj}+\delta_{aj}\delta_{bi}, 
\quad
(A_{ab})_{ij}\equiv\mathrm{i}\left(\delta_{ai}\delta_{bj}-\delta_{aj}\delta_{bi}\right)
\end{align}
and
\begin{align}
D_{n}\equiv\textrm{diag}(1,\cdots,1,-n,0,\cdots,0)\, .
\end{align}

The structure constant $f_{\alpha\beta\gamma}$ is defined by
\begin{align}
    [\tau_\alpha,\tau_\beta]=\mathrm{i}\sum_\gamma f_{\alpha\beta\gamma}\tau_\gamma\, . 
\end{align}
Equivalently, 
\begin{align}
f_{\alpha\beta\gamma}
=
-\mathrm{i}\cdot
\mathrm{Tr}\left(
    [\tau_\alpha,\tau_\beta]\tau_\gamma
    \right)\, . 
\end{align}
Combining this expression and trace cyclicity, we can see that $f_{\alpha\beta\gamma}$ is totally antisymmetric.

%%%%%%%%%%%%%%%%%%%%%%%
%%%%%%%%%%%%%%%%%%%%%%%
\section{Some aspects of gauge symmetry}\label{sec:gauge_symmetry}
%%%%%%%%%%%%%%%%%%%%%%%
%%%%%%%%%%%%%%%%%%%%%%%
In this appendix, we discuss a few issues associated with the gauge symmetry, mostly following ref.~\cite{Hanada2025}. 

When our universal framework is applied to gauge theories, specifically the matrix models and orbifold lattice, the extended Hilbert space that contains is used. The use of the extended Hilber space often causes unnecessary concern due to a widespread misunderstanding, ``physical states must be gauge invariant." We will address this point in Appendix~
\ref{sec:extended_vs_singlet}. There is no problem as long as the Hamiltonian is gauge invariant. The truncation on the coordinate and momentum bases breaks the gauge symmetry. We argue that the restoration of gauge symmetry can be achieved efficiently in the universal framework. 

%%%%%%%%%%%%%%%%%%%%%%%
%%%%%%%%%%%%%%%%%%%%%%%
\subsection{Extended Hilbert space and gauge-invariant Hilbert space}\label{sec:extended_vs_singlet}
%%%%%%%%%%%%%%%%%%%%%%%
%%%%%%%%%%%%%%%%%%%%%%%
In the Hamiltonian formulation in the $A_t=0$ gauge, it is often said that ``physical states must be gauge invariant." In fact, however, the gauge-invariant Hilbert space is only one of many equivalent ways to describe physical states~\cite{Hanada:2020uvt,Hanada:2021ipb,Fliss:2024don}.

Firstly, let us see how the singlet condition arises. Let us take the canonical partition function as an example, although essentially the same argument holds for real-time evolution. As demonstrated in refs.\cite{Rinaldi:2021jbg,Gautam:2022exf} for the cases of Hermitian and unitary matrix models, the path integral formalism can be rewritten in terms of the operator formalism on the extended Hilbert space as
\begin{align}
    Z(T)
    =
    \int [d\phi][dA_t] e^{-S[\phi,A_t]}
    =
    \frac{1}{\mathrm{vol}\, \mathcal{G}}
    \int_{\mathcal{G}}dg
    \mathrm{Tr}_{\mathcal{H}_{\rm ext}}\left(
    \hat{g}e^{-\hat{H}/T}
    \right)\, . 
\end{align}
Here, $A_t$ is the temporal component of gauge field and $\phi$ stands for all other fields including the spatial component of the gauge field. $\mathcal{G}$ is the local gauge transformation, which is $\mathcal{G}=\prod_{\vec{x}:\mathrm{sites}}[\mathrm{SU}(N)]_{\vec{x}}$ for the case of lattice gauge theory, $\mathrm{vol}\, \mathcal{G}$ is the volume of $\mathcal{G}$, and $\hat{g}$ acts on the Hilbert space as a gauge transformation corresponding to a group element $g\in\mathcal{G}$. The integral $\int_{\mathcal{G}}dg$ is taken using the Haar measure. 
$\frac{1}{\mathrm{vol}\, \mathcal{G}}\int_{\mathcal{G}}dg\, \hat{g}$ is a projector from the extended Hilbert space. If one follows the derivation~\cite{Rinaldi:2021jbg,Gautam:2022exf} carefully, one can see that $g$ is the Polyakov line and the integration of $g$ is the integration of $A_t$. 
The same expression holds for discrete groups, replacing $\mathrm{vol}\, \mathcal{G}$ with the number of elements and $\int dg$ with the sum over all elements. Historically the first example is the system of $N$ indistinguishable bosons, which can be seen as a gauge theory with the $\mathrm{S}_N$ gauge symmetry. The projector $\frac{1}{\mathrm{vol}\, \mathcal{G}}\int_{\mathcal{G}}dg\, \hat{g}$ forces us to identify the states related by gauge transformation, like we do in path integral or in classical electrodynamics, or, as Bose and Einstein did, \textit{bosons are not distinguishable}. We can take any state on a gauge orbit~\cite{Hanada:2020uvt,Hanada:2021ipb,Fliss:2024don}, and we can also take the linear combination of all the states on the orbit and obtain a single state. The latter leads to another, equivalent expression of the partition function, 
\begin{align}
    Z(T)
    =
    \mathrm{Tr}_{\mathcal{H}_{\rm inv}}\left(
    e^{-\hat{H}/T}
    \right)\, . 
\end{align}

There is no reason that one must use $\mathcal{H}_{\rm inv}$. By properly avoiding the double counting of euqivalent states, we can describe the same physics by using $\mathcal{H}_{\rm ext}$. 

For quantum simulations, the gauge-invariant Hilbert space has a serious issue that must be addressed before it is used for large-scale simulations that can lead to quantum advantage~\cite{Hanada2025}. Namely, due to the nonlocal nature of the basis, the coding of the states and Hamiltonian is complicated and requires significant computations on classical computer that becomes the bottle neck due to the exponential growth of the dimension of the Hilbert space with respect to the number of qubits. Furthermore, the depth of the circuit grows exponentially. These are in s stark contrast to our universal approach. This makes it hard to take lattice size sufficiently large and lattice spacing sufficiently small. 

They key role of gauge symmetry is to introduce massless vector field compatible with Lorentz symmetry without having negative-norm states~\cite{Weinberg_book,Peskin:1995ev,Kugo:1979gm}. On a lattice, Lorentz symmetry is broken by construction. In lattice Hamiltonian formulation, Lorentz symmetry is restored when truncation in the Hilbert space is removed \textit{and} continuum limit is taken. The gauge-invariant Hilbert space approach makes it difficult to take these limits, and hence, the original motivation of gauge symmetry could be destroyed. Somewhat counter-intuitively, by giving up the exact gauge symmetry at truncated level, one can respect the original motivation of gauge symmetry.  
%%%%%%%%%%%%%%%%%%%%%%%
%%%%%%%%%%%%%%%%%%%%%%%
\subsection{Truncation effect on gauge symmetry}
%%%%%%%%%%%%%%%%%%%%%%%
%%%%%%%%%%%%%%%%%%%%%%%
Because our universal framework allows to take the truncation level $\Lambda$ large, we can remove the truncation effects, including but not limited to those on the gauge symmetry. To demonstrate it, let us study a one-matrix model with the SU(2) gauge group. The coordinate and momebntum operators are $\hat{X}=\frac{1}{\sqrt{2}}\sum_{\alpha=1}^3\hat{X}_\alpha\sigma^\alpha$ and $\hat{P}=\frac{1}{\sqrt{2}}\sum_{\alpha=1}^3\hat{P}_\alpha\sigma^\alpha$, respectively. Because there are only three bosons, we can use the exact diagonalization to study the property of this model.\footnote{
Unlike multi-matrix models, gauge-fixed approach is efficient for this model. However, we use the extended Hilbert space here because our motivation is to know the truncation effect in this approach. 
}  
For the Hamiltonian, we use
\begin{align}
    \hat{H}
    =
    \mathrm{Tr}\left(
            \frac{1}{2}\hat{P}^2
            +
            \frac{1}{2}\hat{X}^2
            +
            \frac{1}{4}\hat{X}^4
    \right)\, . 
\end{align}
The generators of SU(2) gauge transformation is
\begin{align}
    \hat{G}_\alpha
    =
    \sum_{\beta,\gamma}
    f^{\alpha\beta\gamma}
    \hat{X}_\beta\hat{P}_\gamma, 
\end{align}
or written more explicitly, 
\begin{align}
    \hat{G}_1
    &=
    \sqrt{2}
    \left(
    \hat{X}_2\hat{P}_3
    -
    \hat{X}_3\hat{P}_2
    \right)\, , 
    \nonumber\\
    \hat{G}_2
    &=
    \sqrt{2}
    \left(
    \hat{X}_3\hat{P}_1
    -
    \hat{X}_1\hat{P}_3
    \right)\, , 
    \nonumber\\    
    \hat{G}_3
    &=
    \sqrt{2}
    \left(
    \hat{X}_1\hat{P}_2
    -
    \hat{X}_2\hat{P}_1
    \right)\, .
\end{align}

To estimate the truncation effect, we use the ground-state energy $E_0$ and the vacuum expectation value of $\sum_\alpha\hat{G}_\alpha^2$. We take the coordinate basis, so that $\hat{X}_\alpha$ take the form of \eqref{eq:x_as_sum_of_z} with  \eqref{eq:x_pbc}. For $\hat{P}_\alpha$, we use \eqref{p-option-2}, but we use the Fourier transform to express them in the coordinate basis. Therefore, while $\hat{X}_\alpha$ are diagonal, $\hat{P}_\alpha$ are dense matrices. 

\begin{figure}[htbp]
    \centering
   \includegraphics{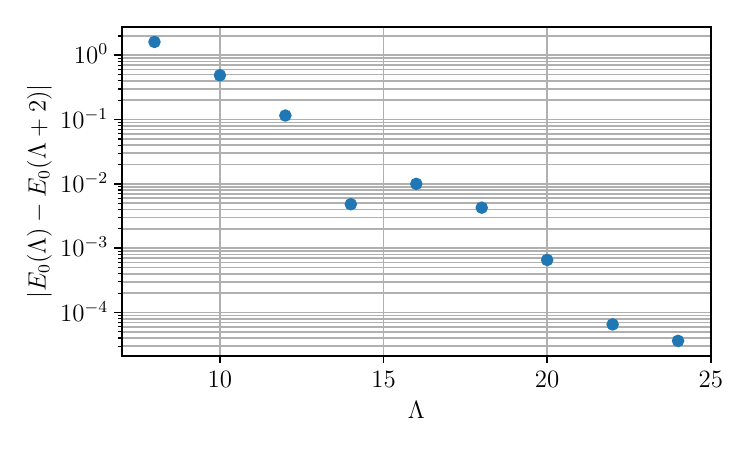}  
    \includegraphics{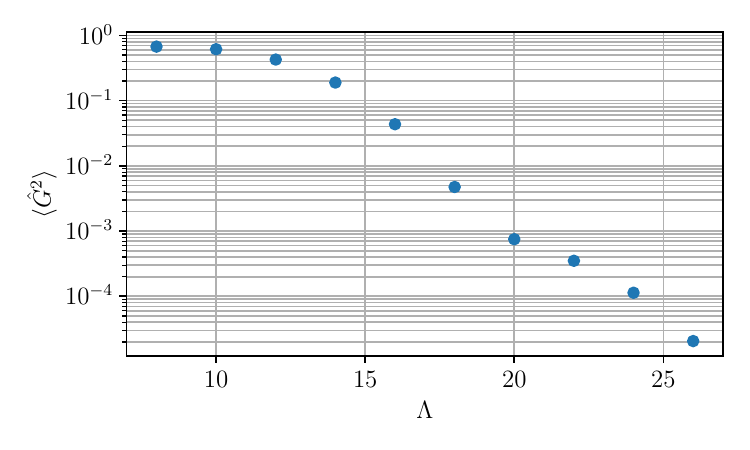} 
    \caption{
    [Top] $|E_0(\Lambda)-E_0(\Lambda+2)|$, $R=8$ (fix) and various values of $\Lambda$. Note that $E_0(\Lambda)-E_0(\Lambda+2)$ positive at $\Lambda\le 14$ and negative at $\Lambda\ge 16$. 
    [Bottom] $\langle\hat{G}^2\rangle$ obtained using the ground state, $R=8$ (fix) and various values of $\Lambda$.}\label{fig:gauge_violation}
\end{figure}

Fig.~\ref{fig:gauge_violation} shows $|E_0(\Lambda)-E_0(\Lambda+2)|$ and $\langle\hat{G}^2\rangle$ obtained using the ground state, $R=8$ (fix) and even values of $\Lambda$. We can see exponentially fast convergence to the ground state, including an exponential decay of $\langle\hat{G}^2\rangle$ indicating the exponentially fast restoration of SU(2) symmetry. (Note that the exact gauge symmetry requires $R\to\infty$ as well and hence $\langle\hat{G}^2\rangle$ does not become exactly zero at finite $R$.)

%%%%%%%%%%%%%%%%%%%%%
\bibliographystyle{utphys}
\bibliography{ref-orbifold}

\providecommand{\href}[2]{#2}\begingroup\raggedright\begin{thebibliography}{10}

\bibitem{Georgescu:2013oza}
I.~M. Georgescu, S.~Ashhab, and F.~Nori, ``{Quantum Simulation},''
  \href{http://dx.doi.org/10.1103/RevModPhys.86.153}{{\em Rev. Mod. Phys.}
  {\bfseries 86} (2014) 153}, \href{http://arxiv.org/abs/1308.6253}{{\ttfamily
  arXiv:1308.6253 [quant-ph]}}.

\bibitem{Dalmonte:2016alw}
M.~Dalmonte and S.~Montangero, ``{Lattice gauge theory simulations in the
  quantum information era},''
  \href{http://dx.doi.org/10.1080/00107514.2016.1151199}{{\em Contemp. Phys.}
  {\bfseries 57} no.~3, (2016) 388--412},
  \href{http://arxiv.org/abs/1602.03776}{{\ttfamily arXiv:1602.03776
  [cond-mat.quant-gas]}}.

\bibitem{Banuls:2019bmf}
M.~C. Ba\~nuls {\em et~al.}, ``{Simulating Lattice Gauge Theories within
  Quantum Technologies},''
  \href{http://dx.doi.org/10.1140/epjd/e2020-100571-8}{{\em Eur. Phys. J. D}
  {\bfseries 74} no.~8, (2020) 165},
  \href{http://arxiv.org/abs/1911.00003}{{\ttfamily arXiv:1911.00003
  [quant-ph]}}.

\bibitem{Zohar:2015hwa}
E.~Zohar, J.~I. Cirac, and B.~Reznik, ``{Quantum Simulations of Lattice Gauge
  Theories using Ultracold Atoms in Optical Lattices},''
  \href{http://dx.doi.org/10.1088/0034-4885/79/1/014401}{{\em Rept. Prog.
  Phys.} {\bfseries 79} no.~1, (2016) 014401},
  \href{http://arxiv.org/abs/1503.02312}{{\ttfamily arXiv:1503.02312
  [quant-ph]}}.

\bibitem{Aidelsburger:2021mia}
M.~Aidelsburger {\em et~al.}, ``{Cold atoms meet lattice gauge theory},''
  \href{http://dx.doi.org/10.1098/rsta.2021.0064}{{\em Phil. Trans. Roy. Soc.
  Lond. A} {\bfseries 380} (2021) 20210064},
  \href{http://arxiv.org/abs/2106.03063}{{\ttfamily arXiv:2106.03063
  [cond-mat.quant-gas]}}.

\bibitem{Zohar:2021nyc}
E.~Zohar, ``{Quantum simulation of lattice gauge theories in more than one
  space dimension\textemdash{}requirements, challenges and methods},''
  \href{http://dx.doi.org/10.1098/rsta.2021.0069}{{\em Phil. Trans. A. Math.
  Phys. Eng. Sci.} {\bfseries 380} no.~2216, (2021) 20210069},
  \href{http://arxiv.org/abs/2106.04609}{{\ttfamily arXiv:2106.04609
  [quant-ph]}}.

\bibitem{Klco:2021lap}
N.~Klco, A.~Roggero, and M.~J. Savage, ``{Standard model physics and the
  digital quantum revolution: thoughts about the interface},''
  \href{http://dx.doi.org/10.1088/1361-6633/ac58a4}{{\em Rept. Prog. Phys.}
  {\bfseries 85} no.~6, (2022) 064301},
  \href{http://arxiv.org/abs/2107.04769}{{\ttfamily arXiv:2107.04769
  [quant-ph]}}.

\bibitem{Bauer:2023qgm}
C.~W. Bauer, Z.~Davoudi, N.~Klco, and M.~J. Savage, ``{Quantum simulation of
  fundamental particles and forces},''
  \href{http://dx.doi.org/10.1038/s42254-023-00599-8}{{\em Nature Rev. Phys.}
  {\bfseries 5} no.~7, (2023) 420--432}.

\bibitem{Halimeh:2023ca}
J.~C. Halimeh, M.~Aidelsburger, F.~Grusdt, P.~Hauke, and B.~Yang, ``Cold-atom
  quantum simulators of gauge theories,''
  \href{http://arxiv.org/abs/2310.12201}{{\ttfamily arXiv:2310.12201
  [cond-mat.quant-gas]}}. \url{https://arxiv.org/abs/2310.12201}.

\bibitem{Bauer:2022hpo}
C.~W. Bauer {\em et~al.}, ``Quantum simulation for high-energy physics,''
  \href{http://dx.doi.org/10.1103/PRXQuantum.4.027001}{{\em PRX Quantum}
  {\bfseries 4} (May, 2023) 027001}.
  \url{https://link.aps.org/doi/10.1103/PRXQuantum.4.027001}.

\bibitem{diMeglio:2023qc}
A.~Di~Meglio {\em et~al.}, ``Quantum computing for high-energy physics: State
  of the art and challenges,''
  \href{http://dx.doi.org/10.1103/PRXQuantum.5.037001}{{\em PRX Quantum}
  {\bfseries 5} (Aug, 2024) 037001}.
  \url{https://link.aps.org/doi/10.1103/PRXQuantum.5.037001}.

\bibitem{Byrnes:2003gg}
T.~M.~R. Byrnes, M.~Loan, C.~J. Hamer, F.~D.~R. Bonnet, D.~B. Leinweber, A.~G.
  Williams, and J.~M. Zanotti, ``{The Hamiltonian limit of (3+1)-D SU(3)
  lattice gauge theory on anisotropic lattices},''
  \href{http://dx.doi.org/10.1103/PhysRevD.69.074509}{{\em Phys. Rev. D}
  {\bfseries 69} (2004) 074509},
  \href{http://arxiv.org/abs/hep-lat/0311014}{{\ttfamily
  arXiv:hep-lat/0311014}}.

\bibitem{Weinberg_book}
S.~Weinberg, {\em The Quantum Theory of Fields}.
\newblock Vol. 2: Modern Applications. Cambridge University Press, 1995.
\newblock \url{https://books.google.de/books?id=doeDB3\_WLvwC}.

\bibitem{Su:2024uuc}
G.-X. Su, J.~J. Osborne, and J.~C. Halimeh, ``{Cold-Atom Particle Collider},''
  \href{http://dx.doi.org/10.1103/PRXQuantum.5.040310}{{\em PRX Quantum}
  {\bfseries 5} no.~4, (2024) 040310},
  \href{http://arxiv.org/abs/2401.05489}{{\ttfamily arXiv:2401.05489
  [cond-mat.quant-gas]}}.

\bibitem{Maldacena:2024qlf}
J.~Maldacena, \href{http://dx.doi.org/10.1007/978-981-19-3079-9_65-1}{{\em {The
  AdS/CFT Correspondence}}}.
\newblock Springer, 2024.

\bibitem{tHooft:1973alw}
G.~'t~Hooft, ``{A Planar Diagram Theory for Strong Interactions},''
  \href{http://dx.doi.org/10.1016/0550-3213(74)90154-0}{{\em Nucl. Phys. B}
  {\bfseries 72} (1974) 461}.

\bibitem{Lucini:2001ej}
B.~Lucini and M.~Teper, ``{SU(N) gauge theories in four-dimensions: Exploring
  the approach to N = infinity},''
  \href{http://dx.doi.org/10.1088/1126-6708/2001/06/050}{{\em JHEP} {\bfseries
  06} (2001) 050}, \href{http://arxiv.org/abs/hep-lat/0103027}{{\ttfamily
  arXiv:hep-lat/0103027}}.

\bibitem{Troyer:2024kra}
M.~Troyer, E.~V. Benjamin, and A.~Gevorkian, ``{Quantum for Good and the
  Societal Impact of Quantum Computing},''  (3, 2024) ,
  \href{http://arxiv.org/abs/2403.02921}{{\ttfamily arXiv:2403.02921
  [physics.soc-ph]}}.

\bibitem{Kogut:1974ag}
J.~B. Kogut and L.~Susskind, ``{Hamiltonian Formulation of Wilson's Lattice
  Gauge Theories},'' \href{http://dx.doi.org/10.1103/PhysRevD.11.395}{{\em
  Phys. Rev. D} {\bfseries 11} (1975) 395--408}.

\bibitem{Wilson:1974sk}
K.~G. Wilson, ``{Confinement of Quarks},''
  \href{http://dx.doi.org/10.1103/PhysRevD.10.2445}{{\em Phys. Rev. D}
  {\bfseries 10} (1974) 2445--2459}.

\bibitem{Peter-Weyl-original}
F.~Peter and H.~Weyl, ``{Die Vollständigkeit der primitiven Darstellungen
  einer geschlossenen kontinuierlichen Gruppe},'' {\em Mathematische Annalen}
  {\bfseries 97} no.~1, (1927) 737--755.

\bibitem{Zohar:2014qma}
E.~Zohar and M.~Burrello, ``{Formulation of lattice gauge theories for quantum
  simulations},'' \href{http://dx.doi.org/10.1103/PhysRevD.91.054506}{{\em
  Phys. Rev. D} {\bfseries 91} no.~5, (2015) 054506},
  \href{http://arxiv.org/abs/1409.3085}{{\ttfamily arXiv:1409.3085
  [quant-ph]}}.

\bibitem{Byrnes:2005qx}
T.~Byrnes and Y.~Yamamoto, ``{Simulating lattice gauge theories on a quantum
  computer},'' \href{http://dx.doi.org/10.1103/PhysRevA.73.022328}{{\em Phys.
  Rev. A} {\bfseries 73} (2006) 022328},
  \href{http://arxiv.org/abs/quant-ph/0510027}{{\ttfamily
  arXiv:quant-ph/0510027}}.

\bibitem{Garofalo:2023zkd}
M.~Garofalo, T.~Hartung, T.~Jakobs, K.~Jansen, J.~Ostmeyer, D.~Rolfes,
  S.~Romiti, and C.~Urbach, ``{Testing the $\mathrm{SU}(2)$ lattice Hamiltonian
  built from $S_3$ partitionings},''
  \href{http://arxiv.org/abs/2311.15926}{{\ttfamily arXiv:2311.15926
  [hep-lat]}}.

\bibitem{Fontana:2024rux}
P.~Fontana, M.~M. Riaza, and A.~Celi, ``{An efficient finite-resource
  formulation of non-Abelian lattice gauge theories beyond one dimension},''
  \href{http://arxiv.org/abs/2409.04441}{{\ttfamily arXiv:2409.04441
  [quant-ph]}}.

\bibitem{Murairi:2024xpc}
E.~M. Murairi, M.~Sohaib~Alam, H.~Lamm, S.~Hadfield, and E.~Gustafson,
  ``{Highly-efficient quantum Fourier transformations for some nonabelian
  groups},''  (7, 2024) , \href{http://arxiv.org/abs/2408.00075}{{\ttfamily
  arXiv:2408.00075 [quant-ph]}}.

\bibitem{Hayata:2023puo}
T.~Hayata and Y.~Hidaka, ``{String-net formulation of Hamiltonian lattice
  Yang-Mills theories and quantum many-body scars in a nonabelian gauge
  theory},'' \href{http://dx.doi.org/10.1007/JHEP09(2023)126}{{\em JHEP}
  {\bfseries 09} (2023) 126}, \href{http://arxiv.org/abs/2305.05950}{{\ttfamily
  arXiv:2305.05950 [hep-lat]}}.

\bibitem{Hayata:2023bgh}
T.~Hayata and Y.~Hidaka, ``{q deformed formulation of Hamiltonian SU(3)
  Yang-Mills theory},'' \href{http://dx.doi.org/10.1007/JHEP09(2023)123}{{\em
  JHEP} {\bfseries 09} (2023) 123},
  \href{http://arxiv.org/abs/2306.12324}{{\ttfamily arXiv:2306.12324
  [hep-lat]}}.

\bibitem{Zache:2023dko}
T.~V. Zache, D.~Gonz\'alez-Cuadra, and P.~Zoller, ``{Quantum and Classical
  Spin-Network Algorithms for q-Deformed Kogut-Susskind Gauge Theories},''
  \href{http://dx.doi.org/10.1103/PhysRevLett.131.171902}{{\em Phys. Rev.
  Lett.} {\bfseries 131} no.~17, (2023) 171902},
  \href{http://arxiv.org/abs/2304.02527}{{\ttfamily arXiv:2304.02527
  [quant-ph]}}.

\bibitem{Ciavarella:2024fzw}
A.~N. Ciavarella and C.~W. Bauer, ``{Quantum Simulation of SU(3) Lattice
  Yang-Mills Theory at Leading Order in Large-Nc Expansion},''
  \href{http://dx.doi.org/10.1103/PhysRevLett.133.111901}{{\em Phys. Rev.
  Lett.} {\bfseries 133} no.~11, (2024) 111901},
  \href{http://arxiv.org/abs/2402.10265}{{\ttfamily arXiv:2402.10265
  [hep-ph]}}.

\bibitem{Yang2020observation}
B.~Yang, H.~Sun, R.~Ott, H.-Y. Wang, T.~V. Zache, J.~C. Halimeh, Z.-S. Yuan,
  P.~Hauke, and J.-W. Pan, ``Observation of gauge invariance in a 71-site
  {Bose--Hubbard} quantum simulator,''
  \href{http://dx.doi.org/10.1038/s41586-020-2910-8}{{\em Nature} {\bfseries
  587} no.~7834, (2020) 392--396}.
  \url{https://doi.org/10.1038/s41586-020-2910-8}.

\bibitem{Zhou2022thermalization}
Z.-Y. Zhou, G.-X. Su, J.~C. Halimeh, R.~Ott, H.~Sun, P.~Hauke, B.~Yang, Z.-S.
  Yuan, J.~Berges, and J.-W. Pan, ``Thermalization dynamics of a gauge theory
  on a quantum simulator,''
  \href{http://dx.doi.org/10.1126/science.abl6277}{{\em Science} {\bfseries
  377} no.~6603, (2022) 311--314}.

\bibitem{Cochran:2024rwe}
T.~A. Cochran {\em et~al.}, ``{Visualizing Dynamics of Charges and Strings in
  (2+1)D Lattice Gauge Theories},''
  \href{http://arxiv.org/abs/2409.17142}{{\ttfamily arXiv:2409.17142
  [quant-ph]}}.

\bibitem{Gyawali:2024hrz}
G.~Gyawali {\em et~al.}, ``{Observation of disorder-free localization and
  efficient disorder averaging on a quantum processor},''
  \href{http://arxiv.org/abs/2410.06557}{{\ttfamily arXiv:2410.06557
  [quant-ph]}}.

\bibitem{gonzalezcuadra2024observationstringbreaking2}
D.~Gonzalez-Cuadra, M.~Hamdan, T.~V. Zache, B.~Braverman, M.~Kornjaca,
  A.~Lukin, S.~H. Cantu, F.~Liu, S.-T. Wang, A.~Keesling, M.~D. Lukin,
  P.~Zoller, and A.~Bylinskii, ``Observation of string breaking on a (2 + 1)d
  rydberg quantum simulator,''
  \href{http://arxiv.org/abs/2410.16558}{{\ttfamily arXiv:2410.16558
  [quant-ph]}}. \url{https://arxiv.org/abs/2410.16558}.

\bibitem{Buser:2020cvn}
A.~J. Buser, H.~Gharibyan, M.~Hanada, M.~Honda, and J.~Liu, ``{Quantum
  simulation of gauge theory via orbifold lattice},''
  \href{http://dx.doi.org/10.1007/JHEP09(2021)034}{{\em JHEP} {\bfseries 09}
  (2021) 034}, \href{http://arxiv.org/abs/2011.06576}{{\ttfamily
  arXiv:2011.06576 [hep-th]}}.

\bibitem{Bergner:2024qjl}
G.~Bergner, M.~Hanada, E.~Rinaldi, and A.~Sch{\" a}fer, ``{Toward QCD on
  quantum computer: orbifold lattice approach},''
  \href{http://dx.doi.org/10.1007/JHEP05(2024)234}{{\em JHEP} {\bfseries 05}
  (2024) 234}, \href{http://arxiv.org/abs/2401.12045}{{\ttfamily
  arXiv:2401.12045 [hep-th]}}.

\bibitem{Kaplan:2002wv}
D.~B. Kaplan, E.~Katz, and M.~{\" U}nsal, ``{Supersymmetry on a spatial
  lattice},'' \href{http://dx.doi.org/10.1088/1126-6708/2003/05/037}{{\em JHEP}
  {\bfseries 05} (2003) 037},
  \href{http://arxiv.org/abs/hep-lat/0206019}{{\ttfamily
  arXiv:hep-lat/0206019}}.

\bibitem{KU-private-communication}
D.~B. Kaplan and M.~{\" U}nsal, ``{Privare communication},''.

\bibitem{Arkani-Hamed:2001kyx}
N.~Arkani-Hamed, A.~G. Cohen, and H.~Georgi, ``{(De)constructing dimensions},''
  \href{http://dx.doi.org/10.1103/PhysRevLett.86.4757}{{\em Phys. Rev. Lett.}
  {\bfseries 86} (2001) 4757--4761},
  \href{http://arxiv.org/abs/hep-th/0104005}{{\ttfamily arXiv:hep-th/0104005}}.

\bibitem{Maldacena:2023acv}
J.~Maldacena, ``{A simple quantum system that describes a black hole},''  (3,
  2023) , \href{http://arxiv.org/abs/2303.11534}{{\ttfamily arXiv:2303.11534
  [hep-th]}}.

\bibitem{Dirac:1925jy}
P.~A.~M. Dirac, ``{The fundamental equations of quantum mechanics},''
  \href{http://dx.doi.org/10.1098/rspa.1925.0150}{{\em Proc. Roy. Soc. Lond. A}
  {\bfseries 109} (1925) 642--653}.

\bibitem{Coppersmith:2002skh}
D.~Coppersmith, ``{An approximate Fourier transform useful in quantum
  factoring},'' \href{http://arxiv.org/abs/quant-ph/0201067}{{\ttfamily
  arXiv:quant-ph/0201067}}.

\bibitem{Jordan:2012xnu}
S.~P. Jordan, K.~S.~M. Lee, and J.~Preskill, ``{Quantum Algorithms for Quantum
  Field Theories},'' \href{http://dx.doi.org/10.1126/science.1217069}{{\em
  Science} {\bfseries 336} (2012) 1130--1133},
  \href{http://arxiv.org/abs/1111.3633}{{\ttfamily arXiv:1111.3633
  [quant-ph]}}.

\bibitem{Hanada:2022pps}
M.~Hanada, J.~Liu, E.~Rinaldi, and M.~Tezuka, ``{Estimating truncation effects
  of quantum bosonic systems using sampling algorithms},''
  \href{http://dx.doi.org/10.1088/2632-2153/ad035c}{{\em Mach. Learn. Sci.
  Tech.} {\bfseries 4} no.~4, (2023) 045021},
  \href{http://arxiv.org/abs/2212.08546}{{\ttfamily arXiv:2212.08546
  [quant-ph]}}.

\bibitem{Somma:2015bcw}
R.~D. Somma, ``{Quantum simulations of one dimensional quantum systems},'' {\em
  Quantum Info. Comput.} (3, 2015) ,
  \href{http://arxiv.org/abs/1503.06319}{{\ttfamily arXiv:1503.06319
  [quant-ph]}}.

\bibitem{Peskin:1995ev}
M.~E. Peskin and D.~V. Schroeder,
  \href{http://dx.doi.org/10.1201/9780429503559}{{\em {An Introduction to
  quantum field theory}}}.
\newblock Addison-Wesley, Reading, USA, 1995.

\bibitem{Fradkin:2021zbi}
E.~Fradkin, {\em {Quantum Field Theory: An Integrated Approach}}.
\newblock Princeton University Press, 3, 2021.

\bibitem{Halimeh2020reliability}
J.~C. Halimeh and P.~Hauke, ``Reliability of lattice gauge theories,''
  \href{http://dx.doi.org/10.1103/PhysRevLett.125.030503}{{\em Phys. Rev.
  Lett.} {\bfseries 125} (Jul, 2020) 030503}.
  \url{https://link.aps.org/doi/10.1103/PhysRevLett.125.030503}.

\bibitem{Halimeh2022gauge}
J.~C. Halimeh, H.~Lang, and P.~Hauke, ``Gauge protection in non-abelian lattice
  gauge theories,'' \href{http://dx.doi.org/10.1088/1367-2630/ac5564}{{\em New
  Journal of Physics} {\bfseries 24} no.~3, (Mar, 2022) 033015}.
  \url{https://dx.doi.org/10.1088/1367-2630/ac5564}.

\bibitem{VanDamme2023reliability}
M.~Van~Damme, H.~Lang, P.~Hauke, and J.~C. Halimeh, ``Reliability of lattice
  gauge theories in the thermodynamic limit,''
  \href{http://dx.doi.org/10.1103/PhysRevB.107.035153}{{\em Phys. Rev. B}
  {\bfseries 107} (Jan, 2023) 035153}.
  \url{https://link.aps.org/doi/10.1103/PhysRevB.107.035153}.

\bibitem{Douglas:1996sw}
M.~R. Douglas and G.~W. Moore, ``{D-branes, quivers, and ALE instantons},''
  \href{http://arxiv.org/abs/hep-th/9603167}{{\ttfamily arXiv:hep-th/9603167}}.

\bibitem{Shaw:2020udc}
A.~F. Shaw, P.~Lougovski, J.~R. Stryker, and N.~Wiebe, ``{Quantum Algorithms
  for Simulating the Lattice Schwinger Model},''
  \href{http://dx.doi.org/10.22331/q-2020-08-10-306}{{\em Quantum} {\bfseries
  4} (2020) 306}, \href{http://arxiv.org/abs/2002.11146}{{\ttfamily
  arXiv:2002.11146 [quant-ph]}}.

\bibitem{Childs:2019hts}
A.~M. Childs, Y.~Su, M.~C. Tran, N.~Wiebe, and S.~Zhu, ``{Theory of Trotter
  Error with Commutator Scaling},''
  \href{http://dx.doi.org/10.1103/PhysRevX.11.011020}{{\em Phys. Rev. X}
  {\bfseries 11} no.~1, (2021) 011020},
  \href{http://arxiv.org/abs/1912.08854}{{\ttfamily arXiv:1912.08854
  [quant-ph]}}.

\bibitem{Cowtan:2019loc}
A.~Cowtan, S.~Dilkes, R.~Duncan, W.~Simmons, and S.~Sivarajah, ``{Phase Gadget
  Synthesis for Shallow Circuits},''
  \href{http://dx.doi.org/10.4204/EPTCS.318.13}{{\em EPTCS} {\bfseries 318}
  (2020) 213--228}, \href{http://arxiv.org/abs/1906.01734}{{\ttfamily
  arXiv:1906.01734 [quant-ph]}}.

\bibitem{Coecke:2011gh}
B.~Coecke and R.~Duncan, ``Interacting quantum observables: categorical algebra
  and diagrammatics,''
  \href{http://dx.doi.org/10.1088/1367-2630/13/4/043016}{{\em New Journal of
  Physics} {\bfseries 13} (2011) 043016}.

\bibitem{quantinuum2024}
Q.~H-series, ``System model h1 product sheet.''
  \url{https://www.quantinuum.com/}, 2024.
\newblock Accessed: 2024-10-22.

\bibitem{FowlerSurfaceReview}
A.~G. Fowler, M.~Mariantoni, J.~M. Martinis, and A.~N. Cleland, ``Surface
  codes: Towards practical large-scale quantum computation,''
  \href{http://dx.doi.org/10.1103/PhysRevA.86.032324}{{\em Phys. Rev. A}
  {\bfseries 86} (Sep, 2012) 032324}.
  \url{https://link.aps.org/doi/10.1103/PhysRevA.86.032324}.

\bibitem{Horsman_2012}
D.~Horsman, A.~G. Fowler, S.~Devitt, and R.~V. Meter, ``Surface code quantum
  computing by lattice surgery,''
  \href{http://dx.doi.org/10.1088/1367-2630/14/12/123011}{{\em New Journal of
  Physics} {\bfseries 14} no.~12, (Dec., 2012) 123011}.
  \url{http://dx.doi.org/10.1088/1367-2630/14/12/123011}.

\bibitem{Herr_2017}
D.~Herr, F.~Nori, and S.~J. Devitt, ``Optimization of lattice surgery is
  np-hard,'' \href{http://dx.doi.org/10.1038/s41534-017-0035-1}{{\em npj
  Quantum Information} {\bfseries 3} no.~1, (Sept., 2017) }.
  \url{http://dx.doi.org/10.1038/s41534-017-0035-1}.

\bibitem{Litinski2019gameofsurfacecodes}
D.~Litinski, ``A {G}ame of {S}urface {C}odes: {L}arge-{S}cale {Q}uantum
  {C}omputing with {L}attice {S}urgery,''
  \href{http://dx.doi.org/10.22331/q-2019-03-05-128}{{\em {Quantum}} {\bfseries
  3} (Mar., 2019) 128}. \url{https://doi.org/10.22331/q-2019-03-05-128}.

\bibitem{BravyiKitaevMagic}
S.~Bravyi and A.~Kitaev, ``Universal quantum computation with ideal clifford
  gates and noisy ancillas,''
  \href{http://dx.doi.org/10.1103/PhysRevA.71.022316}{{\em Phys. Rev. A}
  {\bfseries 71} (Feb, 2005) 022316}.
  \url{https://link.aps.org/doi/10.1103/PhysRevA.71.022316}.

\bibitem{KnillMagic}
E.~{Knill}, ``{Fault-Tolerant Postselected Quantum Computation: Threshold
  Analysis},'' \href{http://dx.doi.org/10.48550/arXiv.quant-ph/0404104}{{\em
  arXiv e-prints} (Apr., 2004) quant--ph/0404104},
  \href{http://arxiv.org/abs/quant-ph/0404104}{{\ttfamily
  arXiv:quant-ph/0404104 [quant-ph]}}.

\bibitem{Gidney_Cultivation}
C.~{Gidney}, N.~{Shutty}, and C.~{Jones}, ``{Magic state cultivation: growing T
  states as cheap as CNOT gates},''
  \href{http://dx.doi.org/10.48550/arXiv.2409.17595}{{\em arXiv e-prints}
  (Sept., 2024) arXiv:2409.17595},
  \href{http://arxiv.org/abs/2409.17595}{{\ttfamily arXiv:2409.17595
  [quant-ph]}}.

\bibitem{RossTgate}
N.~J. {Ross} and P.~{Selinger}, ``{Optimal ancilla-free Clifford+T
  approximation of z-rotations},''
  \href{http://dx.doi.org/10.48550/arXiv.1403.2975}{{\em arXiv e-prints} (Mar.,
  2014) arXiv:1403.2975}, \href{http://arxiv.org/abs/1403.2975}{{\ttfamily
  arXiv:1403.2975 [quant-ph]}}.

\bibitem{Toshio:2024hut}
R.~Toshio, Y.~Akahoshi, J.~Fujisaki, H.~Oshima, S.~Sato, and K.~Fujii,
  ``{Practical quantum advantage on partially fault-tolerant quantum
  computer},''  (8, 2024) , \href{http://arxiv.org/abs/2408.14848}{{\ttfamily
  arXiv:2408.14848 [quant-ph]}}.

\bibitem{Akahoshi:2024yme}
Y.~Akahoshi, R.~Toshio, J.~Fujisaki, H.~Oshima, S.~Sato, and K.~Fujii,
  ``{Compilation of Trotter-Based Time Evolution for Partially Fault-Tolerant
  Quantum Computing Architecture},''  (8, 2024) ,
  \href{http://arxiv.org/abs/2408.14929}{{\ttfamily arXiv:2408.14929
  [quant-ph]}}.

\bibitem{CampbellETFQC}
E.~T. {Campbell}, ``{Early fault-tolerant simulations of the Hubbard model},''
  \href{http://dx.doi.org/10.1088/2058-9565/ac3110}{{\em Quantum Science and
  Technology} {\bfseries 7} no.~1, (Jan., 2022) 015007},
  \href{http://arxiv.org/abs/2012.09238}{{\ttfamily arXiv:2012.09238
  [quant-ph]}}.

\bibitem{Kliuchnikov:2016ong}
V.~Kliuchnikov, D.~Maslov, and M.~Mosca, ``{Practical approximation of
  single-qubit unitaries by single-qubit quantum Clifford and T circuits},''
  \href{http://dx.doi.org/10.1109/TC.2015.2409842}{{\em IEEE Trans. Comput.}
  {\bfseries 65} no.~1, (2016) 161--172}.

\bibitem{NamAQFT}
Y.~{Nam}, Y.~{Su}, and D.~{Maslov}, ``{Approximate Quantum Fourier Transform
  with $O(n \log(n))$ T gates},''
  \href{http://dx.doi.org/10.48550/arXiv.1803.04933}{{\em arXiv e-prints}
  (Mar., 2018) arXiv:1803.04933},
  \href{http://arxiv.org/abs/1803.04933}{{\ttfamily arXiv:1803.04933
  [quant-ph]}}.

\bibitem{Zemlevskiy:2024vxt}
N.~A. Zemlevskiy, ``{Scalable Quantum Simulations of Scattering in Scalar Field
  Theory on 120 Qubits},''  (11, 2024) ,
  \href{http://arxiv.org/abs/2411.02486}{{\ttfamily arXiv:2411.02486
  [quant-ph]}}.

\bibitem{Hanada2025}
M.~Hanada, S.~M. Matsuura, E.~Mendicelli, and E.~Rinaldi, ``{Exponential
  improvement in quantum simulations of bosons},''.

\bibitem{Erdos:2014zgc}
L.~Erd\H{o}s and D.~Schr\"oder, ``{Phase Transition in the Density of States of
  Quantum Spin Glasses},''
  \href{http://dx.doi.org/10.1007/s11040-014-9164-3}{{\em Math. Phys. Anal.
  Geom.} {\bfseries 17} no.~3-4, (2014) 441--464},
  \href{http://arxiv.org/abs/1407.1552}{{\ttfamily arXiv:1407.1552 [math-ph]}}.

\bibitem{Berkooz:2018qkz}
M.~Berkooz, P.~Narayan, and J.~Simon, ``{Chord diagrams, exact correlators in
  spin glasses and black hole bulk reconstruction},''
  \href{http://dx.doi.org/10.1007/JHEP08(2018)192}{{\em JHEP} {\bfseries 08}
  (2018) 192}, \href{http://arxiv.org/abs/1806.04380}{{\ttfamily
  arXiv:1806.04380 [hep-th]}}.

\bibitem{Baldwin:2019dki}
C.~L. Baldwin and B.~Swingle, ``{Quenched vs Annealed: Glassiness from SK to
  SYK},'' \href{http://dx.doi.org/10.1103/PhysRevX.10.031026}{{\em Phys. Rev.
  X} {\bfseries 10} no.~3, (2020) 031026},
  \href{http://arxiv.org/abs/1911.11865}{{\ttfamily arXiv:1911.11865
  [cond-mat.dis-nn]}}.

\bibitem{Hanada:2023rkf}
M.~Hanada, A.~Jevicki, X.~Liu, E.~Rinaldi, and M.~Tezuka, ``{A model of
  randomly-coupled Pauli spins},''
  \href{http://arxiv.org/abs/2309.15349}{{\ttfamily arXiv:2309.15349
  [hep-th]}}.

\bibitem{Anschuetz:2023igd}
E.~R. Anschuetz, D.~Gamarnik, and B.~T. Kiani, ``{Product states optimize
  quantum $p$-spin models for large $p$},''
  \href{http://arxiv.org/abs/2309.11709}{{\ttfamily arXiv:2309.11709
  [quant-ph]}}.

\bibitem{Swingle:2023nvv}
B.~Swingle and M.~Winer, ``{A Bosonic Model of Quantum Holography},'' {\em
  Phys. Rev. B} {\bfseries 109} (11, 2023) 094206,
  \href{http://arxiv.org/abs/2311.01516}{{\ttfamily arXiv:2311.01516
  [hep-th]}}.

\bibitem{Buividovich:2022jgv}
P.~V. Buividovich, ``{Quantum chaos in supersymmetric quantum mechanics: An
  exact diagonalization study},''
  \href{http://dx.doi.org/10.1103/PhysRevD.106.046001}{{\em Phys. Rev. D}
  {\bfseries 106} no.~4, (2022) 046001},
  \href{http://arxiv.org/abs/2205.09704}{{\ttfamily arXiv:2205.09704
  [hep-th]}}.

\bibitem{Carena:2021ltu}
M.~Carena, H.~Lamm, Y.-Y. Li, and W.~Liu, ``{Lattice renormalization of quantum
  simulations},'' \href{http://dx.doi.org/10.1103/PhysRevD.104.094519}{{\em
  Phys. Rev. D} {\bfseries 104} no.~9, (2021) 094519},
  \href{http://arxiv.org/abs/2107.01166}{{\ttfamily arXiv:2107.01166
  [hep-lat]}}.

\bibitem{Carena:2022hpz}
M.~Carena, E.~J. Gustafson, H.~Lamm, Y.-Y. Li, and W.~Liu, ``{Gauge theory
  couplings on anisotropic lattices},''
  \href{http://dx.doi.org/10.1103/PhysRevD.106.114504}{{\em Phys. Rev. D}
  {\bfseries 106} no.~11, (2022) 114504},
  \href{http://arxiv.org/abs/2208.10417}{{\ttfamily arXiv:2208.10417
  [hep-lat]}}.

\bibitem{Funcke:2022opx}
L.~Funcke, C.~F. Gro\ss{}, K.~Jansen, S.~K\"uhn, S.~Romiti, and C.~Urbach,
  ``{Hamiltonian limit of lattice QED in 2+1 dimensions},''
  \href{http://dx.doi.org/10.22323/1.430.0292}{{\em PoS} {\bfseries
  LATTICE2022} (2023) 292}, \href{http://arxiv.org/abs/2212.09627}{{\ttfamily
  arXiv:2212.09627 [hep-lat]}}.

\bibitem{Loan:2005ff}
M.~Loan, X.-Q. Luo, and Z.-H. Luo, ``{Monte Carlo study of glueball masses in
  the Hamiltonian limit of SU(3) lattice gauge theory},''
  \href{http://dx.doi.org/10.1142/S0217751X06029454}{{\em Int. J. Mod. Phys. A}
  {\bfseries 21} (2006) 2905--2936},
  \href{http://arxiv.org/abs/hep-lat/0503038}{{\ttfamily
  arXiv:hep-lat/0503038}}.

\bibitem{Braunstein:2005zz}
S.~L. Braunstein and P.~van Loock, ``{Quantum information with continuous
  variables},'' \href{http://dx.doi.org/10.1103/RevModPhys.77.513}{{\em Rev.
  Mod. Phys.} {\bfseries 77} (2005) 513--577},
  \href{http://arxiv.org/abs/quant-ph/0410100}{{\ttfamily
  arXiv:quant-ph/0410100}}.

\bibitem{Abel:2024kuv}
S.~Abel, M.~Spannowsky, and S.~Williams, ``{Simulating quantum field theories
  on continuous-variable quantum computers},''
  \href{http://dx.doi.org/10.1103/PhysRevA.110.012607}{{\em Phys. Rev. A}
  {\bfseries 110} no.~1, (2024) 012607},
  \href{http://arxiv.org/abs/2403.10619}{{\ttfamily arXiv:2403.10619
  [quant-ph]}}.

\bibitem{Ale:2024uxf}
V.~Ale, N.~M. Bauer, R.~G. Jha, F.~Ringer, and G.~Siopsis, ``{Quantum
  computation of SU(2) lattice gauge theory with continuous variables},''
  \href{http://arxiv.org/abs/2410.14580}{{\ttfamily arXiv:2410.14580
  [hep-lat]}}.

\bibitem{Myers:1999ps}
R.~C. Myers, ``{Dielectric branes},''
  \href{http://dx.doi.org/10.1088/1126-6708/1999/12/022}{{\em JHEP} {\bfseries
  12} (1999) 022}, \href{http://arxiv.org/abs/hep-th/9910053}{{\ttfamily
  arXiv:hep-th/9910053}}.

\bibitem{Gonzalez-Arroyo:1982hyq}
A.~Gonzalez-Arroyo and M.~Okawa, ``{The Twisted Eguchi-Kawai Model: A Reduced
  Model for Large N Lattice Gauge Theory},''
  \href{http://dx.doi.org/10.1103/PhysRevD.27.2397}{{\em Phys. Rev. D}
  {\bfseries 27} (1983) 2397}.

\bibitem{Gharibyan:2020bab}
H.~Gharibyan, M.~Hanada, M.~Honda, and J.~Liu, ``{Toward simulating
  superstring/M-theory on a quantum computer},''
  \href{http://dx.doi.org/10.1007/JHEP07(2021)140}{{\em JHEP} {\bfseries 07}
  (2021) 140}, \href{http://arxiv.org/abs/2011.06573}{{\ttfamily
  arXiv:2011.06573 [hep-th]}}.

\bibitem{Metropolis:1953am}
N.~Metropolis, A.~W. Rosenbluth, M.~N. Rosenbluth, A.~H. Teller, and E.~Teller,
  ``{Equation of state calculations by fast computing machines},''
  \href{http://dx.doi.org/10.1063/1.1699114}{{\em J. Chem. Phys.} {\bfseries
  21} (1953) 1087--1092}.

\bibitem{Duane:1987de}
S.~Duane, A.~D. Kennedy, B.~J. Pendleton, and D.~Roweth, ``{Hybrid Monte
  Carlo},'' \href{http://dx.doi.org/10.1016/0370-2693(87)91197-X}{{\em Phys.
  Lett. B} {\bfseries 195} (1987) 216--222}.

\bibitem{Hanada:2020uvt}
M.~Hanada, H.~Shimada, and N.~Wintergerst, ``{Color confinement and
  Bose-Einstein condensation},''
  \href{http://dx.doi.org/10.1007/JHEP08(2021)039}{{\em JHEP} {\bfseries 08}
  (2021) 039}, \href{http://arxiv.org/abs/2001.10459}{{\ttfamily
  arXiv:2001.10459 [hep-th]}}.

\bibitem{Hanada:2021ipb}
M.~Hanada, ``{Bulk geometry in gauge/gravity duality and color degrees of
  freedom},'' \href{http://dx.doi.org/10.1103/PhysRevD.103.106007}{{\em Phys.
  Rev. D} {\bfseries 103} no.~10, (2021) 106007},
  \href{http://arxiv.org/abs/2102.08982}{{\ttfamily arXiv:2102.08982
  [hep-th]}}.

\bibitem{Fliss:2024don}
J.~R. Fliss, A.~Frenkel, S.~A. Hartnoll, and R.~M. Soni, ``{Minimal Areas from
  Entangled Matrices},'' \href{http://arxiv.org/abs/2408.05274}{{\ttfamily
  arXiv:2408.05274 [hep-th]}}.

\bibitem{Rinaldi:2021jbg}
E.~Rinaldi, X.~Han, M.~Hassan, Y.~Feng, F.~Nori, M.~McGuigan, and M.~Hanada,
  ``{Matrix-Model Simulations Using Quantum Computing, Deep Learning, and
  Lattice Monte Carlo},''
  \href{http://dx.doi.org/10.1103/PRXQuantum.3.010324}{{\em PRX Quantum}
  {\bfseries 3} no.~1, (2022) 010324},
  \href{http://arxiv.org/abs/2108.02942}{{\ttfamily arXiv:2108.02942
  [quant-ph]}}.

\bibitem{Gautam:2022exf}
V.~Gautam, M.~Hanada, J.~Holden, and E.~Rinaldi, ``{Linear confinement in the
  partially-deconfined phase},''
  \href{http://dx.doi.org/10.1007/JHEP03(2023)195}{{\em JHEP} {\bfseries 03}
  (2023) 195}, \href{http://arxiv.org/abs/2208.14402}{{\ttfamily
  arXiv:2208.14402 [hep-th]}}.

\bibitem{Kugo:1979gm}
T.~Kugo and I.~Ojima, ``{Local Covariant Operator Formalism of Nonabelian Gauge
  Theories and Quark Confinement Problem},''
  \href{http://dx.doi.org/10.1143/PTPS.66.1}{{\em Prog. Theor. Phys. Suppl.}
  {\bfseries 66} (1979) 1--130}.

\end{thebibliography}\endgroup

\end{document}